# Surface Screening Mechanisms in Ferroelectric Thin Films and its Effect on Polarization Dynamics and Domain Structures

Sergei V. Kalinin,[1] Yunseok Kim,[2] Dillon Fong,[3] and Anna Morozovska[4]

[1] The Center for Nanophase Materials Sciences, Oak Ridge National Laboratory, Oak Ridge, TN 37922

[2] School of Advanced Materials Science and Engineering, Sungkyunkwan University (SKKU), Suwon 16419, Republic of Korea

[3] Materials Science Division, Argonne National Laboratory, 9700 S. Cass Avenue, Argonne, IL 60439

[4] V. Lashkaryov Institute of Semiconductor Physics, National Academy of Science of Ukraine, 41, pr. Nauki, 03028 Kiev, Ukraine

**Outline:**

I. Introduction

II. Theory and macroscopic studies of polarization screening on surfaces

    II.1. Internal and external screening on surfaces

    II.2. Theory of surface screening

        II.2.1 Surface screening effects on phase stability

        II.2.2 Role of screening on the structure of wall-surface junction

        II.2.3. Role of surface screening on wall dynamics

    II.3. Coupling between physics and electrochemical behaviors

    II.4. Experimental macroscopic studies of polarization screening at surfaces

III. Local studies of polarization screening

    III.1. SPM studies of domain structures

    III.2. Static studies of work function

    III.3. Domain dynamics under constant conditions

    III.4. Lateral fields

        III.4.1. Lateral switching

        III.4.2. Lateral charge injection

        III.4.3. Charge dynamics on ferroelectric surfaces

    III.5. Variable temperature experiments

        III.5.1. Surface potential evolution on heating

        III.5.2. Temperature induced domain potential inversion

        III.5.3. Thermodynamics of screening

    III.6. Environmental effects

    III.7. Interplay between screening and size effects in ferroelectric nanostructures

IV. Tip induced switching

    IV.1. Charge injection on non-ferroelectric surfaces

    IV.2. Potential evolution during switching

    IV.3. Backswitching

    IV.4. Charge collection phenomena

**Abstract:**

For over 70 years, ferroelectric materials have been remaining one of the central research topics for condensed matter physics and material science, the interest driven both by fundamental science and applications. However, ferroelectric surfaces, the key component of ferroelectric films and nanostructures, still present a significant theoretical and even conceptual challenge. Indeed, stability of ferroelectric phase *per se* necessitates screening of polarization charge. At surfaces, this can lead to coupling between ferroelectric and semiconducting properties of material, or with surface (electro) chemistry, going well beyond classical models applicable for ferroelectric interfaces. In this review, we summarize recent studies of surface screening phenomena in ferroelectrics. We provide a brief overview of the historical understanding of physics of ferroelectric surfaces, and existing theoretical models that both introduce screening mechanisms and explore the relationship between screening and relevant aspects of ferroelectric functionalities starting from phase stability itself. Given that majority of ferroelectrics exist in multiple-domain states, we focus on local studies of screening phenomena using scanning probe microscopy techniques. We discuss recent studies of static and dynamic phenomena on ferroelectric surfaces, as well as phenomena observed under lateral transport, light, chemical, and pressure stimuli. We also note that the need for ionic screening renders polarization switching a coupled physical-electrochemical process, and discuss the non-trivial phenomena such as chaotic behavior during domain switching that stem from this.

## I. Introduction

For over 70 years, ferroelectric materials have been remaining one of the central research topics for condensed matter physics and material science, the interest driven both by fundamental science and applications.[1-3] Throughout this time, the focus of research has shifted from bulk crystals and ceramics to thin films and nanostructures, both as perspective materials for information technology and data storage applications and the playground of interesting physics. Many aspects of ferroelectric behavior including soft mode behavior,[4-7] multiferroic couplings,[8-12] as well as new functionalities emerging at topological defects[13-23] are now being extensively studied, the advances made possible by advances in instrumental characterization techniques such as focused X-ray[24-27] and neutron scattering and electron microscopy,[28-34] and progress in atomistic theory.[35-38] However, ferroelectric surfaces, the key component of ferroelectric films and nanostructures, still present a significant theoretical and even conceptual challenge.

Indeed, since the early days of ferroelectricity it has been recognized that the discontinuities of polarization at surfaces and interfaces create a bound charge, as a consequence of fundamental Maxwell electrostatics. The latter gives rise to a depolarization field opposite to the polarization direction. To reduce the energy of the depolarization field and stabilize the ferroelectric film, either 180° domain structures with antiparallel polarization stripes, or the surface compensating free charges to the polarization bound charges should be created. In the absence of the charge compensating process, the ferroelectric phase is absolutely unstable.[1] However, the nature of these screening charges was traditionally excluded from consideration and analysis – rather, they were assumed to be (a) always present and (b) irrelevant to the macroscopic physics of these materials. This assumption is well justified for bulk ferroelectrics close to equilibrium, where the role of surface effects can be expected to be minor.

This postulate, however, is no longer true on the nanoscale, when the free energies of surface ionic and electronic screening become comparable to the bulk free energy of the ferroelectric. In fact, both a range of highly unusual phenomena ranging from hot electron[39-42] and X-ray[43] emission and fusion[44] to ferroelectric states observed on the nanoscale[45] can be traced to this surface charge dynamics. Furthermore, any processes involving polarization switching in system with boundary conditions other than ideal metal electrodes (which provide unlimited source of screening charges spatially coinciding with polarization bound charges) will now be ruled by the interplay between the polarization and screening dynamics, leading to coupling between ferroelectric and charge (electronic or ionic) transport behaviors, as necessitated by charge and mass conservation laws. In other words, polarization cannot be switched unless screening charges redistribute. These considerations in turn necessitate studies of the thermodynamic and kinetic behavior of the screening charges, and their role on ferroelectric functionalities in thin films.

In this review, we summarize recent studies of surface screening phenomena in ferroelectrics. We provide a brief overview of the historical understanding of phenomena on ferroelectric surfaces, and existing theoretical models that both introduce screening mechanisms and explore the relationship between screening and relevant aspects of ferroelectric functionalities starting from phase stability per se. Given that majority of ferroelectrics exist in multiple-domain states, in this review we focus on local studies of screening phenomena using scanning probe microscopy techniques. We discuss recent studies of static and dynamic phenomena on ferroelectric surfaces, as well as phenomena observed under lateral transport, light, chemical, and pressure stimuli. We also note that the need for ionic screening renders polarization switching a coupled physical-electrochemical process, and discuss the non-trivial phenomena such as chaotic behavior during domain switching that stem from this.

In the light of growing interest to nanoscale ferroelectrics and ferroelectric surfaces, as well as recent advances in theoretical methods that can capture surface behaviors of ferroelectrics, we believe that this review will be of interest to a broad range of condensed matter physicists, theorists, and materials scientist.  We further believe that  many of these mechanisms will be applicable to other polar oxide surfaces such as celebrated LAO-STO system,[46, 47] and can be directly coupled to observed electrochemical behaviors[48-52] and polar responses.[52, 53]

**II. Theory and macroscopic of polarization screening on surfaces**
In this section, we provide general overview of the theoretical aspects of the surface screening on ferroelectrics under band bending and electrochemical control, and results of the macroscopic surface science studies of surface chemistry of ferroelectrics.

**II.1. Internal and external screening on surfaces**

The fundamental aspect of ferroelectric materials is the presence of surface and interface bound charge due to the discontinuity of polarization fields. Simple estimates illustrate that if this charge is uncompensated, it will provide bulk-like contribution to free energy of materials, *i.e.* corresponding surface energy diverges with system size. These electrostatic considerations were well recognized since the early days of ferroelectricity, and necessitated the search for mechanisms for lowering of this depolarization energy. One such mechanism is formation of ferroelectric domains, extensively analyzed in classical textbooks[54] and further expanded recently.[55, 56] However, even in the presence of domain structures additional lowering of depolarization energy can be achieved if polarization charge is compensated, or screened, by external charges, were it mobile carriers in metallic electrodes or charged ionic species adsorbed from atmosphere. The energy gain in screening process can be roughly estimated based on the physical separation between polarization and screening charges from the energy of capacitor with fixed charge density. While this separation is comparable to system size for bulk uncompensated ferroelectrics, it is comparable to domain periodicity for domain structures. For screened systems, it is proportional to physical separation between average location of polarization (within ferroelectrics) and screening (outside ferroelectrics for adsorption and metallic electrode, potentially inside ferroelectrics for vacancy screening) charges.

Practically established state of the system is determined both by the kinetics and thermodynamics of available screening mechanisms. Consider the rapid cooling of ferroelectric materials below Curie temperature in the medium with low concentration of mobile charges (*e.g.* in vacuum). In this case, domain formation is the fastest energy lowering process available to the system, yielding domain structure with periodicities governed by Kittell law. The slower screening by mobile surface charges will additionally lower the system energy; however, reconfiguration of domain systems is hindered by high kinetic barriers to nucleate new domains. In comparison, very slow cooling of ferroelectric material in the presence of abundant electronic or ionic screening charges can lead to single domain states, or large periodicity domain systems (since effective polarization charge density and, consequently, depolarization energy are now much smaller). It is further important to note that relative preference of screening process is determined by the medium. In the presence of (inert) conductive metallic electrodes, the screening will be by electrons and will be associated with relatively large physical separation between screening and polarization charges that can be represented by dielectric (dead) layer effect. Screening by adsorbed ions on free surfaces can be more favorable energetically. The important aspect of screening process is the fact that crystal lattice of *e.g.* oxide ferroelectrics can change composition, *e.g.* develops oxygen non-stoichiometry. The vacancies can effectively screen polarization charge.[57, 58] Similarly, oxidation can happen on metal-ferroelectric interfaces.[59]

For future discussion, we will distinguish the external and internal screening. The former is defined as charge compensation phenomena outside the physical boundary of ferroelectrics for a given surface termination, and may include screening by conductive electrons in metallic electrodes, adsorbed charged species, or mobile ionic and ionic species in second (external) component. In the ideal case of external screening, the polarization distribution in the ferroelectrics is that of ideal material, *i.e.* polarization is uniform and changes abruptly at the interface. The internal screening can be defined as that by electrons or oxygen or cationic vacancies within the ferroelectric material, with associated changes in polarization, carrier, and ionic species concentration profiles in the vicinity of the surface. Practically, both scenarios can be realized simultaneously, with the relative contributions controlled both by the thermodynamics and kinetics of respective screening processes.

## II.2. Theory of surface screening

Surface screening of spontaneous polarization in ferroelectrics is typically realized by the mobile charges adsorbed from the ambient in the case of high or normal humidity.[60-64] In a specific case of the very weak screening, or its artificial absence due to the experimental conditions (cleaned surface in dry atmosphere, ultra-high vacuum or thick dielectric layer at the surface) the screening charges can be localized at surface states caused by the strong band-bending by depolarization field.[65-69] For both aforementioned cases the screening charges are at least partially free (mobile), in the sense that the spatial distribution of their quasi two-dimensional density is ruled by the polarization distribution near the surface. When the screening charges follow the polarization changes almost immediately, the screening density can be calculated theoretically in the adiabatic approximation.[66, 68] In the opposite case, one should solve dynamic relaxation type equations for the spatial-temporal distribution of screening charges.[68] Due the long-range nature of the depolarization effects, the incomplete surface screening of ferroelectric polarization strongly influences the domain structure and leads to its pronounced changes both near and relatively far from the surface, and affects phenomena such as domain wall pinning mechanisms at surfaces and interfaces, as well as nucleation dynamics, domain shape and period control in thin film under the open-circuited conditions,[70, 71] in films placed between imperfect "real" electrodes with finite Tomas-Fermi screening length[72] or separated from the electrodes by ultra-thin dead layers[73] or physical gaps,[74] leads to the formation of closure domains near free surfaces in multiaxial ferroelectrics,[70, 75, 76] domain wall broadening in both uniaxial and multiaxial ferroelectrics[77, 78] and crossover between different screening regimes of the moving domain wall - surface junctions.[79, 80]

**II.2.1 Surface screening effects on phase stability**

Dead (or screening) layers [figure II.1(a, c)] are responsible for the imperfect screening of the film ferroelectric polarization that in turn leads to the emergence of multidomain ferroelectric phase. Typically the phase stability region is between the paraelectric and homogeneous ferroelectric phase [figure II.1(b, d)]. At that the critical thickness $l_{cr}$ of the size-induced phase transition into a paraelectric phase can vary in a wide range, from nanometers to micrometers, depending on the geometry (gap or dead layer thickness), ferroelectric material parameters, temperature and screening charges concentration [compare figure II.1(b) and II.1(d)]. For the case of dead layers (or gaps) presence [shown in figure II.1(b)] the line marked $l_d$ delineates the paraelectric and ferroelectric domain phases, while the one marked $l_{ms}$ indicates the boundary of the metastability regions of the monodomain state. The unusual "re-entrant" type shape of monodomain ferroelectric-paraelectric phase transition boundary [shown in figure II.1(c) for the open-circuited boundary conditions) originated from the temperature dependence of the screening length. Further the transition to poly-domain state occurs with either temperature or film thickness decrease. Note that the approximation $l_{cr} \gg \rho k$ (k is the screening length) is rather rigorous for the open-circuit conditions.

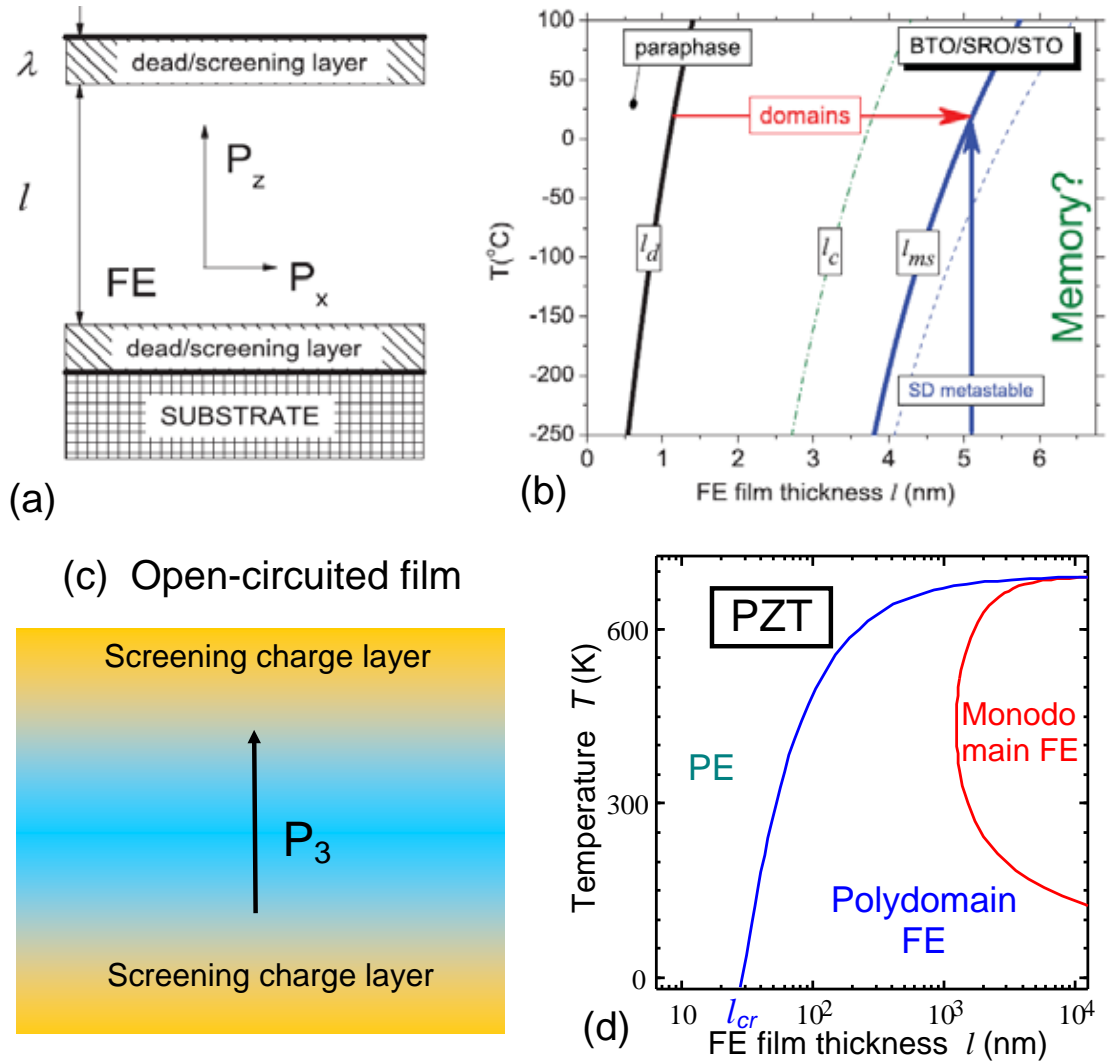

**Figure II.1.** (a) Schematic of the ferroelectric film with thickness $l$ sandwiched between metal electrodes with screening length $\lambda$ on a misfit substrate. The misfit between the film and substrate makes the film a uniaxial ferroelectric with a spontaneous polarization along the $z$ axis. (b) The phase diagram for BaTiO$_3$/SrRuO$_3$/SrTiO$_3$ films in the coordinates "temperature – thickness". Panels (a) and (b) are reprinted with permission from [Bratkovsky A M and Levanyuk A P 2011 *Phys. Rev. B* **84** 045401]. Copyright (2011) by The American Physical Society [73]. (c) Open-circuited film without electrodes and with diffuse screening charge layer in the vicinity of film surfaces. (d) Phase diagram in coordinates "film thickness – temperature" calculated for open-circuited PbZr$_{40}$Ti$_{60}$O$_3$ (PZT) film (without electrodes). Panels (c) and (d) are reprinted with permission from [Eliseev E A, Kalinin S V and Morozovska A N 2015 *J. Appl. Phys.* **117** 034102]. Copyright 2015, AIP Publishing LLC[71].

## II.2.2 Role of screening on the structure of wall-surface junction

Among the effects considered in Refs.[68, 70-81], the domain wall broadening seems universal and especially illustrative for the physical understanding of the role of surface screening charges on ferroelectric structures.[77] The broadening occurs because bound polarization and surface screening charges form an electric double layer, and the breaking of this layer by the domain wall induces stray depolarization field, which in turn changes the domain wall structure. Power law decay of the stray field results in the power law of polarization saturation near the surface, as compared to exponential saturation in the bulk. The relative wall broadening is the strongest in weak uniaxial ferroelectrics and the most noticeable near Curie temperature and does not fully relax under the temperature decrease allowing for wall pinning effects.

Below we analyze the typical model situations of the surface screening charge either separated from the ferroelectric polarization break by an ultra-thin dead layer or physical gap of thickness $H$, as well as the case of strong polarization gradient at the surface [see figure II.2].

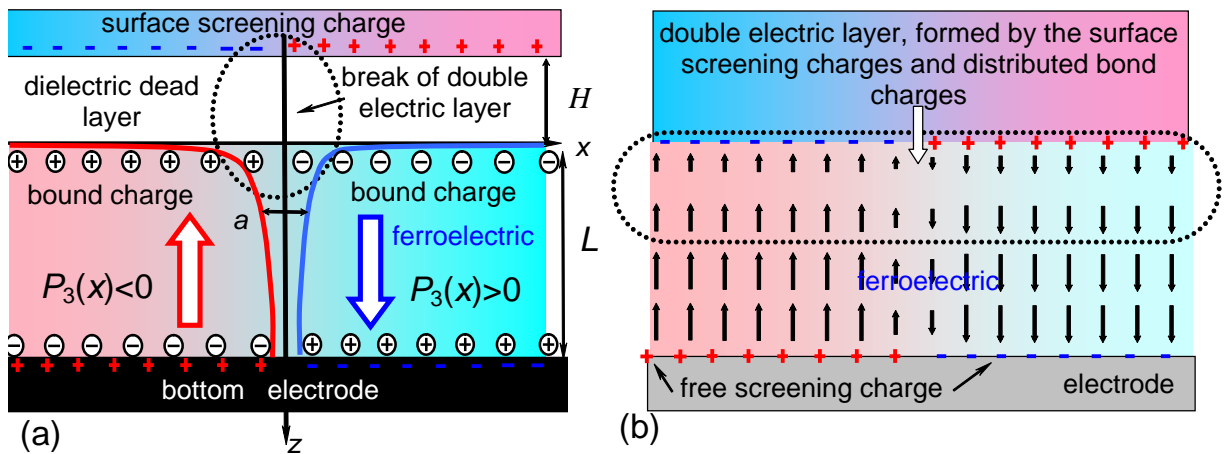

**Figure II.2.** 180°-domain wall structure near the film surface. The double electric layer is formed due to either the physical dead layer (a) or intrinsic surface effect leading to diminished polarization at interface (b). Discontinuity of the double electric layer of screening and bound changes results in an additional stray depolarization field. Reprinted from [Eliseev E A, Morozovska A N, Kalinin S V, Li Y, Shen J, Glinchuk M D, Chen L Q and Gopalan V 2009 *J. Appl. Phys.* **106** 084102]. Copyright 2009, AIP Publishing LLC [78].

Within the framework of LGD theory spatial distribution of the ferroelectric polarization component $P_3$ in the uniaxial ferroelectric obeys nonlinear time-dependent equation:

$$\Gamma\frac{\partial P_3}{\partial t} + \left(\alpha(T) - 2q_{ij33}u_{ij}\right)P_3 + b_u P_3^3 + d P_3^5 - \zeta\frac{\partial^2 P_3}{\partial x_3^2} - \eta\left(\frac{\partial^2 P_3}{\partial x_2^2} + \frac{\partial^2 P_3}{\partial x_1^2}\right) = E_3^d \ . \quad \text{(II.1)}$$

Kinetic coefficient $\Gamma$ is defined by phonon relaxation and hence the Landau-Khalatnikov time is small $\tau_K = \Gamma/|a_3|$ far from immediate vicinity of the ferroelectric phase transition (where the critical slowing down can occur). Coefficient $\alpha(T) = \alpha_T(T - T_C)$ explicitly depends on temperature $T$, gradient terms coefficients are $\zeta > 0$ and $\eta > 0$, expansion coefficients $d > 0$, while $b_u < 0$ for the first order phase transitions or $b_u > 0$ for the second order ones. The spontaneous polarization $P_S$ of mechanically free bulk system is determined from the equation $\alpha(T) + b P_S^2 + d P_S^4 = 0$. However, it was shown that inhomogeneous stress always exists in the vicinity of domain wall due to the electrostriction coupling. Thus $u_{jk}$ are the strain tensor components, $q_{ijkl}$ and $c_{ijkl}$ are the components of electrostriction and elastic stiffness tensor correspondingly related with the elastic stress tensor components $\sigma_{ij}$ via equation of state $c_{ijkl}u_{lk} - q_{ij33}P_3^2 = \sigma_{ij}$. These equations should be supplemented the conditions of mechanical equilibrium,[82, 83] $\partial\sigma_{ij}/\partial x_i = 0$, and compatibility relations for elastic field, $e_{ikl}e_{jmn}(\partial^2 u_{ln}/\partial x_k \partial x_m) = 0$, where $e_{ikl}$ is the permutation symbol or anti-symmetric Levi-Civita tensor. The homogeneous strain causes the renormalization of the expansion coefficient

$$b_u = b + 4\frac{(q_{11} - q_{12})^2}{3(c_{11} - c_{12})} + 2\frac{(q_{11} + 2q_{12})^2}{3(c_{11} + 2c_{12})}. \quad \text{(II.2)}$$

The boundary conditions for polarization in Eq.(1) have the form:[84]

$$\left(P_3 - \lambda_1 \frac{\partial P_3}{\partial x_3}\right)\bigg|_{x_3=0} = 0, \quad \left(P_3 + \lambda_2 \frac{\partial P_3}{\partial x_3}\right)\bigg|_{x_3=L} = 0. \quad \text{(II.3)}$$

Extrapolation length $\lambda_{1,2}$ may be different for $x_3=0$ and $x_3=L$. Reported experimental values are $\lambda = 2 - 50$nm.[29, 85] In the right-hand-side of equation (II.1) stands the depolarization field $E_3^d$, caused by imperfect screening and inhomogeneous polarization distribution. The field is determined from Maxwell equations.

Assuming, that all uncompensated polarization bond charges are localized in thin near-surface layer (i.e. $\partial P_3(x, z > 0)/\partial z \approx 0$), and the stray field and polarization distributions have the form:

$$E_3(x,z) = -\frac{(P_S/\varepsilon_0)}{\varepsilon_{33}^b + \gamma \varepsilon_g} \frac{2}{\pi}\left(\arctan\left(\frac{x}{z/\gamma}\right) - \arctan\left(\frac{x}{z/\gamma + 2H}\right)\right), \quad \text{(II.4a)}$$

$$P_3(x,z) \approx P_S\left(\text{sign}(x) - \frac{\varepsilon_{33}^f - \varepsilon_{33}^b}{\varepsilon_{33}^f + \gamma \varepsilon_g} \frac{4\pi xH}{(\pi|x| + 2z/\gamma)(\pi|x| + 2(z/\gamma + 2H))}\right). \quad \text{(II.4b)}$$

Here $\varepsilon_{33}^f(T) = \varepsilon_{33}^b + 1/(\varepsilon_0(\alpha(T) + 3\beta P_S^2 + 5\delta P_S^4))$ is the dielectric permittivity of ferroelectric. The dielectric anisotropy factor $\gamma = \sqrt{\varepsilon_{33}^b/\varepsilon_{11}}$; $\varepsilon_{33}^b$ is the dielectric permittivity of the background or reference state[86] (typically $\varepsilon_{33}^b \leq 10$); $\varepsilon_0$ is the dielectric constant, $\varepsilon_g$ is the isotropic dielectric permittivity of dead layer with thickness $H$. Ferroelectric film thickness $L$ is much higher than $H$ [see figure II.2(a)].

From equations (II.4), the emerging depolarization field induces polarization saturating as slow as $\sim 1/x$ even for abrupt initial domain wall. Since the factor of order $\sim P_S/(\varepsilon_0 \varepsilon_{33}^f)$ is much higher than thermodynamic coercive field, the stray field (equation (II.4(a))) would influence the polarization distribution at distances $z$ much higher than dead layer thickness $H$. In fact, for $z \gg H$ the length $H$ is no longer the characteristic scale of polarization (equation (II.4(b))) and depolarization field (equation (II.4(a))), but rather determines the amplitude.

The typical view of the domain wall broadening induced by the incomplete surface screening via dead layer is shown in figure II.3(a). Coordinate x in the figure is measured in the correlation length units, where the correlation length $R_\wedge$ is given by expression:

$$R_\wedge(T) = \sqrt{\frac{\eta}{\alpha(T) + (3\beta + 4q^2)P_S^2 + 5\delta P_S^4}}, \quad q^2 \circ \frac{2(q_{11} - q_{12})^2}{3(c_{11} - c_{12})} + \frac{(q_{11} + 2q_{12})^2}{3(c_{11} + 2c_{12})} - \frac{q_{12}^2}{c_{11}}. \quad \text{(II.5)}$$

Discontinuity of the double electric layer of screening and bound changes results in an additional stray depolarization field [see figure II.3(b)]. For the case the domain wall broadening is incorporated in the polarization profile as [75]:

$$P_3(x,z) \approx \left(\begin{array}{c}\left(1 - \dfrac{\exp\left(-z/\sqrt{\varepsilon_{33}^b \varepsilon_0 z}\right)}{1 + 1/\sqrt{\varepsilon_{33}^b \varepsilon_0 z}}\right) P_S \tanh\left(\dfrac{x}{2R_\wedge}\right) - \\ \dfrac{d}{4R_\wedge}\sqrt{\dfrac{\varepsilon_{11}}{\varepsilon_{33}^f}} \ln\left(\dfrac{(2R_\wedge + x)^2 + (\varepsilon_{11}/\varepsilon_{33}^f)(z+d)^2}{(2R_\wedge - x)^2 + (\varepsilon_{11}/\varepsilon_{33}^f)(z+d)^2}\right) P_S\end{array}\right) \quad \text{(II.6)}$$

where the distance $d = z/\left((\alpha + (3\beta + 4q^2)P_S^2)\left(\sqrt{\varepsilon_{33}^b \varepsilon_0 z} + 1\right)\right)$ corresponds to the apparent separation between the surface screening charge and bulk polarization value (i.e. "gradient scale"). Hence, the distance $d$ plays the role of effective thickness of screening double layer. At distances $|x| \gg R_\wedge$ the second logarithmic term in equation (II.6) behaves as $2x/\left(x^2 + (\varepsilon_{11}/\varepsilon_{33}^f)(z+d)^2\right)$ after corresponding series expansion, defining the distribution of stray depolarization field far from the break of double electric layer. Thus the second term in equation (II.6) proves the power law of the of domain wall profile saturation near the surface, which is much slower compared to exponential saturation of bulk profile $P_S \tanh(x/2R_\wedge)$ [see figure II.3(b)].

The domain wall width calculated analytically as a function of its depth from the surface of uniaxial ferroelectric LiTaO$_3$ is shown in figure II.3(c) (solid curves) in comparison with experimental data in 500 nm thick stoichiometric LiTaO$_3$ (squires, SLT)[87] and 50nm thick congruent LiTaO$_3$ (triangles, CLT)[88]. Note that the domain wall profiles are strongly asymmetric, namely at the surface $z=0$ the width is several times bigger than saturated "bulk" value near the surface $z=L$. Dotted curves in figure II.3(c) are numerical calculations by phase field method.[89] Thus the figure II.3(c) demonstrates that the double electric layers formed by the bound and surface screening charges can be responsible for the domain broadening effect. Considering the alternative explanation of the domain walls broadening near the surface in LiTaO$_3$ due to their interaction with defects having larger density near the surface, we note the following. Daimon and Cho[29, 85] fabricated and examined two types of LiTaO$_3$ films: congruent (with numerous defects) and stoichiometric (almost without defects). Results for both types of samples demonstrated domain walls surface broadening, proving that the effect is defect-independent.

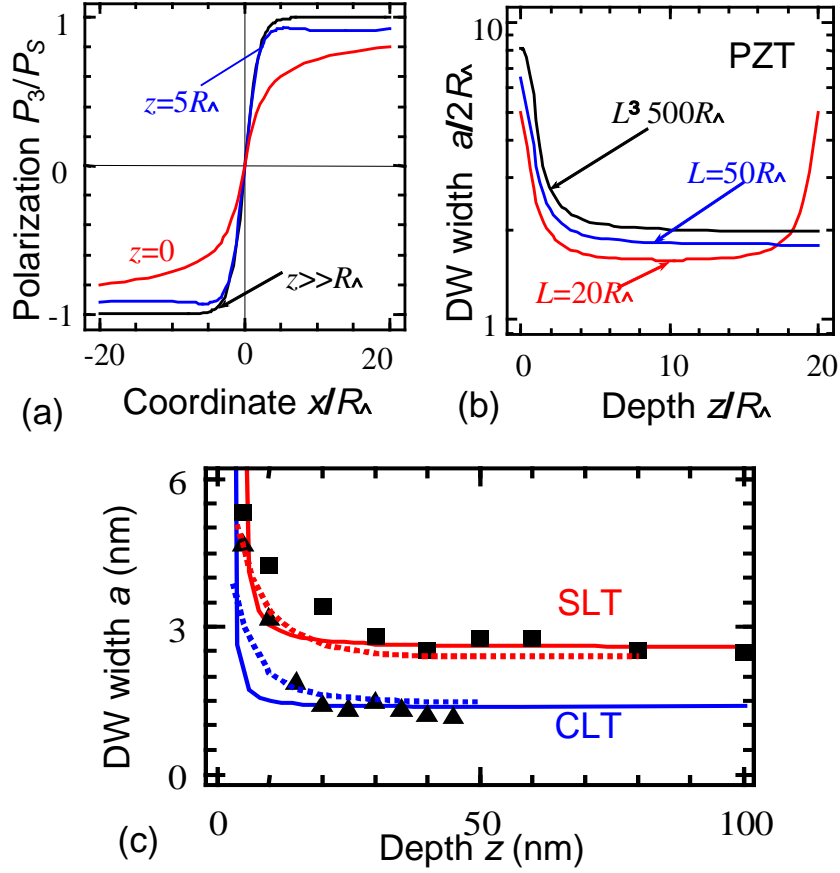

**Figure II.3.** (a) Normalized domain wall profile $P_3(x,z)/P_S$ calculated at fixed temperature and extrapolation length $\lambda = 0.5R_\wedge$. Different curves correspond to different distances $z/R_\wedge = 0, 5, \infty$ from the surface of thick ferroelectric film. (b) The domain wall width $a/2R_\wedge$ calculated at level 0.76 as a function of depth $z/R_\wedge$ from the surface, extrapolation length $\lambda = 0.5R_\wedge$. Material parameters are for PbZr$_{0.5}$Ti$_{0.5}$O$_3$ (PZT). (c) Thickness of domain wall at level 0.76 as a function of its depth from the surface of LiTaO$_3$. Squires are experimental data from Refs.[29, 85] for 500nm thick stoichiometric LiTaO$_3$ (SLT), triangles correspond to 50nm thick congruent LiTaO$_3$ (CLT). Solid curves are analytical calculations for fitting parameters $R_\wedge = 1.3$ nm and $\lambda_1 = 0.1$ nm for SLT; while $R_\wedge = 0.7$ nm and $\lambda_1 = 0.1$ nm for CLT, $\lambda_2 \gg 30$nm. Corresponding dotted curves are numerical calculations by phase field modelling for the same fitting parameters. Reprinted from [Eliseev E A, Morozovska A N, Kalinin S V, Li Y, Shen J, Glinchuk M D, Chen L Q and Gopalan V 2009 *J. Appl. Phys.* **106** 084102]. Copyright 2009, AIP Publishing LLC [78].

## II.2.3. Role of surface screening on wall dynamics

We further discuss the influence of surface screening charge relaxation on the dynamics of the domain walls near the surface. Eliseev et al[77] consider the planar 180º-domain wall uniformly moving under homogeneous external field in ferroelectric capacitor with a special attention to the dynamics of depolarization electric field induced by the bound charge at the wall-surface junction near the dead layer. The wall dynamics is analyzed within the full Ginzburg-Landau-Devonshire model, thus providing insight into mesoscopic structure of moving domain wall.

The schematic representation of ferroelectric capacitor with natural or artificially deposited dielectric dead layer and surface screening charge located at the interface between the ferroelectric layer and the dead layer is shown in figure II.4. The structure consists of conducting top and bottom electrodes; wide-band-gap semi-conductor ferroelectric film (*f*) of thickness *L* with dielectric permittivity tensor $\varepsilon_{ij}^f$ [90] and interfacial surface screening charge layer that density $\sigma(x,t)$ depends on time *t* and coordinate x; dielectric dead layer of thickness *H* with isotropic dielectric permittivity $\varepsilon_g$. Note, that due to the effects of "reflections" in bottom electrode this asymmetric system is equivalent to symmetric capacitor with two dead and screening charge layers and thickness of ferroelectric doubled.[24] The spontaneous polarization vector $\mathbf{P}_0 = (0,0,P_0(x,t))$ is pointed either along or opposite the polar axis *z* and depending on coordinate *x* and time *t* allowing for the domain wall motion, at that $\text{div}\mathbf{P}_0 = 0$ inside a ferroelectric layer. Quasi-stationary Maxwell equations should be satisfied for each layer. Inside the dead layer and outside the screening layer electric potential $\varphi$ satisfies the Laplace's equation. The boundary conditions of fixed top and bottom electrode potentials and continuous potential and normal component (*n*) of displacement on the boundaries between dead layer (d) and ferroelectric (f) are imposed:

$$\varphi(x,z=-H)=0, \quad \varphi(x,z=+0)=\varphi(x,z=-0), \quad \varphi(x,z=L)=U, \quad D_{fn} - D_{dn} = \sigma(x,t) \quad \text{(II.7)}$$

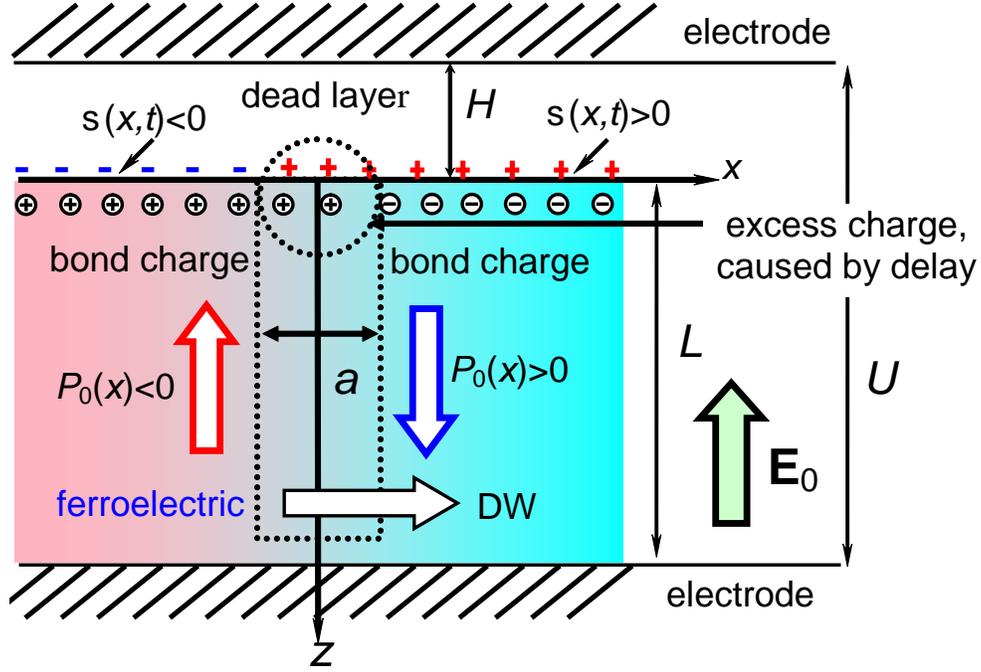

**Figure II.4.** $\mathbf{P}_0(x,z,t)$ is spontaneous polarization, $E_f$ is electric field *inside the ferroelectric*, voltage *U* is applied between the electrodes. Screening charge layer, originated from bend bending, is localized inside ferroelectric near the dead layer boundary. Dotted line indicates the moving boundary of 180°-domain wall (*DW*). The normal vector n is pointed from media 1 to media 2. Reprinted with permission from [Eliseev E A, Morozovska A N, Svechnikov G S, Rumyantsev E L, Shishkin E I, Shur V Y and Kalinin S V 2008 *Phys. Rev. B* **78** 245409]. Copyright (2008) by The American Physical Society [80].

Let us consider the typical case of the uniformly moving 180-degree domain wall with the spatially invariant shape, $P_0(x,t) \approx P_0(x-vt)$. The unknown domain wall velocity, *v*, should be found self-consistently. This approximation is justified given that shape fluctuations in the z-directions are associated with significant depolarization fields.[91]

The relaxation equation for effective surface charge density $\sigma$ is [77, 80]:

$$\tau \frac{\partial \sigma(x,t)}{\partial t} = P_0(x-vt) - \sigma(x,t) \qquad (II.8)$$

where the relaxation time $\tau = \varepsilon_0 \varepsilon_{11}^f / \rho$, and $\rho$ is the electric conductivity of the ultra-thin screening layer. Under the condition of full screening $\sigma(x,0) = P_0(x)$ at the initial moment of time *t*=0, the solution of equation (II.8) is

$$\sigma(x,t) = P_0(x-vt) + v\int_0^t dt' \frac{\partial P_0(x-vt')}{\partial x} \exp\left(-\frac{t-t'}{\tau_m}\right). \tag{II.9}$$

Hereinafter the following approximations for the polarization distribution within the domains wall will be used

$$P_0(x) = P_S \begin{cases} 1 - \exp(-x/a), & x > 0; \\ -1 + \exp(x/a), & x < 0. \end{cases} \tag{II.10}$$

Here $a$ is the effective half-width of the domain wall, e.g. 0.5-2 nm. Here, $P_S$ is spontaneous polarization.

Substitution of equation (II.10) into equation (II.9) admits exact analytical expression for the surface charge density (see equation 12 in [77, 80]). Under the condition $t \gg \tau$, i.e. in the stationary regime, the approximate expression for the surface charge density is:

$$\sigma(x,t) \approx P_0(x-vt) + P_S \begin{cases} \dfrac{v\tau}{a+v\tau} \exp\left(-\dfrac{x-vt}{a}\right), & x > vt; \\ \dfrac{v\tau}{a-v\tau}\left(\exp\left(\dfrac{x-vt}{a}\right) - \dfrac{2v\tau}{a+v\tau}\exp\left(\dfrac{x-vt}{v\tau}\right)\right), & x < vt. \end{cases} \tag{II.11}$$

Excess charge density, defined as the difference of surface screening and bond charge densities, $\delta\sigma(x,t) = \sigma(x,t) - P_0(x-vt)$, is caused by the delay in screening. For the very slow moving wall ($\tau v \to 0$) excess charge is absent.

The distinctive feature is the fact that the maximum of surface screening charge is located behind the moving wall at finite width, $a$, and exactly at the wall in the case $a \to 0$. Hence, the surface screening charge effect on domain wall dynamics can be described only in the case of finite wall width (see figure II.5(a-c)). Per equation II.11, charge density $\delta\sigma(x,t)$ depends only on $x-vt$, i.e. describes the charge wave accompanying moving domain wall at times $t \gg \tau$. In the limiting case of ultra-thin (or rapidly moving) domain wall, $(a/v\tau) \ll 1$, approximation (equation II.11) works with high accuracy at distances $x > vt - a$ (i.e. in front of the domain wall) even starting from the small times $t \ll \tau$. Reasonable estimations for the domain wall intrinsic width $a = 5 - 0.5$ nm, velocity $v = 10^{-6} - 10^{-3}$ m/s and relaxation time $\tau = 10^{-3} - 1$ s lead to the interval $(a/v\tau) = 1 - 10^{-6}$, justifying the limit of $(a/v\tau) \ll 1$. Thus, one can use approximation (equation II.11) in the region $x > vt - a$.

The normal component of the electric field inside the ferroelectric layer has the form:

$$E_3(x,z,t) = E_{d3}(x,z,t) + E_0, \qquad E_0 = -\frac{U \varepsilon_g}{H \varepsilon_{33}^f + L \varepsilon_g}, \qquad \text{(II.12(a))}$$

$$E_{d3}(x,z,t) \approx -\int_{-\infty}^{+\infty} dk \frac{\exp(-ikx)\left(\tilde{P}_0(k,t) - \tilde{s}(k,t)\right)\tanh(kH)\cosh(k(L-z)/\gamma)}{\varepsilon_0 \sqrt{2\pi}\; \varepsilon_{33}^f \cosh(kL/\gamma)\tanh(kH) + \gamma \varepsilon_g \sinh(kL/\gamma)}. \qquad \text{(II.12(b))}$$

Here $\gamma = \sqrt{\varepsilon_{33}^f/\varepsilon_{11}^f}$ is the dielectric anisotropy factor. $\tilde{P}_0(k,t)$ and $\tilde{s}(k,t)$ are Fourier images of $P_0(x,t)$ and $s(x,t)$ over coordinate $x$. The term $E_{d3}(x,z,t)$ in equation (II.12(a)) is the internal electric field produced by the wall and partially screened by the free charges on the top electrode. The screening is realized by the surface screening charges $s(x,t)$ with delay determined by finite relaxation time, $\tau$. The last term in equation (II.12(a)) is the external electric field induced by applied bias, $U$. Far from the wall $E_{d3} \rightarrow \frac{s - P_S}{\varepsilon_0}\frac{H}{H\varepsilon_{33}^f + L\varepsilon_g}$ as anticipated for the stationary case.

The $x$-distribution of the depolarization field $E_{d3}(x,z,t)$ calculated at different depths $z$ is shown in figure II.5(d). The field is maximal behind the moving wall for finite intrinsic width $a$, and achieves maximum exactly at the wall in the limiting case $a \rightarrow 0$ (corresponding to the jump at the wall). The field maximum decreases and diffuses with increasing penetration depth $z$.

The effect of depolarization field on domain velocity, $v$, can be found in two limiting cases, when the velocity of the domain wall is determined as the nucleation rate of at the wall – surface junction (i) or averaged over the wall surface (ii), which in turn is determined by the full electric field in this point. Using linear approximation the domain wall velocity dependence on the electric field is

$$n = \begin{cases} 0, & E < E_{th}, \\ \mu(E(v) - E_{th}), & E > E_{th}. \end{cases} \qquad \text{(II.13)}$$

where $\mu$ is the wall mobility and $E_{th}$ is the threshold value of electric field, $E$ is the superposition of external filed $E_0$ and depolarization field value at the domain wall surface, $E_{dm}(z) \equiv E_{d3}(x = vt, z)$. In particular $E(v) = E_0 + E_{dm}(0)$ for the case (i), while $E(v) = E_0 + \langle E_{dm}(z)\rangle$ for the case (ii). The dependence of $E_{dm}$ on the wall velocity $v$ allows two-point Pade approximation that quantitatively reproduces the behavior of exact expressions equation II.12(b) in the entire region of parameters:

$$E_{dm} \approx \frac{P_S}{\varepsilon_0(\varepsilon_{33}^f + \gamma\varepsilon_g)}\frac{vt}{(a+vt)}\left(1 + \frac{\rho\gamma\varepsilon_g}{2(\varepsilon_{33}^f + \gamma\varepsilon_g)}\left[\ln\left(\frac{a}{vt}\right)\right]^{-1}\left[\frac{a-vt}{H}\right]^{-1}\right) \quad (II.14)$$

The derived dependence of field on wall velocity is at first glance unphysical, since the length of "tail" of uncompensated charge moving after the wall increases with velocity increase. However, with the increase of tail length (about $vt$) the electric field tends to zero at $z > 0$ (similarly to the vanishing of the electric field outside the flat capacitor).[92] Thus, the damping role of the internal field at the surface $z=0$ may be negligible for rapidly moving walls. In other words, there is a crossover between (I) polarization screening by low-mobility charges $\sigma(x,t)$ and (II) polarization screening by electrode free charges with giant (in theory infinite) mobility, appeared when the sluggish charges are belated. Below, we demonstrate the "local" effect on the domain wall velocity $v$.

Obtained results are presented in figure II.5(e). At small values of external field the slope of the velocity dependence is from several times to several orders of magnitude smaller than at high fields (initial or ideal mobility). The dotted parts of curves correspond to the unstable regime when the velocity decreases with field increase. The result means possible self-acceleration of the wall near the top surface, while the effect should become negligibly small with increasing depth, z. Given that the field strength is maximal at the wall-surface junction and decreases with the depth increase (see curves 1-3 in figure II.5(e)), this depth dependence is expected to lead to the domain wall bending near the surface. This effect will be compensated by the depolarization field produced by the charge excess, and the interplay between the two will yield the equilibrium domain wall geometry.

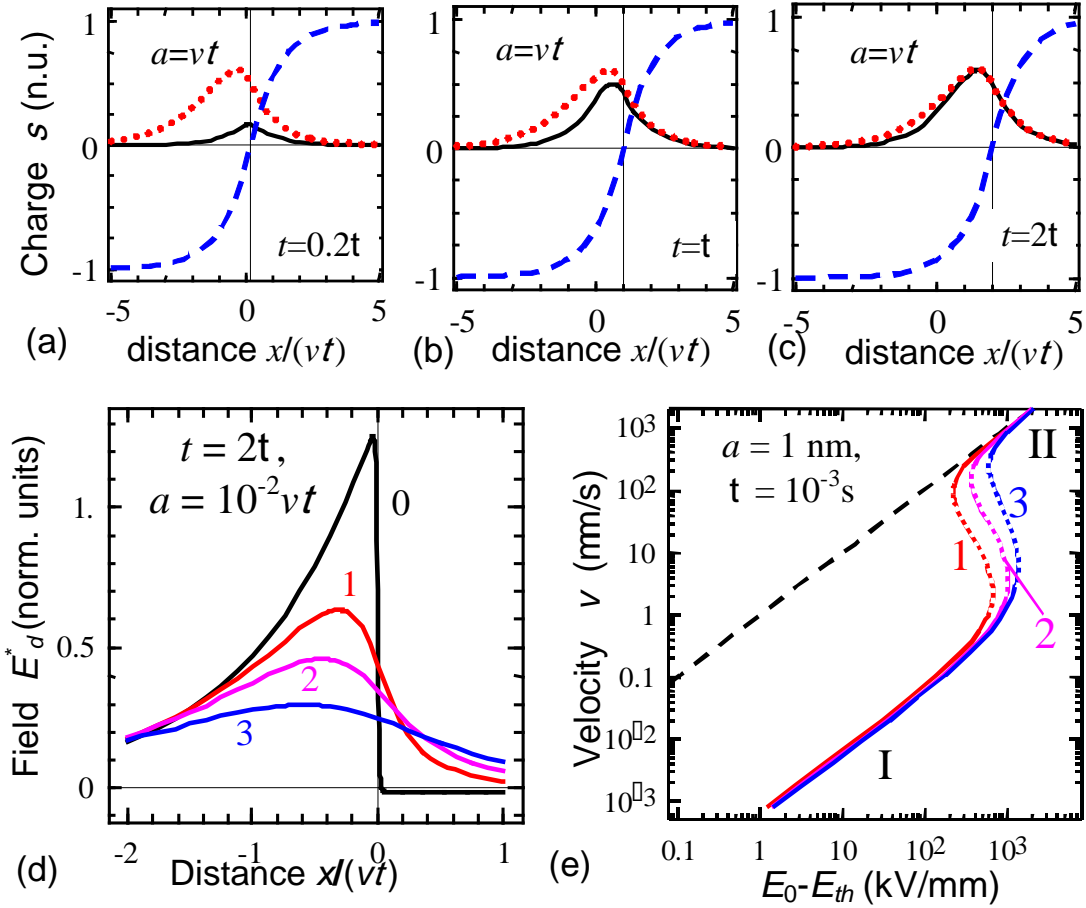

**Figure II.5.** (a-c) Solid curves are exact distribution of the excess charge $\delta\sigma(x,t)$, dotted curves are approximate distribution of $\delta\sigma(x,t)$ calculated from Eq.(11), dashed curves are the polarization distribution $P_0(x-vt)$ at different moments of time $t/\tau = 0.2, 1, 2$ (from (a) to (c)), $(a/v\tau)=1$. (d) The dependence of the normalized depolarization field at the domain wall $E_d^* = \varepsilon_0 \varepsilon_{33}^f E_{dm}/P_S$ on the distance $x/vt$ calculated for parameters $\gamma\varepsilon_g/\varepsilon_{33}^f=0.5$, $L/(\gamma v\tau)=100$ and $(a/v\tau)=10^{-2}$. Curves 0 - 3 corresponds to the different depth $z/(\gamma v\tau)=0, 0.25, 0.5, 1$. (e) Dependence of the domain wall velocity $v$ on the applied electric field calculated for mobility $\mu= 10^{-6}$ mm$^2$/(V sec), $P_S = 0.75$ C/m$^2$, $\varepsilon_{33}^f=30$, $\gamma\varepsilon_g/\varepsilon_{33}^f = 0.5$, $\tau = 10^{-3}$ sec; $a=1$ nm. Solid curves 1, 2, 3 corresponds to the $H = 1, 3, 10$ nm, dashed curve for $H = 0$ corresponds to the case without delay in screening. Roman numbers I and II designate two different screening regimes separated by the unstable region (dashed curves). Reprinted with permission from [Eliseev E A, Morozovska A N, Svechnikov G S, Rumyantsev E L, Shishkin E I, Shur V Y and Kalinin S V 2008 *Phys. Rev. B* **78** 245409]. Copyright (2008) by The American Physical Society [80].

Hence the crossover between two screening regimes: the first one corresponds to the low domain wall velocity, when the wall drags the sluggish surface screening charges, while the second regime appeared for high domain wall velocity, when the delay of sluggish screening charges are essential and the wall depolarization field is screened by the instant free charges located at the electrode. The "local" motion at the wall-surface junction can be unstable for nonzero dead layer thickness, since the internal (stray) field at the wall-surface junction decreases with domain wall velocity increase. The instability may lead to the domain wall surface bending and actual broadening in thick samples.

**II.3. Coupling between physics and electrochemical behaviors**

Mixed ionic-electronic conductors, such as solids with rechargeable ions or vacancies, which also can be mobile, free electrons and/or holes, can display a reversible dynamics of the space charge layers that leads to a pronounced resistive switching between meta-stable states with high and low resistance[93-98] and unique dynamic properties (including hysteretic) electro-mechanical response.[99-106] Though ferroelectrics with mixed type conductivity (FeMIECs) are promising candidates for the nonvolatile and memristive [107-109] memory devices, the physical principles of the charge transfer phenomena at their surfaces and interfaces are far from clear.[102]

Space charge dynamics in ferroelectric thin films were studied theoretically primarily in the framework of linear drift-diffusion Poisson-Planck-Nernst theory and diluted species approximation.[105, 107, 110, 111] However one can expect a strong correlation between the electrochemical and electromechanical response in FeMIECs, because the spatial gradient of mobile species (ions, vacancies and electrons) concentration near the surface caused by electromigration (electric field-driven for charged species) and diffusion (concentration gradient-driven for both charged and neutral species) mechanisms can change the lattice molar volume. The changes in the volume result in local electrochemical stresses, so called "Vegard stress" or "chemical pressure".[112, 113] The Vegard mechanism plays a decisive role in the origin and evolution of local strains caused by the point defect kinetics in solids.[114, 115, 116]

To understand the coupling between ferroelectric and electrochemical phenomena, one can consider the classical tetragonal ferroelectric for which the polarization components depend on the inner field $E$ as $P_1 = \varepsilon_0(\varepsilon_{11}^f - 1)E_1$ and $P_2 = \varepsilon_0(\varepsilon_{11}^f - 1)E_2$ and $P_3(\mathbf{r}, E_3) = P_3^f(\mathbf{r}, E_3) + \varepsilon_0(\varepsilon_{33}^b - 1)E_3$, where a relative background permittivity $\varepsilon_{ij}^b \leq 10$ is introduced.[90] The ferroelectric permittivity $\varepsilon_{33}^f >> \varepsilon_{33}^b$. The spatial distribution of the ferroelectric polarization $P_3^f(z)$ is given by the time-dependent relaxation type Landau-Ginzburg-Devonshire (T-LGD) equation (II.1) with the boundary conditions (II.3). The electrochemical surface charges that partially screen the film polarization is localized at the ferroelectric – dead layer interface. The problem geometry is the same as shown in figure II.2(a), with the only simplification that we consider the situation far from a domain wall [117].

The system of electrostatic equations of each of the medium (dead layer and ferroelectric film) acquires the form of Laplace equation in a dead layer and anisotropic Poisson equation in a ferroelectric. Boundary conditions (BCs) to the electrostatic equations are the equivalence of the electric potential to the tip electrode potential $U$ at the dead layer/electrode interface $z = -H$ and the equivalence of the normal component of displacements to the chemical surface charge density $\sigma(\xi)$ at $z = 0$; the continuity of the electric potential and normal component of displacements $D_3 = \varepsilon_0 E_3 + P_3$ and $D_3 = \varepsilon_0 \varepsilon_d E_3$ at dead layer / ferroelectric interface $z = 0$; and zero potential at the bottom conducting electrode $z = L$.

To quantify the coupling between ferroelectric phenomena and surface chemistry and to use a relevant model for $\sigma(\xi)$, one can be based on the approach developed by Highland and Stephenson (S&H).[118, 119] Thermodynamic theory of S&H is developed for the ferroelectric phase transition of an ultrathin film in equilibrium with a chemical environment that supplies ionic species to compensate its polarization bound charge at the surface. Equations of state and free energy expressions are developed based on Landau-Ginzburg-Devonshire theory, using electrochemical equilibria to provide ionic compensation boundary conditions. Calculations are presented for a monodomain ferroelectric film that top surface is exposed to a controlled oxygen partial pressure in equilibrium with electronically conducting bottom electrode [see schematic in figure II.6(a)]. The stability and metastability boundaries of phases of different polarization orientations ("positive" or "negative") are determined as a function of temperature and oxygen partial pressure in dependence on the film thickness [figure II.6(b)]. At temperatures below the thickness-dependent critical one ($T_{cr}(h)$), high or low oxygen partial pressure stabilizes positive or negative polarization, respectively ("chemical screening effect"). In equilibrium with an environment the chemical screening causes new peculiarities in the phase diagram of ultrathin films, namely a stable nonpolar phase can occur between the ferroelectric phases with positive and negative polarizations at fixed temperature, under the conditions of partial chemical screening (when charged surface species concentration is not sufficient to stabilize polar phase).

A synchrotron x-ray study of the equilibrium polarization structure of ultrathin $PbTiO_3$ films on $SrRuO_3$ electrodes epitaxially grown on $SrTiO_3$ substrates, as a function of temperature and the external oxygen partial pressure ($pO_2$) controlling their surface charge compensation revealed that the $T_{cr}(h)$ varies with $pO_2$ and has a minimum at the intermediate $pO_2$, where the polarization below TC changes sign. The experiments are in semi-quantitative agreement with the S&H model [figure II.6(c)].

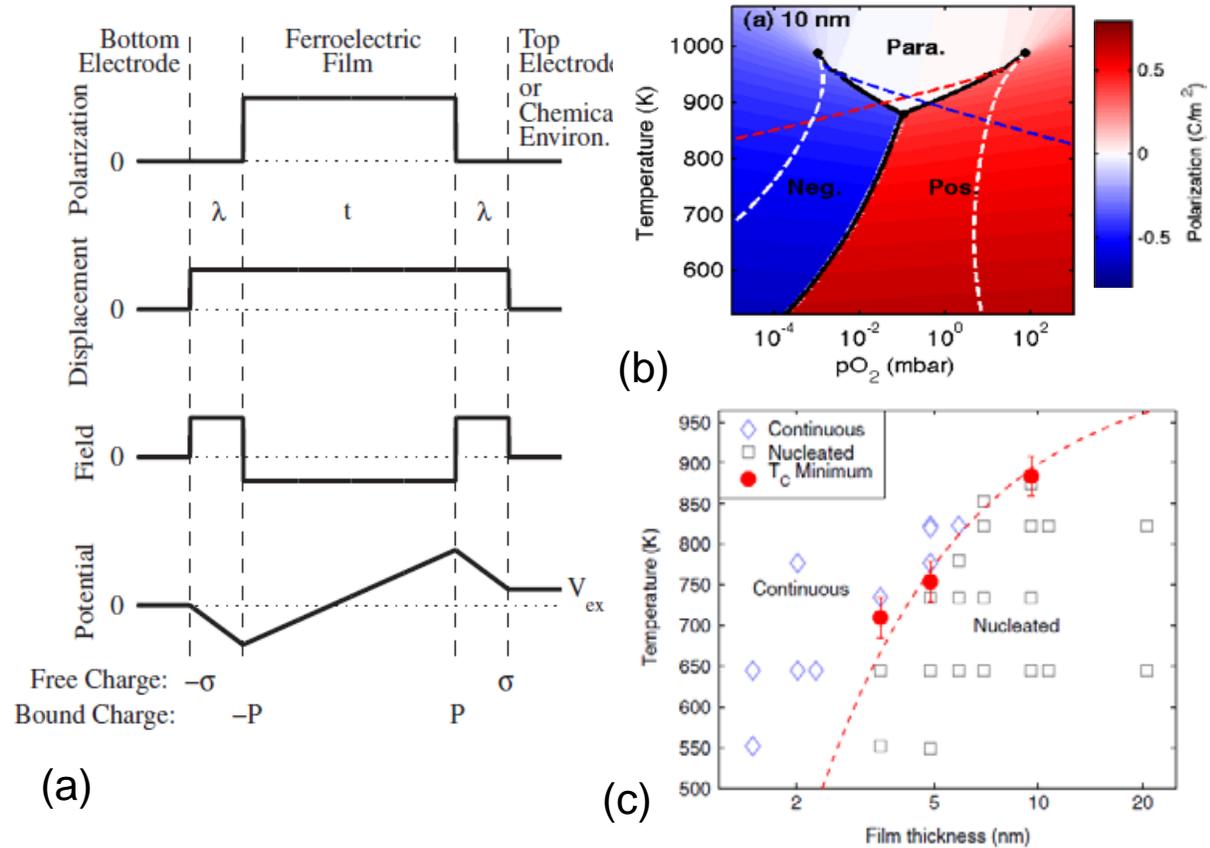

**Figure II.6.** (a) Schematic of polarization, displacement, electric field, and electric potential in the bulk and at the interfaces of a ferroelectric film of thickness $t$ and polarization $P$. Compensating planes of charge density $\sigma$ can be considered to reside at a separation $\lambda$ equal to the effective screening length in the electrodes. Panel (a) is reprinted with permission from [Stephenson G B and Highland M J 2011 *Phys Rev. B* **84** 064107]. Copyright (2011) by The American Physical Society [119]. (b) Predicted polarization phase diagrams as a function of temperature and $p_{O2}$ for 10 nm PbTiO$_3$ film on SrRuO$_3$ coherently strained to SrTiO$_3$ (001), from theory for ferroelectric films with ionic compensation. Solid curves show equilibrium phase boundaries, and dashed red, blue, and white curves show metastability limits for the positive, negative, and paraelectric phases, respectively. (c) Correspondence between regions of continuous and nucleated switching mechanisms found previously (open symbols) and the minimum of observed $T_{cr}(h)$ (solid circles). The dashed curve is a fit. Panels (b) and (c) are reprinted with permission from [Highland M J, Fister T T, Fong D D, Fuoss P H, Thompson C, Eastman J A, Streiffer S K and Stephenson G B 2011 *Phys. Rev. Lett.* **107** 187602]. Copyright (2011) by The American Physical Society [118].

In the theoretical formalism evolved by SH the equilibrium chemical surface charge density is controlled by the concentration of surface ions $\theta_i$ as

$$\sigma_0[\varphi] = \sum_i \frac{eZ_i \theta_i(\varphi)}{A_i}, \qquad \text{(II.15(a))}$$

where $e$ is an elementary charge, $Z_i$ is the charge of the surface ions/electrons, $\theta_i$ are relative concentration of surface ions, $A_i$ is the saturation densities of the surface ions, at that $i \geq 2$ to reach the charge compensation. The surface charge is controlled by virtual potential $\varphi$ acting at the interface $z = 0$:

$$\theta_i(\varphi) = \frac{a(\varphi)}{1 + a(\varphi)}, \qquad a(\varphi) = (p_{O2})^{1/n_i} \exp\left(\frac{\Delta G_i^{00} - eZ_i\varphi}{k_B T}\right). \qquad \text{(II.15(b))}$$

Here $p_{O2}$ is the oxygen partial pressure (relative to atmospheric), $n_i$ is the number of surface ions created per oxygen molecule, $\Delta G_i^{00}$ is the standard free energy of the surface ion formation at $p_{O2} = 1$ bar and $U=0$, and $e$ is the magnitude of the electron charge. Equation (II.15(b)) is analogous to the Langmuir adsorption isotherm used in interfacial electrochemistry for adsorption of neutral species onto a conducting electrode exposed to ions in a solution.[81] Equations (II.15) can be incorporated in the numerical or phase-field formalism, providing chemical BCs to complement classical fixed potential or fixed charge physical BCs.

**II.4. Experimental macroscopic studies of polarization screening at surfaces**

Due to the strong interaction between extrinsic charges and the ferroelectric surface, great care is necessary to investigate screening mechanisms at clean ferroelectric surfaces. Even after following cleaning procedures in an ultrahigh vacuum (UHV) environment (which itself may lead to unintentional effects including surface reduction and cation unmixing), charges from hydrogen and other foreign atoms are difficult to avoid and cannot be eliminated entirely. For this reason, it is advantageous to study the ferroelectric surface *in situ*, after growth and within the deposition chamber, such that exposure to uncontrolled species is limited. This was the strategy employed by different groups studying the behavior of $PbTiO_3$ and $BaTiO_3$ (001) epitaxial thin films.[121-125]. The $PbTiO_3$ films were grown by metalorganic chemical vapor deposition (MOCVD) in a chamber installed at the Advanced Photon Source for *in situ* surface X-ray diffraction (SXRD) measurements. The $BaTiO_3$ films were grown by pulsed laser deposition (PLD) or oxide molecular beam epitaxy (MBE) and transferred in vacuum to an adjacent chamber for low-energy electron diffraction (LEED) studies.

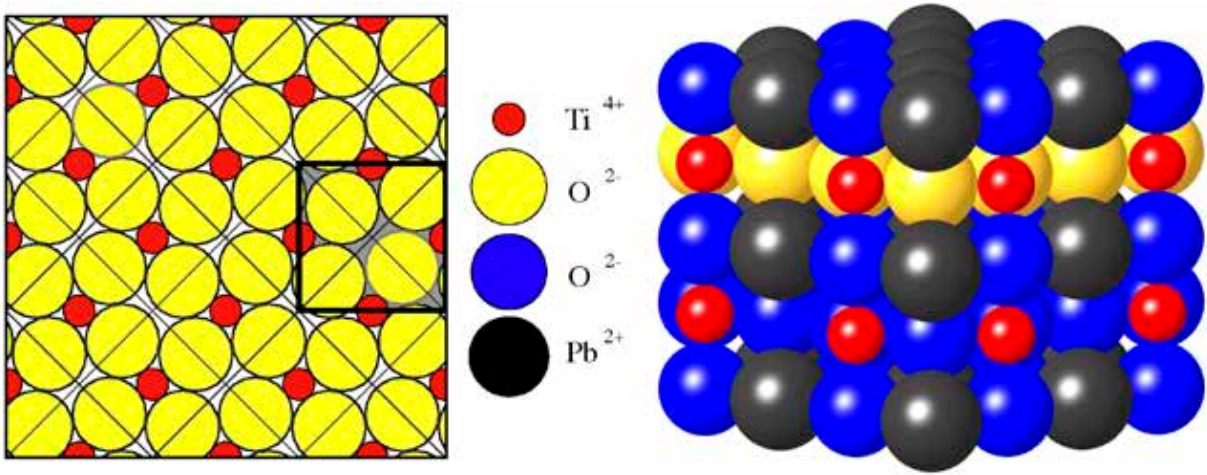

**Figure II.7**. Atomic structure of the equilibrium $PbTiO_3$ surface. View at left is a section parallel to the surface through the uppermost $TiO_2$ plane, showing the oxygen ions which form the c(2×2) reconstruction (lighter shade). The outermost unit cell has an antiferrodistortive structure, obtained by 10° rotations of the oxygen octahedra around the titanium ions. Reprinted with permission from [Stephenson G B, Fong D D, Ramana Murty M V, Streiffer S K, Eastman J A, Auciello O, Fuoss P H, Munkholm A, Aanerud M E M and Thompson C 2003 *Physica. B.* **336** 81-9]. Copyright (2003) with permission from Elsevier [126].

Regardless of ferroelectric polarization, solid surfaces and interfaces often restructure to minimize energy and stress.[127, 128] An antiferrodistortive reconstruction occurs at the PbO-terminated surface of PbTiO$_3$ (001), resulting in a pattern of counter-rotated TiO$_2$ octahedra in the layer below the topmost PbO (figure II.7). The $c(2\times2)$ symmetry of these surface octahedra gives rise to sharp half-order surface rods in reciprocal space that can be easily detected by SXRD[129] and are present in both the paraelectric and ferroelectric phases for PbTiO$_3$ films grown on SrTiO$_3$ (001). Note that similar antiferrodistortive reconstructions may be expected at the PbTiO$_3$ / SrTiO$_3$ interface,[130] but the $c(2\times2)$ reflections rapidly disappear when the PbO surface termination is lifted, eventually giving way to a 1×6 reconstruction.[126] In this particular case, the in-plane surface reconstruction shows little interference with the out-of-plane ferroelectric polarization favored by the compressive misfit strain, allowing ferroelectric order to persist down to a thickness of three PbTiO$_3$ unit cells.[131]

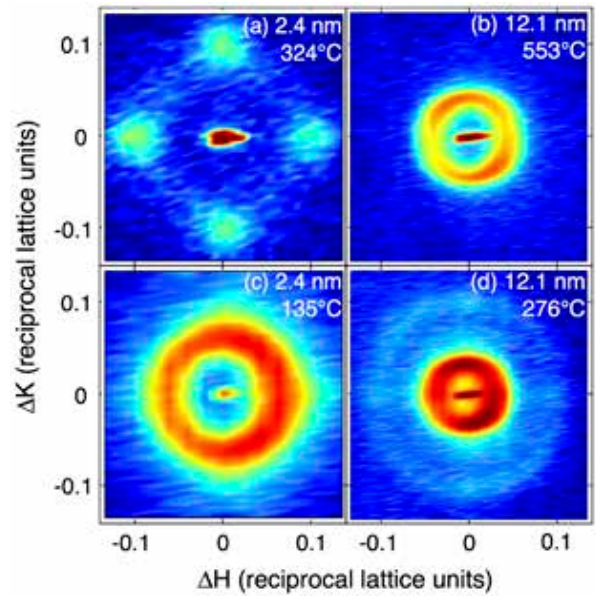

**Figure II.8.** In-plane reciprocal space maps around the PbTiO$_3$ 304 peak for 2.4- and 12.1-nm-thick films at temperatures above and below the $F_\alpha$ to $F_\beta$ transition. Both films were grown on < 0.1° miscut substrates. Redder hues indicate higher intensity (log scale). Panels (a) and (c) are reprinted from [Thompson C, Fong D D, Wang R V, Jiang F, Streiffer S K, Latifi K, Eastman J A, Fuoss P H and Stephenson G B 2008 *Appl. Phys. Lett.* **93** 182901]. Copyright 2008, AIP Publishing LLC [45]. Panels (b) and (d) are reprinted from [Streiffer S K, Eastman J A, Fong D

D, Thompson C, Munkholm A, Ramana Murty M V, Auciello O, Bai G. R and Stephenson G B 2002 *Phys. Rev. Lett.* **89** 067601]. Copyright (2002) by the American Physical Society [27].

For the low vacuum/PbTiO$_3$/SrTiO$_3$ (001) system, 180° stripe domains form as the film is cooled below $T_C$, with an in-plane period that scales with thickness according to the Kittel law.[27] In reciprocal space, the ordered stripe domains generate diffuse intensity (satellites) adjacent to the PbTiO$_3$ Bragg peaks, as shown in figure II.8, which depicts results for a 2.4-nm-thick film on the left and a 12.1-nm-thick film on the right.[45] The stripe period is inversely proportional to the radial distance from the central Bragg peak (at ΔH=ΔK=0) to the diffuse intensity, and the average stripe morphology can be determined by the inverse Fourier transform of the diffuse intensity pattern, with figure II.8(a) indicating that the domain walls at 324°C are preferentially aligned along the 100 directions. However, as the 2.4 nm film is cooled to 135°C (figure II.8b), the satellites move in towards the Bragg peak and form a torus, demonstrating that the stripe domains have increased their in-plane period and that their walls lose their preferential orientation. Apart from the change in morphology, the 12.1-nm-thick PbTiO$_3$ behaves similarly, as shown in figures II.8(b) and (d). In all cases, the increase in stripe period upon cooling corresponds to a factor of 2, suggesting that charge compensation has occurred at one of the two PbTiO$_3$ interfaces,[132] most likely at the surface due to presence of various chemical species in the MOCVD environment. This was called the $F_α$ to $F_β$ transition in Refs.[27, 45, 121, 131, 132] At even lower temperatures, the interface with SrTiO$_3$ is also screened, leading to a mostly monodomain state (called $F_γ$) with upward polarization. Depth resolved atomic positions measured through the thickness of these $F_γ$ films show no evidence of a dead layer at the PbTiO$_3$ surface.[133] These studies demonstrate that even in the absence of screening charges, nanoscale stripe domains can stabilize the ferroelectric phase in ultrathin PbTiO$_3$ films, with half-domain periods similar to that of the film thickness.

The behavior changes dramatically with the PbTiO$_3$/SrRuO$_3$/SrTiO$_3$ (001) system, as the presence of conducting SrRuO$_3$ at the bottom interface means that the ferroelectric structure depends almost entirely on the availability of screening charge at the PbTiO$_3$ surface.[24, 118, 134] As noted in section II.3, the oxygen partial pressure can be used to chemically switch the out-of-plane polarization from monodomain up at high pO$_2$ to monodomain down at low pO$_2$. Scans along the 30L crystal truncation rod as the pO$_2$ is decreased (from right to left) are shown in figure II.9(a), for a 5-nm-thick PbTiO$_3$ film at 740 K. The diffuse scatter from up/down domains is seen in figure II.9(b). It is observed to reach a maximum in intensity at ~$10^{-3}$ mbar, where the intensity from the 304 Bragg reflection is minimized. This stems from destructive interference between the up and domain domains at the Bragg reflection, while the diffuse intensity results from the incoherent addition of intensities from the up and down domains.

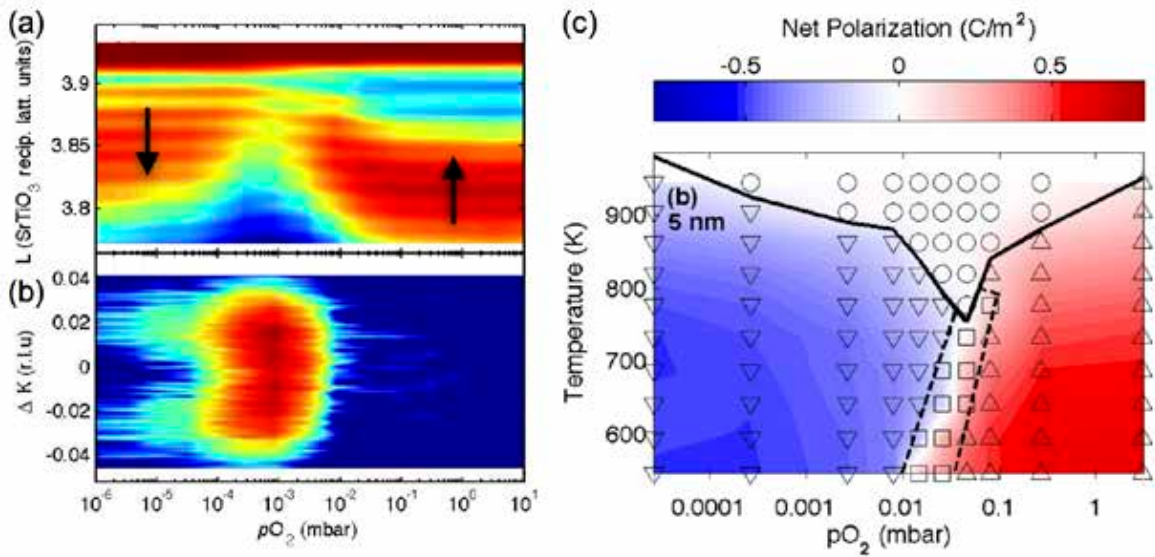

**Figure II.9.** X-ray scattering distribution near the PbTiO$_3$ 304 peak of a 5 nm film at 740 K, during switching by decreasing pO$_2$. (a) Distribution in the L (out-of-plane) direction along the Bragg rod. (b) Change in diffuse scattering around the Bragg rod in the K (in-plane) direction, relative to high pO$_2$ value, integrated from L = 3.82 to 3.91. Redder hues indicate higher intensity (log scale). Panels (a) and (b) are reprinted from [Wang R V, Fong D D, Jiang F, Highland M J, Fuoss P H, Thompson C, Kolpak A M, Eastman J A, Streiffer S K, Rappe A M and Stephenson G B 2009 *Phys. Rev. Lett.* **102** 047601]. Copyright (2009) by the American Physical Society [134]. (c) Temperature vs pO$_2$ phase diagram for 5-nm-thick PbTiO$_3$ on SrRuO$_3$ coherently strained to SrTiO$_3$ (001). Color scale indicates net polarization. Symbols show points measured and phase observed. Solid lines indicate $T_C$, and dashed lines are boundaries of 180° stripe domain regions. Panel (c) is reprinted from [Highland M J, Fister T T, Fong D D, Fuoss P

Temperature vs pO$_2$ phase diagrams may be constructed when the paraelectric film is first equilibrated at elevated temperatures in a fixed pO$_2$ and then cooled through the phase transition. Figure II.9(c) provides the experimental counterpart to figure II.6(b) but for a 5-nm-thick film. In this case, it is seen that $T_C$ depends strongly on pO$_2$, decreasing by ~200 K depending on whether or not screening can take place at the PbTiO$_3$ surface.[18] The effect is largest for ultrathin films, as expected and discussed in section II.3. These equilibrium phase diagrams show that while decreasing pO$_2$ at 740 K will lead to stripe domain formation, as observed in figure II.9(b), a similar pO$_2$ sweep at 780 K would result in stabilization of the paraelectric phase at intermediate oxygen partial pressures. Chemical switching can therefore change from the more typical nucleated behavior to continuous behavior, with a boundary line that depends on film thickness (figure II.6(c)). For films that undergo continuous switching, the internal electric field reaches that of the intrinsic coercive field due to the absence of a nucleation barrier.[135]

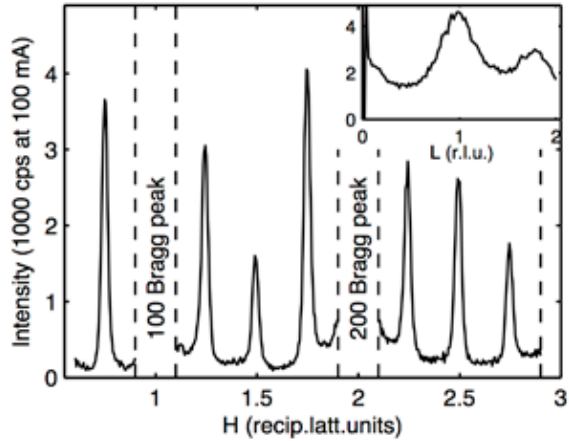

**Figure II.10.** Characteristic fractional-order diffraction peaks along the H 0 0 azimuth from the 4×1 reconstruction that forms under low-pO$_2$ conditions. Inset shows L scan through the 7/4, 0, 0 peak. Reprinted from [Wang R V, Fong D D, Jiang F, Highland M J, Fuoss P H, Thompson C, Kolpak A M, Eastman J A, Streiffer S K, Rappe A M and Stephenson G B 2009 *Phys. Rev. Lett.* **102** 047601]. Copyright (2009) by the American Physical Society [134].

While changes in pO$_2$ clearly switch the ferroelectric polarization, the nature of the compensating species at the surface is more difficult to ascertain. To maintain the observed polarization, the surface requires roughly ½ an electronic charge per unit cell area. In Ref.[24], density functional theory (DFT) calculations were used to investigate the effect of OH$^-$, O$^{2-}$, and H$^+$ adsorbates on the PbO-terminated PbTiO$_3$ surface. They found that an overlayer of OH$^-$ or O$^{2-}$ bonded to the surface Pb cations favored upward polarization, while H$^+$ bonded to the surface oxygens led to downward polarization. (On the other hand, undissociated H$_2$O and CO$_2$ resulted in a non-polar state.) Due to their ease of formation in reducing environments and elevated temperatures,[136] oxygen vacancies are more likely candidates for the positive screening charge. Since they are typically charged 2+ relative to the lattice site, only one vacancy is necessary per four surface unit cells. Wang et al. in fact discovered the appearance of a 4×1 reconstruction at the PbTiO$_3$ surface in reducing environments (with domains along both [100] and [010] directions), giving rise to the quarter-integer reflections shown in figure II.10.[134] The inset of the figure shows an L-scan through one of the reconstruction peaks; its behavior shows that the oxygen vacancies are arranged in a two-unit-cell-thick surface layer with long-range lateral order.

In the case of PLD-grown high vacuum/BaTiO$_3$/SrRuO$_3$/SrTiO$_3$(001) heterostructures, the BaO-terminated surface shows no change in the in-plane symmetry,[123] although out-of-plane relaxation and rumpling may occur regardless of the ferroelectric state. Quantitative analysis of the out-of-plane structure at room temperature by LEED was performed by fitting a set of diffracted intensity vs. voltage (*I-V*) curves for different in-plane reflections. Although low energy electrons cannot penetrate very far (roughly the top three unit cells), the BaTiO$_3$ thicknesses were similar to the probe depth (4 and 10 unit cells). Shin et al. found the BaTiO$_3$ films to be up polarized with bulk-like atomic displacements but that the terminating BaO layer exhibited only slight upward distortions, possibly due to a competition between surface relaxation (favoring a downward pointing surface dipole) and the polar field.[123]

Upon exposure to H$_2$O ( 6×10$^4$ L), the situation reverses. As seen in figure II.11(a-c), there are no substantial changes in the LEED patterns from the BaO-terminated surface prior to exposure (a), after exposure to 6×10$^4$ L (b), and after exposure to 3×10$^5$ L, indicating that at most a single (ordered) layer of H$_2$O adsorbs onto the surface. Changes in the *I-V* curves with $3\times10^5$ L are shown in figure II.11(d). Several different structural models with full oxygen occupancy were explored; a topview of the models along with the associated Pendry factor ($R_P$) are shown in figure II.12(a). The relatively large Pendry factors indicate poor agreement, and the best fits allow for a large concentration of oxygen vacancies in the topmost BaO layer (~ 70% according to the best fit shown in figure II.12(c), with the out-of-plane displacements shown in figure II.12(b). The primary difference with that of the clean surface is the appearance of downward polarization in the TiO$_2$ layer.

To provide greater insight into the adsorbed structure, Shin et al.[124] conducted a set of DFT calculations, finding that H$_2$O is energetically favored to dissociate on the BaO surface, forming Ba(OH)$_2$. The LEED results, however, suggest that the hydroxide layer is disordered, leading to oxygen vacancies in the rumpled BaO plane and downward polarization in the underlying unit cells. This is not unlike the results for PbTiO$_3$/SrRuO$_3$/SrTiO$_3$(001), where surface oxygen vacancies promote downward polarization.[134]

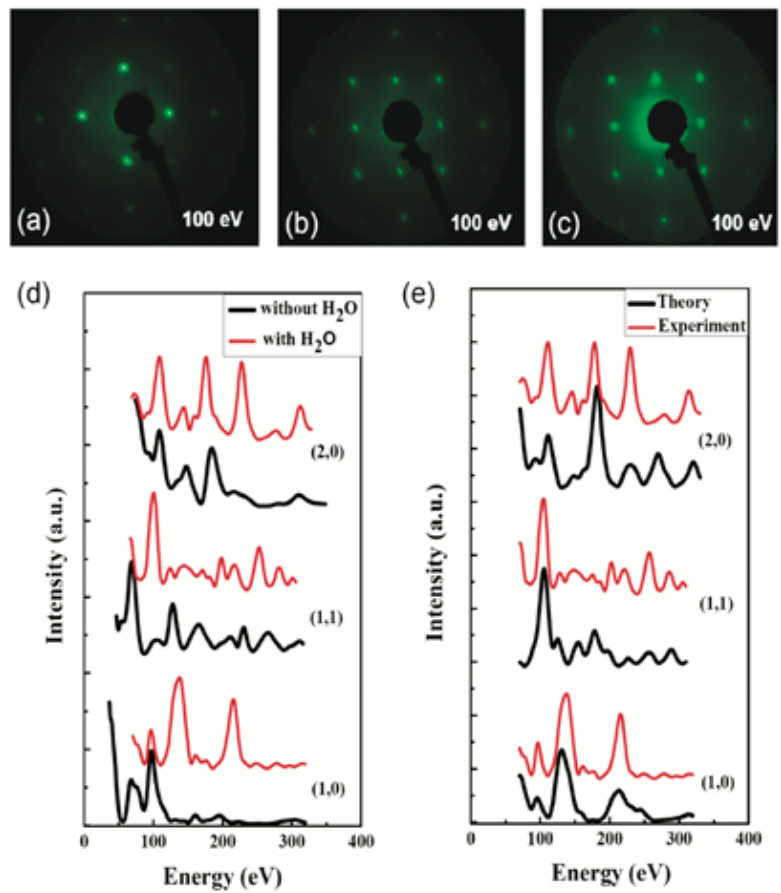

**Figure II.11.** LEED patterns for BaTiO$_3$ films (a) before exposure, and after exposures to H$_2$O of (b) $6 \times 10^4$ L, and (c) $3.6 \times 10^5$ L. (d) Experimental LEED *I-V* curves of 10 ML thick BaTiO$_3$ films before and after exposure to $3.6 \times 10^5$ L H$_2$O. (e) Comparison between several (of eight) experimental and theoretical *I(V)* curves obtained for the best fit model, which has 30% average oxygen occupancy and an inverted polarization dipole. Reprinted with permission from [Shin J, Nascimento V B, Geneste G, Rundgren J, Plummer E W, Dkhil B, Kalinin S V and Baddor A P 2009 *Nano Lett.* **9** 3720-5]. Copyright (2009) ACS Publications [124].

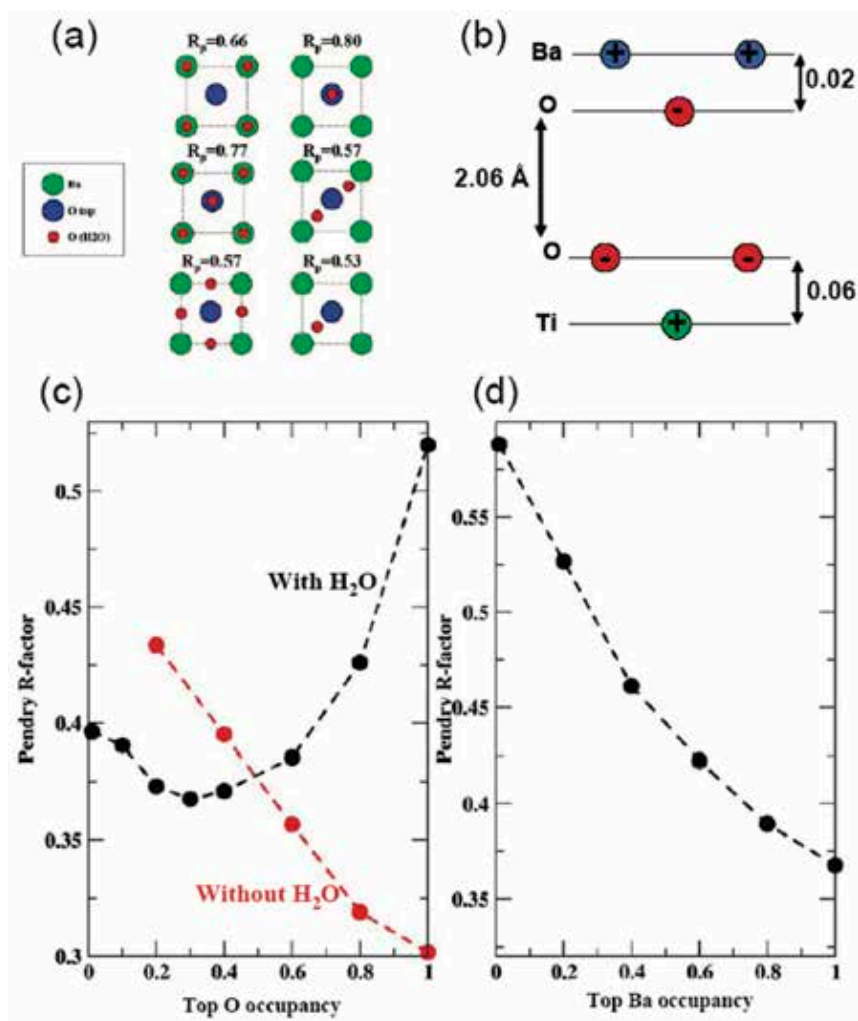

**Figure II.12.** (a) Explored structural models for adsorbed $H_2O$ on the $BaTiO_3$ (001) surface with associated $R_p$ factors. (b) Layer spacings in the suggested structural model. Optimized Pendry $R_p$ factors showing dependence on (c) oxygen occupancy in the top layer for surfaces with and without exposure to $H_2O$, and (d) barium occupancy in the top layer after water exposure. Reprinted with permission from [Shin J, Nascimento V B, Geneste G, Rundgren J, Plummer E W, Dkhil B, Kalinin S V and Baddor A P 2009 *Nano Lett.* **9** 3720-5]. Copyright (2009) ACS Publications [124].

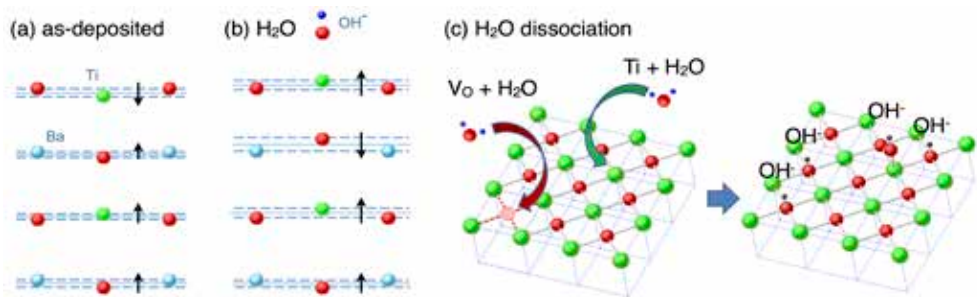

**Figure II.13.** Schematic sections of the optimized surface structure of BaTiO$_3$ (001) (a) before and (b) after water adsorption, as determined from minimization of the R$_p$ factor by out-of-plane relaxation and in (b) OH$^−$ on-top Ti coverage. (c) Schematic diagram showing the two adsorption processes leading to chemisorbed OH$^−$ at the surface of TiO$_2$-terminated BaTiO$_3$, either at a vacant lattice oxygen site or on-top of a surface Ti. The smaller (red) atoms are O, and the larger (green) atoms are Ti. Reprinted with permission from [Wang J L, Gaillard F, Pancotti A, Gautier B, Niu G, Vilquin B, Pillard V, Rodrigues G. L. M. P and Barrett N, 2012 *J. Phys. Chem. C* **116** 21802-9]. Copyright (2012)ACS Publications [125].

Similar studies were conducted by Wang et al. for TiO$_2$-terminated BaTiO$_3$ thin films grown directly on Nb-doped SrTiO$_3$ (001).[125] Here, the films were deposited by molecular beam epitaxy in a system permitting in-vacuum transfer to another chamber allowing LEED and XPS. They also found upward polarization for the as-deposited state using *I-V* fitting, although the top TiO$_2$ was slightly downward polarized (figure II.13(a)). After $3.6\times10^5$ L of H$_2$O was introduced into the chamber, the BaTiO$_3$ remained up polarized (figure II.13(b)), unlike the case of the BaO-terminated BaTiO$_3$ film discussed above. However, the surface oxygen vacancy concentration was here determined to be relatively small (~1%) both for the as-grown film and after water exposure, although the LEED results suggest only ~20% of dissociated H$_2$O. As may be expected, oxygen vacancies serve as active sites for water dissociation (as confirmed by a separately performed temperature programmed desorption study), but the XPS results suggest that most of the OH$^-$ groups are bonded to the surface Ti ions, as depicted in figure II.13(c). This chemisorption is hypothesized to be responsible for reversing the direction of the dipoles in the surface TiO$_2$ and underlying BaO layers shown in figure II.13(b). Interestingly, by analysis of the so-called O$_{II}$ and O$_{III}$ features in the O 1s spectra at higher binding energies (~530.5 eV and 531.5 eV, respectively), Wang et al. found that the as-grown BaTiO$_3$ films exhibited only slightly less OH$^-$ concentrations than H$_2$O-exposed samples, suggesting that hydrogen can easily be incorporated during the thin film growth of oxides.

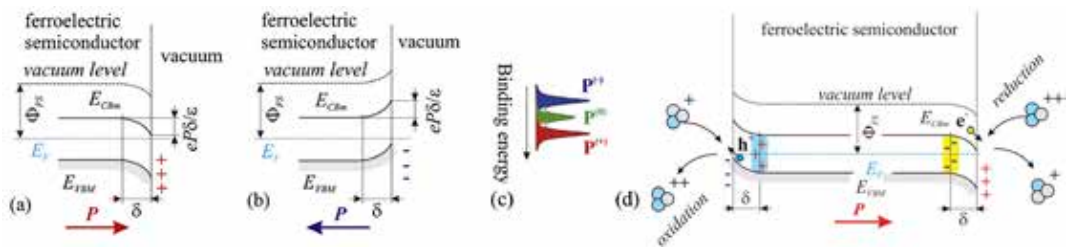

**Figure II.14.** Band diagrams at ideal ferroelectric surfaces: (a) upward polarization; (b) downward polarization. For PZT, band bending on the order of 1 eV is expected. (c) Expected effect of polarization on core-level spectra. Panels (a-c) are reprinted with permission from [Apostol N G, Stoflea L E, Lungu G A, Tache C A, Popescu D G, Pintilie L and Teodorescu C M 2013 *Mater. Sci. Eng. B-Adv.* **178** 1317-22]. Copyright (2013) with permission from Elsevier [137]. (d) Charge transport towards surfaces and possible catalytic activities of both surfaces (reduction for the upward polarization, oxidation for the downward polarization). Panel (d) is reprinted with permission from [Ştoflea L E, Apostol N G, Trupină L and Teodorescu C M 2014 *J. Mater. Chem. A* **2** 14386-92]. Copyright (2014) by The Royal Society of Chemistry [138].

Even in the absence of defects or absorbates, charge doping of the surfaces may occur in ferroelectric films, possibly leading to conductive surfaces.[139-141] Up and down polarizations naturally result in the type of band bending shown in figures II.14(a) and (b), respectively, and electrons from the bottom of the conduction band (screening upward polarization) or holes from the top of the valence band (screening downward polarization) can compensate the surface, thereby making the surfaces active for reduction or oxidation reactions, respectively (figure II.14(d)). As XPS can be used to measure band bending (figure II.14(c)), Teodorescu et al. used this technique to study the surfaces of PbO-terminated PbZr$_{0.2}$Ti$_{0.8}$O$_3$ (PZT)/SrRuO$_3$/SrTiO$_3$(001) heterostructures grown by PLD in a separate chamber and brought in air to the XPS system.[137, 138] They compared three types of samples -- 1) "fresh" samples introduced to the XPS system soon after PLD growth, 2) samples brought to the XPS system one week after growth, and 3) samples annealed at 400°C in UHV -- and attempted to rationalize the resulting XPS measurements shown in figure. II.15 by varying levels of carbon contamination. While the Ti 2p and O 1s spectra shown in figures II.15(c) and (d) for the 400°C annealed film appear comparable to others in the literature, [125] there is still a substantial carbon 1s peak, and Teodorescu et al. assign *upward polarization* for this sample. Compared to this, the Ti 2p and O 1s for the fresh sample are shifted to lower binding energies, suggesting downward polarization, and the results for the 1-week-old sample lie somewhere in between.

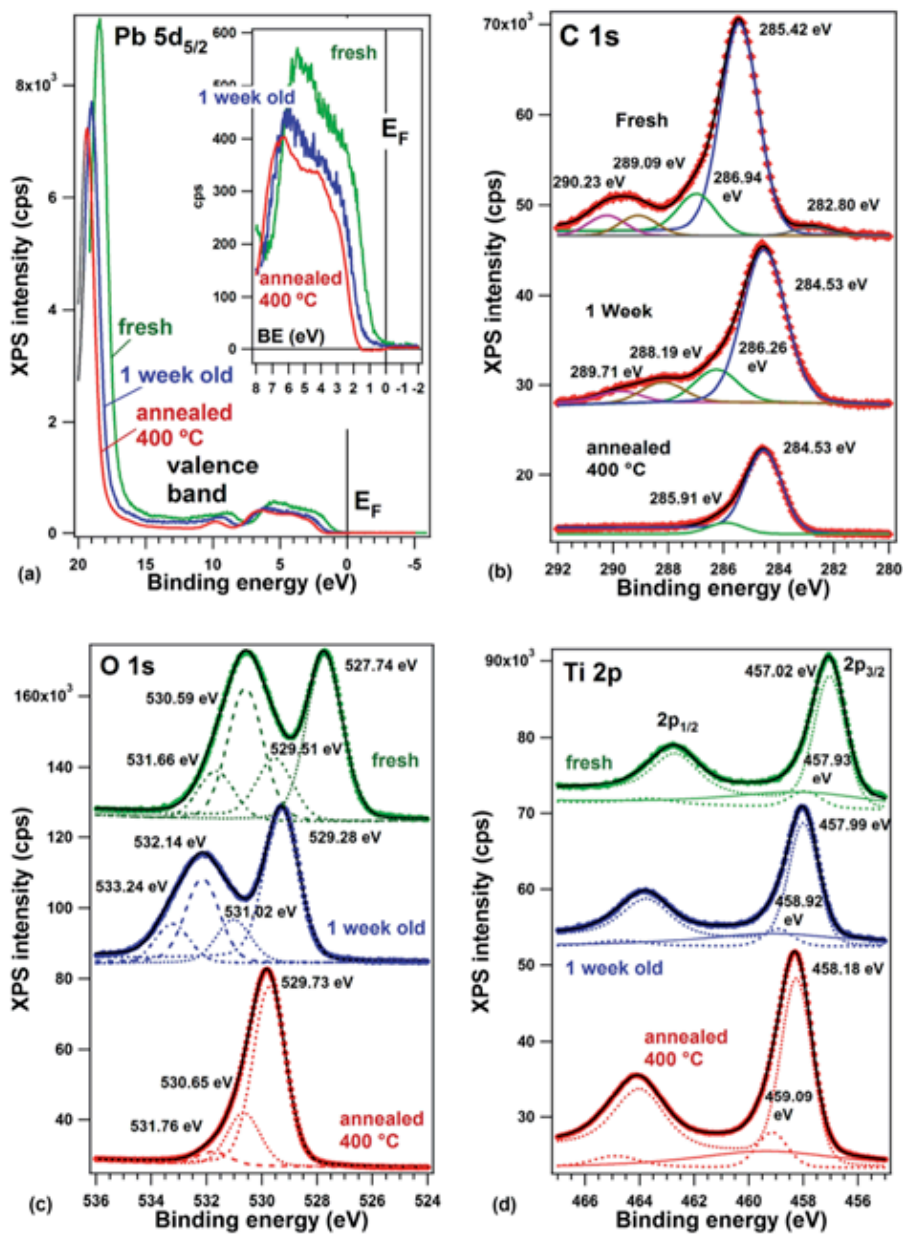

**Figure II.15.** XPS data for the three samples: (a) valence band spectra, with an inset showing the region near the Fermi level inserted; note the rigid shift of the Pb $5d_{5/2}$ level with the valence band onset; (b) C 1s; (c) O 1s; (d) Ti 2p. Reprinted with permission from [Ștoflea L E, Apostol N G, Trupină L and Teodorescu C M 2014 *J. Mater. Chem. A* **2** 14386-92]. Copyright (2014) by The Royal Society of Chemistry [138].

Indeed, the main complication stems from the C 1s spectra shown in figure II.15(b): only the 400°C annealed shows the expected single peak at ~284.6 eV coming from C-C bonds. The extra high binding energy features in the C 1s spectra for the "fresh" and "1-week" samples are also observed in the O 1s spectra (figure II.15(c)), indicating the existence of C-O bonded molecules at the surface. As will be discussed below, such molecules are expected to adsorb mainly to up-polarized domains. If true, attenuation of the photoelectrons by the carbon contaminants would primarily show the behavior of down-polarized domains; only the C 1s spectra would be shifted to higher binding energies, as shown in figure II.15(b). To explain the evolution of the spectra from "fresh" to "1-week", the carbon contaminants must migrate from primarily the up-polarized domains to all domains, and the authors suggest that this results in a loss of out-of-plane polarization.[138] While some details of this study remain unresolved, it demonstrates many of the challenges associated with understanding ferroelectric surfaces exposed to ambient conditions. The simple relationship between XPS binding energies and polarization depicted in figures II.15(c) and (d) may be inverted when screening occurs via ionic charges rather than electronic carriers.

In fact, more recent work by the same group made this clear when it showed that a PZT film totally free of surface carbon forms a *down-polarized* monodomain state (opposite to that above).[142] They achieved this by performing repeated anneals in $5\times10^{-5}$ mbar of $O_2$. However, rather than attributing downward polarization to the presence of surface oxygen vacancies, they suggest the existence of a significant concentration of cation vacancies that give rise to mobile holes that screen the PZT surface. This was consistent with their observation of reduced binding energies for all the core levels (O 1s, Ti 2p, Zr 3d, and Pb 4f) when compared with carbon contaminated PZT samples. After dosing the clean surface with $6\times10^3$ L of CO at room temperature, they observed sub-monolayer adsorption, with roughly 1 C or 1 CO molecule adsorbed per 3-4 surface unit cells. Desorption is most likely in the form of $CO_2$, removing lattice oxygen from the film and changing the polarization state to monodomain up.

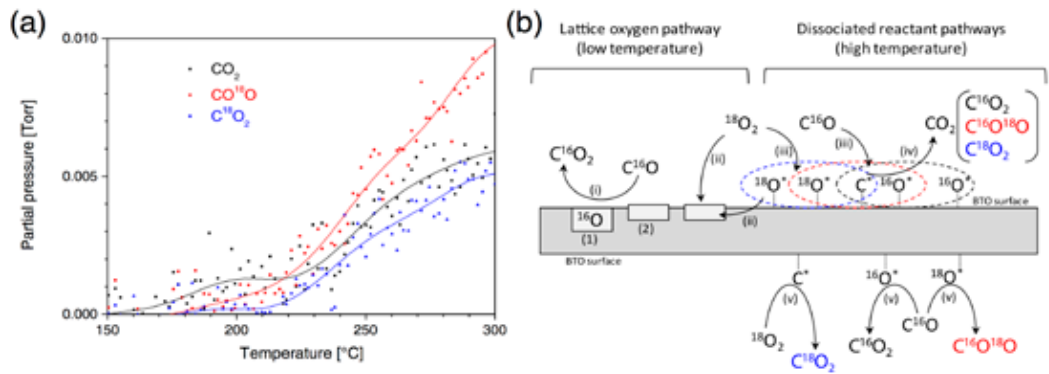

**Figure II.16.** (a) Partial pressures of $CO_2$ isotopes versus sample temperature during CO oxidation on $BaTiO_3/SrTiO_3$(100). Initial pressures: 2 Torr CO, 1 Torr $^{18}O_2$. (b) Scheme showing the possible elementary steps associated to the CO oxidation reaction on $BaTiO_3$ thin films. (i) Reaction of molecular CO with lattice oxygen; (ii) filling of oxygen vacancy by adsorbed or gaseous oxygen; (iii) dissociative adsorption of $O_2$ or CO; (iv) surface reaction be- tween 2O* and C*; and (v) reaction of a molecular species ($O_2$ or CO) with an adsorbed atom (C* or O*, respectively). (1) and (2) depict lattice oxygen and surface oxygen vacancy, respectively. Adsorbates are depicted with *. Reprinted with permission from [Nassreddine S, Morfin F, Niu G, Vilquin B, Gaillarda F and Piccoloa L 2014 *Surf. Interface Anal.* **46** 721-5]. Copyright 2014 Publisher Wiley-VCH Verlag GmbH & Co. KGaA [143].

The issue of carbon monoxide oxidation on ferroelectric surfaces was also studied by Nassreddine et al.[143] Here, they grew $BaTiO_3$ films on Nb:$SrTiO_3$(001) by MBE and transferred them *ex situ* to a surface analysis system attached to a catalytic reaction chamber outfitted with mass spectrometry. The ferroelectric structure of the $BaTiO_3$ was not described, but the surface was $TiO_2$-terminated as deduced from the (2×2) reconstruction observed by RHEED. If similar to the MBE-grown $BaTiO_3$ described by Wang et al.[125], the films are primarily up-polarized in the as-grown state. After transfer to the analysis system, they observed carbon contamination on the surface and trace amounts of sulfur. As expected, the carbon could not be removed by a simple anneal in UHV (even up to 700°C), but annealing at 500°C for 1 hour at $10^{-6}$ Torr $O_2$ left no detectable carbon signal by Auger electron spectroscopy. Figure II.16(a) shows the partial pressure of $CO_2$ measured by mass spectrometry when flowing 2 Torr of CO and 1 Torr of $^{18}O_2$ into the reaction chamber. First, it should be noted that there is very little catalytic activity, with roughly 0.14 $CO_2$ molecules formed per surface Ti atom per second during the 150°C-300°C ramp. (On the other hand, Pt(111) forms ~ 3 $CO_2$ molecules per Pt per second). Second, the same catalytic activity was observed for bare $SrTiO_3$(001), indicating that ferroelectric polarization had no significant effect on reactivity. Third, the use of isotopic $^{18}O_2$ as an input gas shows that at lower temperatures (primarily < 180°C), $CO_2$ is largely made from CO and lattice oxygen, leaving oxygen vacancies at the surface ((**i**) in figure II.16(b)), which is consistent with the Mars-van Krevelen mechanism. At higher temperatures, dissociative adsorption of $^{18}O_2$ and CO helps to fill the vacant sites (**ii**) and leave adsorbed O* and C* (**iii**), permitting $CO_2$ formation via adsorbate reaction (**iv**) or through direct combination of incoming $O_2$ with C* or incoming CO with O* (**v**) (the Eley-Rideal mechanism).[143] Although $BaTiO_3$ did not demonstrate high activity in this study, recent calculations have predicted better results for other ferroelectric materials and reactions when following particular polarization/catalytic cycles.[144, 145]. Furthermore, there have been several theoretical studies on the creation of tunable catalysts using ferroelectric oxides as supports for metal catalysts[146, 147] but experiments along these lines have proven to be highly challenging due to the difficulty in producing a metal film only one monolayer thick.[148, 149] A more achievable route may be growth of ultrathin transition metal oxides onto ferroelectric surfaces.[150, 151] For additional information regarding the activity of ferroelectric surfaces, a review by Khan et al. has recently been published.[152]

As mentioned above, there can be a large effect of ferroelectric polarization on adsorption, particularly for polar molecules. Yun et al. found that for pre-poled $LiNbO_3$ (0001) single crystals, the ferroelectric polarization can strengthen or weaken interaction with the adsorbed molecule.[153, 154] Prior to introduction of the adsorbates, the surfaces were cleaned of carbon by annealing in an oxygen plasma at 300°C for 1 hour. Temperature programmed desorption results are shown in figure II.17(a), for polar acetic acid, and in figure II.17(b), for non-polar dodecane. The different sets of curves for each pre-poled set refer to different adsorbate coverages. As observed, the peak desorption temperature ($T_P$) of acetic acid is roughly 100 K higher for upward polarization than for downward polarization; for dodecane, the desorption temperatures are the same (at ~250 K; the lower peak comes from multilayer desorption). The effect is more visible in the Redhead plot shown in figure II.17(c), where β is the heating rate (1 K/s). The exact relationship between adsorption and polarization is difficult to determine from such plots, however, as the polarization also varies as a function of temperature. Furthermore, one must be careful to distinguish electrostatic effects from those resulting from differences in surface structure, as noted in Ref.[155]

Similar studies by Garra et al. showed that both polarization and adsorbate coverage were important.[156, 157] Again for polar molecules (here water and methanol) on pre-poled $LiNbO_3$ (0001), $T_P$ was higher for the up-polarized state than the down-polarized state, as shown in figures II.17(d) and (e). Exposures ranged from 0.01 to 1.3 L, which were then converted into monolayers assuming a sticking coefficient of 1, as shown in these plots. Clearly the peak desorption temperature decreases with coverage for both water and methanol, regardless of the ferroelectric polarization direction. Since these molecules most likely adsorb without dissociation at these temperatures, this coverage dependence probably originates from a coverage-dependent desorption energy. This can be caused by hydrogen-driven electrostatic interactions between adsorbates weakening the interaction with the oxide surface (regardless of whether or not a ferroelectric material).[156]

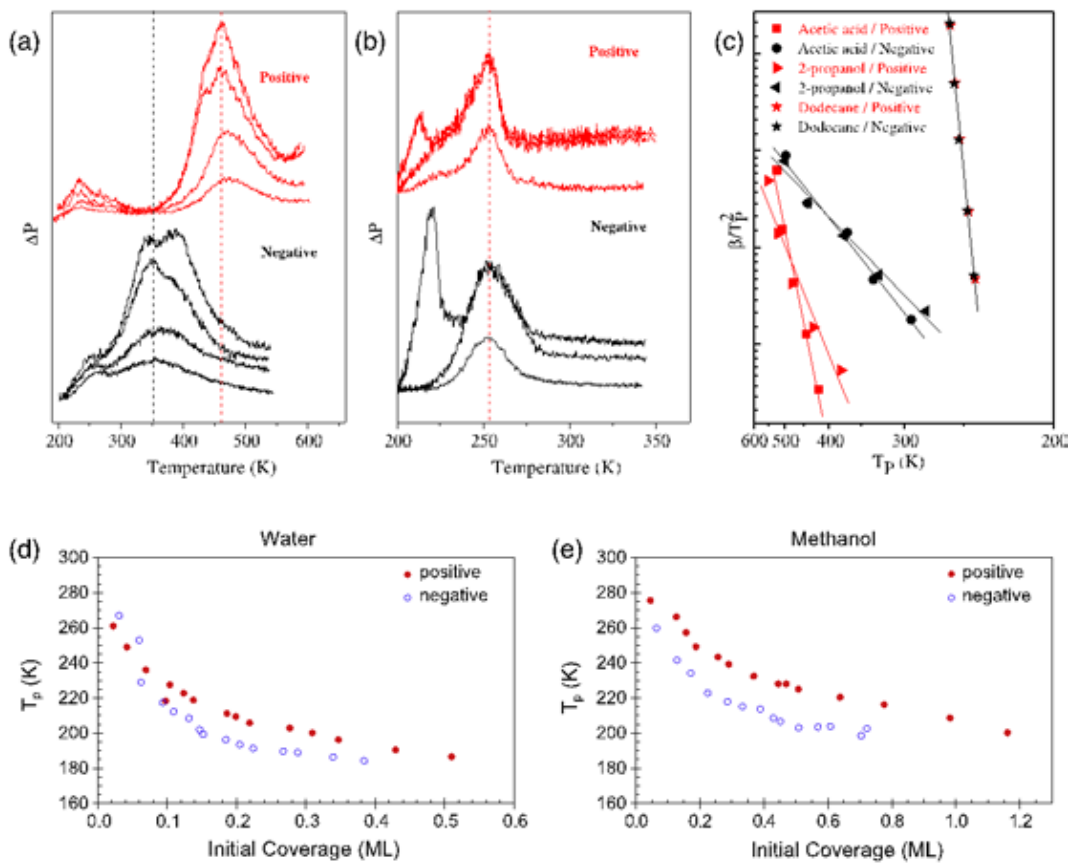

**Figure II.17.** (a) Comparison of acetic acid TPD curves for positively (top, red) and negatively (bottom, black) poled LiNbO$_3$ (0001). The sequence of curves illustrates the coverage dependence on the two surfaces ($1.5\times10^3$ to $1.6\times10^4$ L); the vertical dashed lines highlight the difference in desorption peak temperatures. (b) Dodecane TPD curves for several different coverages (10-80 L) for positively (top, red) and negatively (bottom, black) poled LiNbO$_3$. The low temperature peak is due to multilayer desorption. The vertical dashed line shows that monolayer desorption has a peak at the same temperature on both surfaces, so that differences in the multilayer peak temperatures are due to the higher coverages on the negative surface. (c) Redhead plots for acetic acid, 2-propanol, and dodecane on positively and negatively poled LiNbO$_3$ (0001) where β is the heating rate and T$_P$ the desorption peak temperature. Panels (a-c) are reprinted with permission from [Kakekhani A, Ismail-Beigi S and Altman E I 2016 *Surf. Sci.* **650** 302-16]. Copyright (2016) with permission from Elsevier [155]. Dependence of TPD peak temperature on adsorbate coverage for (d) water and (e) methanol. The initial coverage is given in monolayer units (ML) and was calculated by assuming a sticking coefficient of 1.0 for both molecules on both surfaces. For most coverages, the desorption temperature ($T_P$) for the positive surface (filled circles) is 10–20 K higher than for the negative surface (empty circles). Panels (d) and (e) are reprinted with permission from [Garra J, Vohs J M and Bonnell D A 2009 *Surf. Sci.* **603** 1106-14]. Copyright (2009) with permission from Elsevier [156].

Finally, it should be noted that X-ray and electron probes can themselves affect the behavior of ferroelectric surfaces. Popescu et al. found that the degree of band bending gradually disappears during (focused) soft X-ray irradiation due to the generation of electronic carriers,[158] while Husanu et al. observed increased out-of-plane polarization with time and gradual Pb metal precipitation on PZT(111).[159] Wang et al. found that the high flux from (hard) synchrotron X-rays tend to accelerate the kinetics in evolution of the lattice parameter with changes in $pO_2$.[134] This is not surprising given that X-rays continuously generate photoelectrons during illumination. Rault et al. even used low energy electrons (< 10 eV) to both observe and manipulate the in-plane polarization of $BaTiO_3$ (001).[160] In addition, photons and electrons have been known to stimulate both cation and anion desorption from surfaces.[161, 162]

## III. Local studies of polarization screening

The classical surface science methods are best suited to spatially uniform systems, whereas domain formation necessitates probing surface properties within a single domain. In this section, we discuss the application of scanning probe microscopy to explore domain-controlled functionalities of ferroelectric surfaces.

### III.1. SPM studies of domain structures

The ubiquitous formation of domains, the role of domain structures and their dynamics in ferroelectric materials play in the optical, dielectric, and electromechanical functionalities of these materials which have generated tremendous interest to local exploration of these phenomena. These studies can be roughly divided into two broad categories, namely those aimed at *visualization of domain structures* and study of the dynamics of domains under the application of electric fields, temperature, and other stimuli, and studying local materials properties *within* a domain. Below, we summarize some SPM-based approaches and briefly discuss their potential for imaging and quantitative probing of ferroelectric properties and screening phenomena. The reader is referred to other sources for discussion of electron beam and other imaging methods.[163, 164]

The characteristic feature of domains in multi-axial ferroelectrics is the formation of surface corrugations at the ferroelastic domain walls, directly related to the crystallographic structure. These topographic structures are directly amenable to contact and intermittent contact atomic force microscopy, as reported by a number of authors.[165-174] Beyond imaging domain structures, high-resolution SPM was proposed as a tool to explore the internal structure of the domain wall from direct comparison of topographic profile and predictions of Ginzburg-Landau theory.[175]

While simple to implement, the topographic studies do no not allow (in the absence of build-in electric fields) to distinguish antiparallel domains, since these do not change surface topography. More significantly, observation of domain-related topography requires cooling of ideally flat surface through ferroelectric phase transition. Polishing of material below $T_C$ or polishing material of multi-domain state with subsequent cycling through $T_C$ gives rise to complex surface topographies formed by "ghosts" of pre-existing domains and corrugations at newly formed domains. Similarly, this mode of domain imaging is inapplicable for surfaces with significant polishing damage. Finally, observation of topographic structure generally does not provide insight into the screening mechanism. As an interesting exception, the stripe domain patterns were directly measured by tapping mode AFM and, further, control of stripe domain patterns was found to be controlled by a vicinal surface through temperature dependent measurements by synchrotron x-ray scattering.[45] These domain patterns could be correlated with screening behavior.

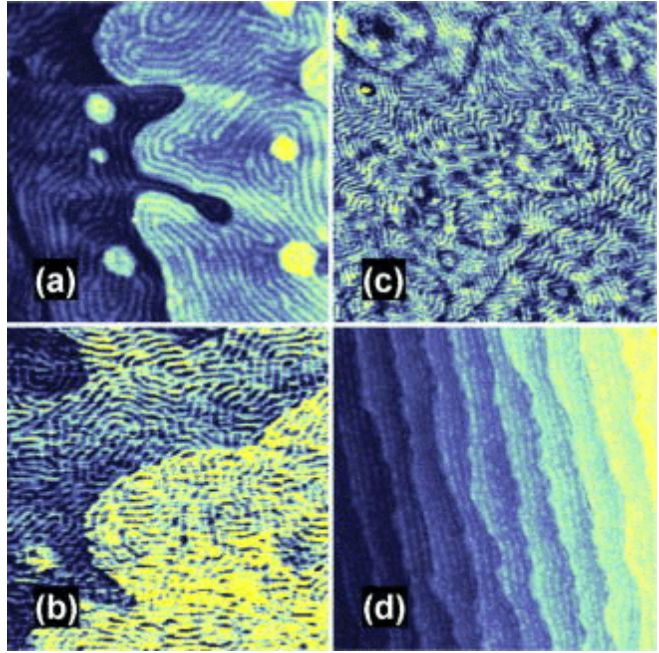

**Figure III.1.** Room-temperature tapping mode AFM images of surface steps and 180° stripe domains for epitaxial PbTiO3 films on SrTiO3 (001). Images are 500 × 500 nm2 with borders aligned close to <100> directions. [(a) and (b)] 10-nm-thick films, height image (color scale range: 0.8 nm) cooled from 735 °C at 0.2 °C/ s and from 655 °C at 1 °C/ s, respectively. (c) 5 nm phase image (3° range). (d) 10 nm film on 0.5° miscut substrate height image (2.0 nm range). Reprinted with permission from [Thompson C, Fong D D, Wang R V, Jiang F, Streiffer S K, Latifi K, Eastman J A, Fuoss P H and Stephenson G B 2008 *Appl. Phys. Lett.* **93** 182901]. Copyright 2008, AIP Publishing LLC [45].

As an additional modality, a number of authors reported observation of ferroelectric domains in the lateral force microscopy modes though the detection of polarization-dependent friction forces.[173, 176-180] While, in semiconductors, friction forces recently were correlated to the surface electronic structure using quantitative data obtained using ultra-high vacuum atomic force microscopy (AFM) on atomically clean surfaces,[181-183] analogous studies in ferroelectrics were performed almost exclusively in ambient environment, hindering quantitative and even qualitative interpretation. Exemplarily, most of friction force studies of ferroelectric surfaces were done in late 90ies, with recent decade seeing rapid growth of piezoresponse force (PFM) and Kelvin probe force microscopy (KPFM) studies as described below.

A number of authors have explored the mechanical properties of ferroelectric surfaces using techniques such as frequency-modulation atomic force acoustic microscopy (AFAM),[184, 185] and amplitude based AFAM.[186-190] However, the fundamental limitation of these studies is the extreme sensitivity of these methods to surface topography, owing to the fact the tip-surface contact stiffness that determines the resonance frequency shift of the cantilever is a product of contact area and mechanical properties of material. These two factors cannot be separated, giving rise to strong direct topographic cross-talk.[191] Furthermore, mechanical properties of ferroelectrics (for ideal surface) are expected to be similar for antiparallel polarization orientations and only ferroelastic domains expected to offer small changes in elastic properties due to the anisotropy of corresponding strain tensors and contribution of piezoelectric and dielectric properties to corresponding indentation moduli.[192, 193] Hence, AFAM and similar studies in certain cases can provide qualitative data on domain configurations, and quantitative information on local mechanical properties. However, inherent error in these studies, relatively weak sensitivity of elastic properties to electrostatic phenomena render these methods poorly suited for exploration of surface screening mechanisms.

The characteristic feature of ferroelectric materials is the broad range of electric functionalities, most notably piezoelectric and electrostrictive electromechanical couplings in the bulk and presence of surface and interface charges associated with polarization discontinuities. These factors naturally allow for probing ferroelectric materials using voltage modulated SPMs, generally based on measuring oscillatory cantilever response to bias applied to the probe. Some of the notable early works include that of Takata, Hong, Franke and others using complex semi-contact voltage and mechanically modulated signals.[194-197] However, it was rapidly recognized that the presence of multiple possible signal sources necessitates decoupling of electromechanical (piezoelectric and electrostrictive) and electrostatic (polarization charges and work function) interactions. In response to this challenge, two families of SPM techniques have emerged - the contact mode based PFM and a family of non-contact voltage modulated techniques including amplitude modulated KPFM (AM-KPFM) (also known as scanning surface potential microscopy),[198, 199] frequency modulated KPFM (FM-KPFM),[200-204] and electrostatic force microscopy (EFM).[205] These two families of technique are presently the mainstay of local ferroelectric characterization.

The PFM and related spectroscopic modes are based on the direct measurement of bias-induced surface deformation (electromechanical strain). Typically, the measurements are performed in the strong indentation regime using stiff cantilevers, minimizing the contribution of electrostatic cantilever-surface interactions, as analyzed by several groups.[206-212] Due to its simplicity and high spatial resolution (on the ~10 nm scale for many materials), PFM has emerged as a primary method for probing nanoscale phenomena in ferroelectrics in the last decade. Applications of PFM for imaging and control of ferroelectric domains is summarized in a number of archival[213-215] and recent reviews.[216-224] The remarkable aspect of PFM is that it is sensitive to electromechanical response of material, and provides only indirect information on the surface state and screening mechanisms. The multitude of available studies now suggest that while direct information of surface screening cannot be obtained from PFM, these processes nevertheless strongly affect or even control dynamics of ferroelectric domains induced by the bias applied to PFM tip, and hence kinetic studies of domain growth and PFM spectroscopy, as will be discussed below.

Complementary to PFM is non-contact voltage modulated techniques sensitive to stray electrostatic fields (polarization and screening charges, work function) and dielectric properties of material, but insensitive to the piezoelectric properties. While these techniques typically offer lower spatial resolution (50-100 nm), they provide direct information on charge dynamics on surfaces[225, 226] and surface screening mechanism. Further opportunities are offered by combinations of voltage modulation approach with other modulation modalities, as exemplified by *e.g.* pyroelectric charge microscopy developed by Groten et al.[227]

The brief overview of the SPM modes for probing ferroelectric materials would be incomplete without mentioning the near-field optical[228-230] and near-field Raman microscopies.[231] These techniques provide high-resolution optical contrast, but generally are more sensitive to the bulk properties of ferroelectrics. Finally, scanning nonlinear dielectric microscopy (SNDM)[232-235] provides high-resolution information on non-linear dielectric properties and has emerged as high resolution domain imaging mode (arguably with below nm resolution). However, relatively limited availability of this mode greatly reduced full exploration of its quantitative potential.

### III.2. Static studies of work function

The physics of screening phenomena on ferroelectric surfaces and interfaces is directly linked to the stray electrostatic fields above the surface, rendering these phenomena directly amenable to KPFM[198, 199] and related non-contact voltage modulated techniques. In AM-KPFM, the electric potential on a conductive SPM probe is modulated as

$$V_{tip} = V_{dc} + V_{ac}\cos(\omega t), \qquad (\text{III.1})$$

where $V_{dc}$ is the static (or slowly changing) potential offset, $V_{ac}$ is the driving voltage, and the driving frequency $\omega$ is typically chosen close to the free cantilever resonance. The tip bias results in the capacitive tip-surface force,

$$F_{el} = C'_z (V_{tip} - V_{surf})^2, \qquad (\text{III.2})$$

where $C'_z$ is the (unknown) tip-surface capacitance gradient, $V_{surf} = V_s + \Delta CPD$ is electrochemical potential, $V_s$ is electrostatic surface potential, and $\Delta CPD$ is the contact potential (work function) difference between the tip and surface. Depending on the experimental configuration, the voltage modulation can be applied either during the interleave scan (*i.e.* when tip retraces a pre-determined surface topography while maintaining constant separation), or during the acquisition of topographic information (at a different frequency from that used in the topographic feedback loop).

For a periodically modulated tip bias, the first harmonic components of electrostatic force between the tip and the surface is

$$F_{el}(1\omega) = C'_z V_{ac}(V_{dc} - V_{surf}), \qquad (\text{III.3})$$

This periodic force acting on the tip results in cantilever oscillations with magnitude proportional to the force. In principle, the displacement of the cantilever can be measured directly (open loop KPFM), but the signal variation in this case is affected both by surface potential and surface topography (through capacitance gradient term). Alternatively, the signal can be measured at each point as a function of $V_{dc}$, and surface potential can be found as an apex of parabola equation (III.1) (if full force is measured), or intercept of line in equation (III.3) (if the first harmonic of the force is measured). However, the elegant way for probing surface potential obviating the contribution of (unknown) surface topography is through the use of the feedback loop to keep the oscillation amplitude equal to zero. From equation (III.3), this condition is achieved for $V_{dc} = V_{surf}$. Hence, recording microscope-control dc potential offset on the tip, $V_{dc}$, provides information on the (unknown) surface potential, $V_{surf}$.

The interpretation of KPFM data strongly depends on materials systems. For metallic or highly conductive surfaces, measured surface potential is a sum of work function (modified by adsorption and screening)[236, 237] and local electrostatic potential. For semi-conductive and dielectric surfaces, the image formation mechanism can be considerably more complex.[238, 239]

Similar principle is employed in FM-KPFM.[200-204] FM-KPFM utilizes a phase-locked loop (PLL) to determine the bias-induced frequency shift, directly related to tip-surface force gradient. The bias-dependence of frequency shift is nullified using a second feedback loop, allowing direct detection of surface potential. The use of two sequential feedback loops results in relatively slow imaging rates. However, higher localization of electrostatic force gradients controlling frequency shift allows for very high spatial resolution compared to force-based signal, and in several cases atomic and molecular level resolution has been reported.[199]

Similar to other SPM techniques, KPFM imaging is non-ideal technique[240] and experimental measurements contain both systematic (feedback related) errors and topographic cross-talk. The resolution in AM-KPFM is controlled by electrostatic interactions between the tip and the surface containing local (tip) and non-local (cantilever components),[205, 241-243] giving rise to a logarithmic dependence of KPFM contrast with tip-surface separation.[244] The non-ideality of the feedback gives rise to $1/V_{ac}$ dependence of the measured signal[245, 246] and strong topographic cross-talk due to variations of tip-surface capacitance on rough surfaces.[247, 248] More subtle cross-talk (indirect cross-talk[249]) will include *e.g.* bias and position dependent variations of the cantilever resonant frequency in AM-KPFM.

Despite these limitations, AM-KPFM and FM-KPFM have emerged as powerful techniques for probing static and dynamic electric and electrochemical phenomena on the nanometer and in some cases to molecular and atomic levels.[199, 238, 250-253] Multiple applications of KPFM to energy conversion and storage materials,[254-257] organic materials,[237, 258-260] low-dimensional systems,[261-265] electroceramic oxides and semiconductors[266-271] have been reported. Below, we discuss in detail application of KPFM for probing screening physics of ferroelectric surfaces.[272-277]

The detailed study of ferroelectric domains on (001) surfaces of prototypical tetragonal BaTiO$_3$ single crystals polished above $T_C$ by combination of topographic AFM and KPFM was reported by Kalinin and Bonnell.[244] The domain structure of this material is formed by out of plane *c* domains with two possible $c^+$ and $c^-$ orientation, and in-plane *a* domains with 4 possible orientations along the (100), (-100), (010), and (0-10) axes. These domains are separated by ferroelectric 180 walls (*e.g.* between $c^+$ and $c^-$ domains) and ferroelastic 90 domain walls. For the latter, the tetragonal symmetry of BaTiO$_3$ unit cell results in characteristic surface corrugations as illustrated on figure III.2. The corrugation angle is directly related to the tetragonality of the unit call as $q = \pi/2 - 2\arctan(a/c)$, where *a* and *c* are the lattice parameters. Given high precision of conventional metrological AFM platforms, the deviation of the angle from ideal (for independently determined lattice parameters) can be used as a measure of local strains or clamping in multi-domain structures.

As discussed in Section II.2.2., the topographic imaging distinguishes *ferroelastic* walls only. At the same time, the difference in electric properties of the surface measured as effective surface potential by KPFM can distinguish *c*-domains of opposite polarity and *c* vs. *a* domains, as illustrated in figure III.2.[278] Note that potential and topographic information is insufficient to distinguish antiparallel *a* domains (invisible in optical microscope) and *a*1-*a*2 domain walls (visible in optical microscope). However, in many cases, the domain structures can be (partially) reconstructed from the topographic and KPFM data based on the knowledge of crystallographic compatibility and charge neutrality of the walls.

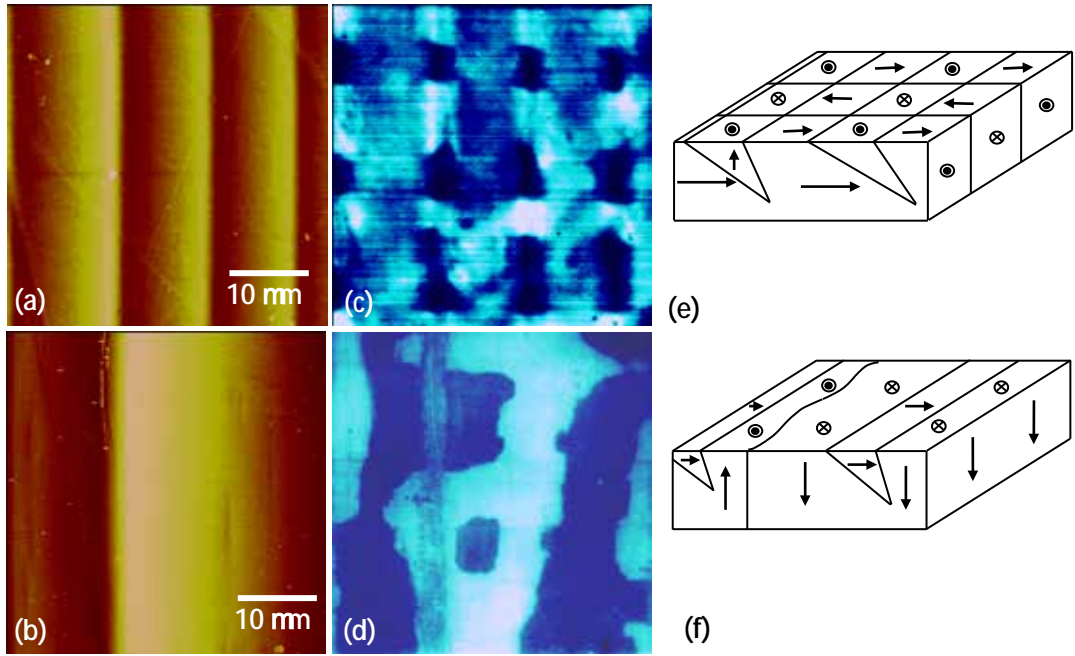

**Figure III.2.** (a,b) Surface topography, (c,d) surface potential and (e,f) schematics of domain structures in (a,c,e) *a*-domain region with *c*-domain wedges and in (b,d,f) *c*-domain region with *a*-domain wedges. Reprinted with permission from [Kalinin S V and Bonnell D A 2001 *Phys. Rev. B* **63** 125411]. Copyright (2001) by The American Physical Society [278].

The directly measurable parameter in KPFM is the surface potential of the surface, directly related to work function and hence structure of surface dipole layer. For BaTiO$_3$ (BTO) (100) surface, KPFM imaging yields potential difference between $c^+$ and $c^-$ domains as ~150 mV, whereas potential imaging between *a* and *c* domains is ~75 mV. Notably, surface potential is virtually uniform within the domains with rapid changes in the vicinity of ferroelectric and ferroelastic walls. The width of the transition is ~100-200 nm depending on imaging conditions and represents the measure of the intrinsic resolution of KPFM, rather than intrinsic wall width.[279], [280] These observations offer two immediate questions regarding the relationship between measured potential and domain structure. The first question is the relationship of measured potential, the polarization charge and screening mechanisms of the BTO surface. The less obvious question is what is the qualitative relationship between domain potential and the polarization direction is, *i.e.* whether domains positive on potential image are $c^+$ or $c^-$.

To address these questions, we consider semi-quantitative model for screening on ferroelectric surface valid for domain size significantly larger then screening layer width. We assume that the surface is characterized by the presence of polarization charge $\sigma_{pol} = \mathbf{P} \cdot \mathbf{n}$ and screening charge equivalent to surface charge density, $\sigma_s$, of the opposite polarity. The following cases can be distinguished:

1. Completely unscreened surface, $\sigma_s = 0$,
2. Partially screened surface, $\sigma_{pol} > -\sigma_s$,
3. Completely screened surface, $\sigma_{pol} = -\sigma_s$,
4. Over-screened surface, $\sigma_{pol} < -\sigma_s$.

As discussed above, completely unscreened surface is extremely unfavorable from an energetic point of view due to the large depolarization energy. An over-screened surface is likely to occur during bias-induced domain switching, *i.e.* under conditions when charge injection on the surface from biased SPM probe is possible or during changes in temperature or chemical properties of the environment. Partially or completely screened surfaces are likely to be the usual state of ferroelectric surfaces in air.

Assuming uniform charge densities, the charge distribution on a ferroelectric surface can be represented as a superposition of a double layer of width, $h$, dipole moment density $h \times min[\sigma_{pol}, \sigma_s]$ and an uncompensated charge component, $d\sigma = \sigma_{pol} - \sigma_s$. For future discussion, it is convenient to introduce degree of screening $a = -\sigma_s/\sigma_{pol}$. It is further assumed that the screening is symmetric, *i.e.* the degree of screening for $c^+$ and $c^-$ domains is the same. This assumption is supported by the experimentally observed near-equality of potential differences between $a$-$c^+$ and $a$-$c^-$ domains. Depending on the relative spatial localization of the polarization and screening charges (*e.g.* on the polarity of dipole layer), surface potential in the completely screened case can have the same sign as $\sigma_{pol}$, or be of the opposite sign.

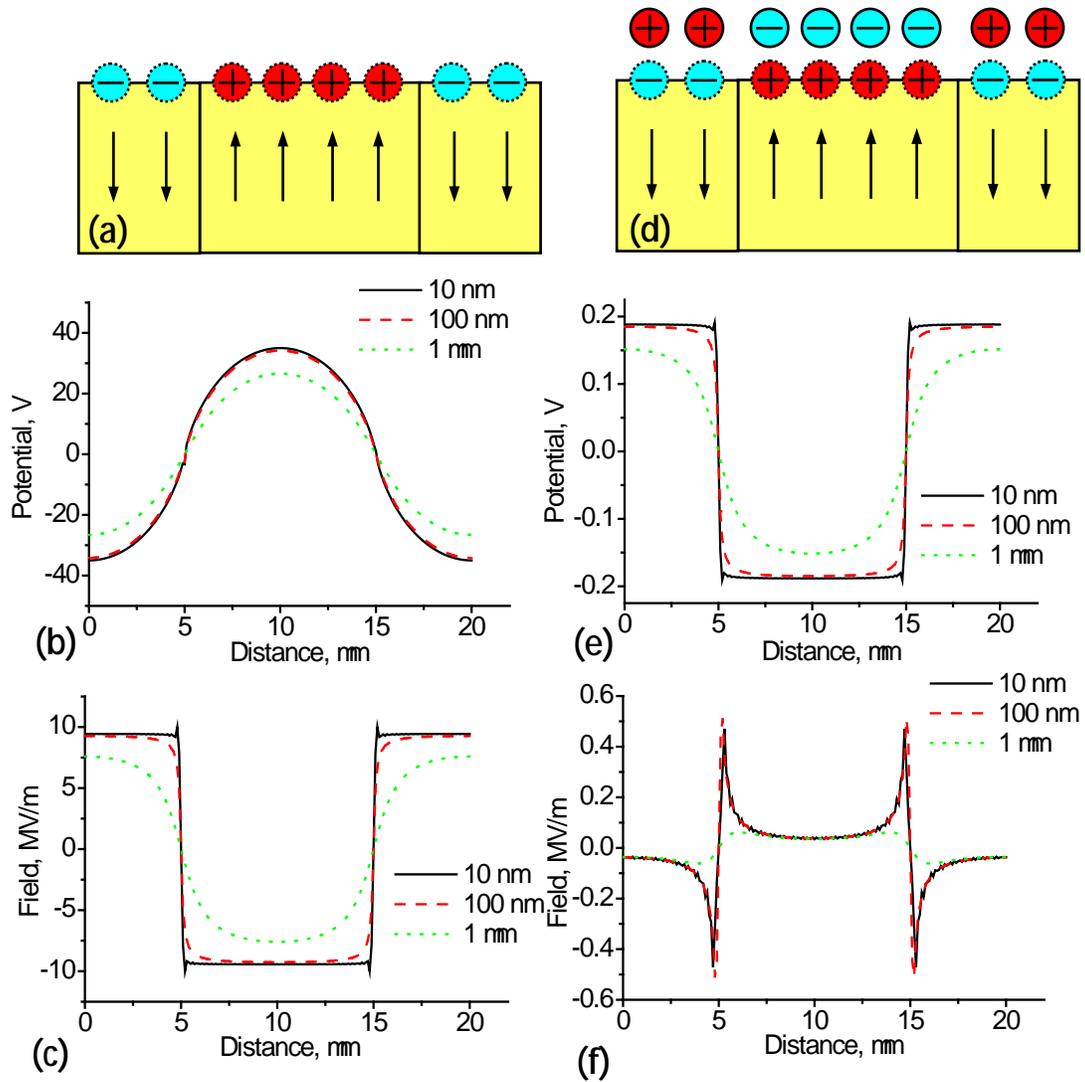

**Figure III.3.** (a,d) Simplified surface charge distribution, (b,e) potential and (c,f) the field in the vicinity of ferroelectric surface for (a,b,c) unscreened and (d,e,f) completely screened cases. Surface charge density is 0.25 C/m$^2$, domain size 10 nm, width of the double layer 2 nm. Reprinted with permission from [Kalinin S V and Bonnell D A 2001 *Phys, Rev. B* **63** 125411]. Copyright (2001) by The American Physical Society [278].

To analyze the origins of image contrast in EFM and KPFM, it is instructive to calculate the potential and field distributions above ferroelectric surface in the completely screened and completely unscreened cases. Usually non-contact measurements are performed at tip-surface separations of 10-100 nm, which are much smaller than typical domain sizes (~1-10 μm). Simple arguments predict that the surface potential above the unscreened surfaces and electric field above the completely screened surfaces scale linearly and reciprocally with domain size, while electric field over the unscreened surfaces and potential over the screened surfaces are virtually domain size-independent.

For anisotropic ferroelectrics with dielectric constants $\varepsilon_{xx} = \varepsilon_{yy} = \varepsilon_x$, $\varepsilon_{zz} = \varepsilon_z$ the domain structure is defined as $P(x) = P\,\mathbf{z}$, ($0<x<L/2$), $-P\,\mathbf{z}$, ($L/2<x<L$), where $L$ is characteristic domain size and is uniform in $y$ direction. The adsorbate charge density is $\sigma(x) = -\sigma$ ($0<x<L/2$) and $\sigma$ ($L/2<x<L$). Potential in the air, in the adsorbate layer and in the ferroelectrics are denoted as $\varphi_1$, $\varphi_2$ and $\varphi_3$ respectively. These potentials can be found as solutions of Laplace equations $\nabla^2\varphi_1 = 0$, for $z > h$, $\nabla^2\varphi_2 = 0$, in the screening layer $h > z > 0$, and

$$\varepsilon_x \frac{\partial^2 \varphi_3}{\partial x^2} + \varepsilon_z \frac{\partial^2 \varphi_3}{\partial z^2} = 0, \quad z < 0 \tag{III.4}$$

In ferroelectrics, the corresponding boundary conditions are the continuity of potential at the interfaces $\varphi_1(h) = \varphi_2(h)$, $\varphi_2(0) = \varphi_3(0)$ and the normal component of the displacement vectors are as

$$\varepsilon_0 \frac{\partial \varphi_1}{\partial z} - \varepsilon_2 \frac{\partial \varphi_2}{\partial z} = \sigma(x) \text{ for } z = h, \quad \varepsilon_2 \frac{\partial \varphi_2}{\partial z} - \varepsilon_z \frac{\partial \varphi_3}{\partial z} = P(x) \text{ for } z = 0 \tag{III.5}$$

Of interest are potentials in the completely screened case, $\sigma = -P$, in which case

$$\varphi_1^s = -\frac{4h\sqrt{\varepsilon_x \varepsilon_z}\,\sigma}{\pi \varepsilon_2 (\varepsilon_0 + \sqrt{\varepsilon_x \varepsilon_z})} \sum_{n=0}^{\infty} \frac{1}{2n+1} \sin\left(\frac{2\pi(2n+1)x}{L}\right) \exp\left(-\frac{2\pi(2n+1)z}{L}\right) \tag{III.6}$$

$$\varphi_3^s = \frac{4h\varepsilon_0 \sigma}{\pi \varepsilon_2 (\varepsilon_0 + \sqrt{\varepsilon_x \varepsilon_z})} \sum_{n=0}^{\infty} \frac{1}{2n+1} \sin\left(\frac{2\pi(2n+1)x}{L}\right) \exp\left(\frac{2\pi(2n+1)z}{L}\sqrt{\frac{\varepsilon_x}{\varepsilon_z}}\right) \tag{III.7}$$

The potential difference between the domains of opposite polarity is therefore:

$$\Delta \varphi_s = \frac{2h\sqrt{\varepsilon_x \varepsilon_z}\,\sigma}{\varepsilon_2 (\varepsilon_0 + \sqrt{\varepsilon_x \varepsilon_z})} \approx \frac{2h\sigma}{\varepsilon_2} \tag{III.8}$$

Depolarization energy density for the screened case is $E_{el}^s = \frac{1}{L}\int_0^L \left(-\varphi_1^s + \varphi_3^s\right)P\,dx$ or $E_{el}^s = \frac{hP^2}{\varepsilon_2}$.

In the unscreened case ($h = 0$, $\sigma = 0$), so

$$j_1^u = \frac{2LP}{\pi^2\left(\varepsilon_0 + \sqrt{\varepsilon_x \varepsilon_z}\right)} \sum_{n=0}^{\infty} \frac{1}{(2n+1)^2} \sin\left(\frac{2\pi(2n+1)x}{L}\right) \exp\left(-\frac{2\pi(2n+1)z}{L}\right) \qquad (\text{III.9})$$

$$j_3^u = \frac{2LP}{\pi^2\left(\varepsilon_0 + \sqrt{\varepsilon_x \varepsilon_z}\right)} \sum_{n=0}^{\infty} \frac{1}{(2n+1)^2} \sin\left(\frac{2\pi(2n+1)x}{L}\right) \exp\left(-\frac{2\pi(2n+1)z}{L}\sqrt{\frac{\varepsilon_x}{\varepsilon_z}}\right) \qquad (\text{III.10})$$

Depolarization energy density in the unscreened case is

$$E_{el}^u = \frac{P^2 L}{\varepsilon_0 + \sqrt{\varepsilon_x \varepsilon_z}} \frac{7\zeta(3)}{8\pi^3} \qquad (\text{III.11})$$

where $\zeta(3) \approx 1.202$ is zeta function.

Potential in the partially screened case is represented as the sum of the potential in the completely screened case with a surface charge density $\sigma$ and in the unscreened case with a polarization, $P-\sigma$. Therefore, depolarization energy is

$$E_{el} = \frac{1}{L}\int_0^L \left\{P(j_3^u + j_3^s) + \sigma(j_1^u + j_1^s)\right\} dx \qquad (\text{III.12})$$

where $j_1^u$ and $j_3^u$ are given by equations (III.9 and III.10) with $P-\sigma$ rather $P$.

Introducing the degree of screening, $\alpha$, such that $\sigma = -\alpha P$, depolarization energy density is calculated as

$$E_{el} = \frac{P^2}{\varepsilon_0 + \sqrt{\varepsilon_x \varepsilon_z}} \left\{(1-\alpha)^2 L \frac{7\zeta(3)}{2\pi^3} + \alpha h \frac{\varepsilon_0}{\varepsilon_2} + \alpha^2 h \frac{\sqrt{\varepsilon_x \varepsilon_z}}{\varepsilon_2}\right\} \qquad (\text{III.13})$$

The experimentally observed non-uniform image contrast within the domain can be attributed either to the potential variation above the surface and corresponding change of the capacitive interaction, or to the variation in the surface charge density and normal electric field that results in additional Coulombic interaction between the tip and the surface. However, the expected (almost) complete screening of polarization suggests that the dipole layer model is far more likely. Experimentally observed potential difference of 0.140 V can be ascribed to a 0.20 nm double layer of a dielectric constant $\varepsilon_1 = 80$ (H$_2$O) on a ferroelectric substrate (external screening) or a 9.5 nm depletion layer in a ferroelectrics with a dielectric constant $\varepsilon_2 = 3000$ (intrinsic screening). While the former estimate is reasonable for a molecular adsorbate layer or occupation/depletion of surface states, the latter is unreasonably small for a depletion layer width in a semiconductor with a low charge carrier concentration (~1 $\mu$m). Moreover, potential differences between $c^+$-$a$ and $c^-$-$a$ domains are almost equal, suggesting that the screening is symmetric. This is not the case if the screening is due to the free carriers in materials with a predominant electron or hole conduction, in which the width of accumulation layer for the polarization charge opposite to the majority carrier charge and width of depletion layer for the polarization charge similar to the majority carrier charge are vastly different. Thus, surface adsorption can be expected to be the dominant mechanism for polarization screening on a ferroelectric surface in ambient conditions, though a minor contribution from intrinsic screening cannot be excluded. The ionic nature of screening charges is consistent with multiple other observations on non-ferroelectric surfaces, including charge retention on electronically conductive doped $SrTiO_3$ surfaces and potential inversion on the grain boundaries in electroceramic materials.[60]

Following initial studies, the surface potential on pristine ferroelectric surfaces was explored by several groups, most notably Rosenman and Rosenwaks and Kitamura. Exemplarily, Shvebelman et al.[275] explored potential contrast on $KTiOPO_4$ surface and observed potential difference between the domains of the order of 40 mV. This contrast was attributed purely by internal screening by mobile ions, with corresponding screening length of the order of 10 nm. In comparison, Kitamura group has explored surface potential of $LiNbO_3$ under different conditions, and observed strong dependence of KPFM contrast on environment, as could be expected for screening by ionic species.[281]

Furthermore, the surface potential dependence on the microstructure of the films was explored by several groups. Choi et al. reported strong dependence of surface potential on grain size in the polycrystalline materials.[282] Kim et al. further have explored the origin of its influence on the surface potential in polycrystalline thin films and found that grain boundary can be responsible for the local surface potential distribution.[283, 284] While the epitaxial films show rather uniform surface potential distribution, the polycrystalline films show local surface potential distribution over the whole film surface. Since grain boundaries can act as conducting paths of current flow, electric charge can easily move through the grain boundaries during and/or after switching procedure.

Finally, surface potential of ferroelectric polymers was explored in a series of work by Matsushige et al.[285, 286] The authors ascribe the experimentally observed negative work function of the in-situ deposited PVDF to the molecule-substrate interactions that results in the preferential dipole orientations. The comparison between PFM and KPFM data illustrated the presence of frozen dipoles in the material, and also illustrate that polarization switching results in injection of additional charges on material surfaces, *i.e.* formation of over-screened regions. This behavior was later found to be universal for tip-induced polarization switching, as discussed in detail in Section II.2.4.

Further, Iwata el al.[287] explored surface potential of domain wall in $Pb(Zn_{1/3}Nb_{2/3})O_3$ (PZN)- $PbTiO_3$ (PT). The surface potential lowering in the a-domain was larger than that in the c-domain, the behavior attributed to the fact that the dielectric constant perpendicular to the ferroelectric polarization is larger than the parallel one.

**III.3. Domain dynamics under constant conditions**

Significant insight in the mechanisms of screening phenomena can be inferred from the observations of surface potential dynamics. Indeed, potential over static domain structures correspond to the thermodynamic minimum, whereas details and kinetics of screening phenomena are not observable. At the same time, perturbation of the domain state though domain wall motion, temperature change, or external field displaces the system form equilibrium state and allows direct observation of the kinetics and spatial localization of screening processes. These observations are summarized in this section.

Insight into screening mechanism can be obtained from observation of potential evolution during domain wall motion driven by wall tension (note that application of external electric fields via tip or lateral electrodes will result in the charge injection and surface electric fields and field-driven propagation of screening charges, as was observed for non-ferroelectric surfaces). Figure III.4 illustrates KPFM images of $c^+$ - $c^-$ domain structures in BTO (100) obtained at a 12 h interval. The shrinking of the negative domain results in a dark rim in the direction of domain wall motion. The formation of the rim can be understood assuming that the screening charge relaxation is a slow process as compared to polarization dynamics. Interestingly, simple considerations imply that a negative rim in the direction of wall motion is possible only if domain related potential features are determined by the screening charges, rather than polarization charges. In other words, for $c^+$ domain with positive polarization charge surface potential is actually negative, whereas for $c^-$ domain with negative polarization charge the surface potential is positive.

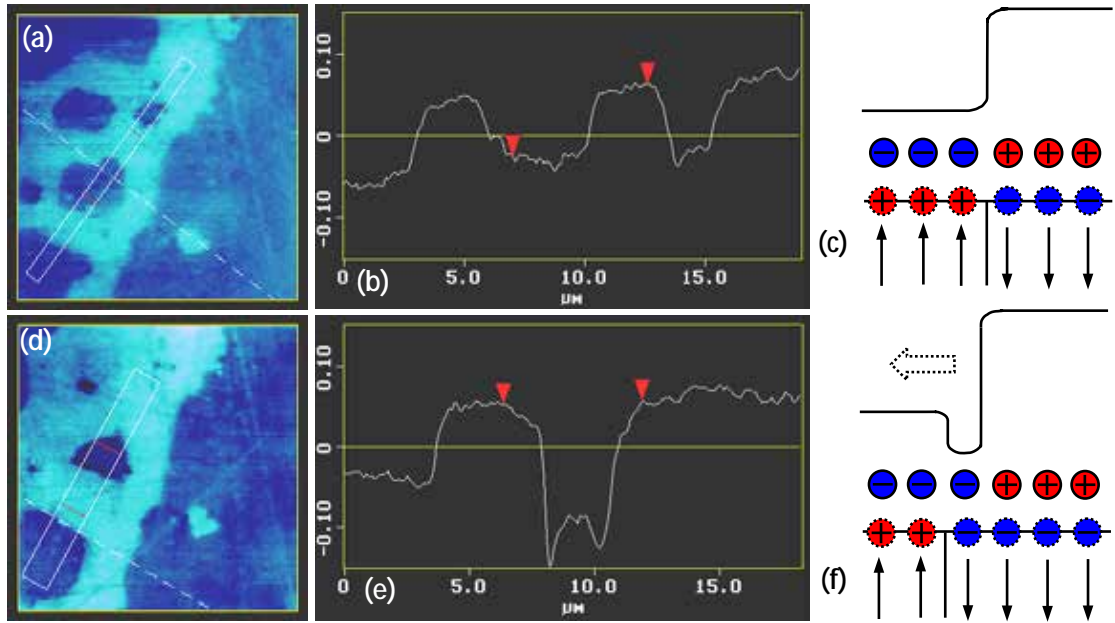

**Figure III.4**. (a,d) Surface potential images of $c^+$-$c^-$ domain region BaTiO$_3$ (100) acquired at 12 h interval, (b,e) average profiles along the boxes and (c,f) the scheme of surface charge distribution. Reprinted with permission from [Kalinin S V and Bonnell D A 2001 *Phys, Rev. B* **63** 125411]. Copyright (2001) by The American Physical Society [278].

While at the first glance surprising, this conclusion is fully consistent with the charge distribution of fully screened surface. Indeed, the screening charge is equal in magnitude to polarization charge and is located outside of the ferroelectric material, leading to the opposite potential of dipole layer compared to sign of polarization charge. Interestingly, this behavior is ubiquitous for oxide surfaces in ambient and this charge inversion was reported to the grain boundaries in electroceramic materials, etc. It is also important to note that this behavior can be observed only for well-equilibrated surface in the absence of charge injection (*e.g.* from in-plane electrodes or SPM tip), since in the latter case the strong Coulombic forces form injected charges dominate the signal.

## III.4. Lateral fields

Many applications of ferroelectric materials are based on the capacitor geometry, referring to the ferroelectric slab between the parallel electrodes. However, these systems are poorly amenable to experimental observations. The need for understanding of these system as well as emergence of ferroelectric race track memory like devices nucleated research effort in studying ferroelectric devices between interdigitated electrodes.

### III.4.1. Lateral switching

Laterally applied fields through in-plane capacitor structures could allow exploring a direct domain wall motion and growth behavior. Balke et al.[288] directly observed nucleation sites as well as forward and lateral growth stages of domain formation by combination of vertical (out-of plane) and lateral (in-plane) PFMs in the planar electrode structures of BiFeO$_3$ (BFO) as presented in figure III.5. The authors found that the location of the nucleation sites is correlated with a switching direction and the nucleation can be controlled by the bias history of the sample. Based on the results, the authors also demonstrated the manipulation of the nucleation through the domain structure engineering. In a similar planar electrode structure, Xi Zou et al[289] explored polarization fatigue mechanism in BFO by combination of PFM and KPFM. The authors found that the charged domain walls were formed due to the interaction between local space charges and polarization charges during the cyclic switching and, further found that the fatigue occurred due to the negative charge accumulation at the electrode/film interfaces.

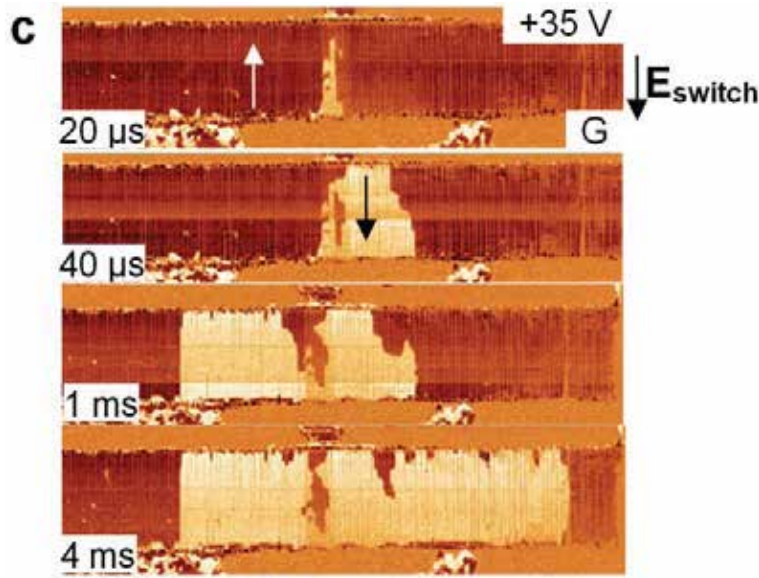

**Figure III.5.** IP PFM images showing the switching sequence for positive switching voltages. Reprinted with permission from [Balke N, Gajek M, Tagantsev A K, Martin L W, Chu Y,

Ramesh R and Kalinin S V 2010 *Adv. Funct. Mater.* **20** 3466-75] Copyright 2010 Publisher Wiley-VCH Verlag GmbH & Co. KGaA [288].

Whereas figure III.5 present rather usual domain wall motion, unusual domain wall motion has been reported in the planar electrode structures as well.[290, 291] McQuaid et al.[291] reported spontaneously formed flux-closure patterns after removal of a uniform applied electric field. Further, Sharma et al.[290] observed a unique domain wall motion of superdomains, which are distinct bundles of a-c domains, in thin single-crystalline lamellae of $BaTiO_3$.

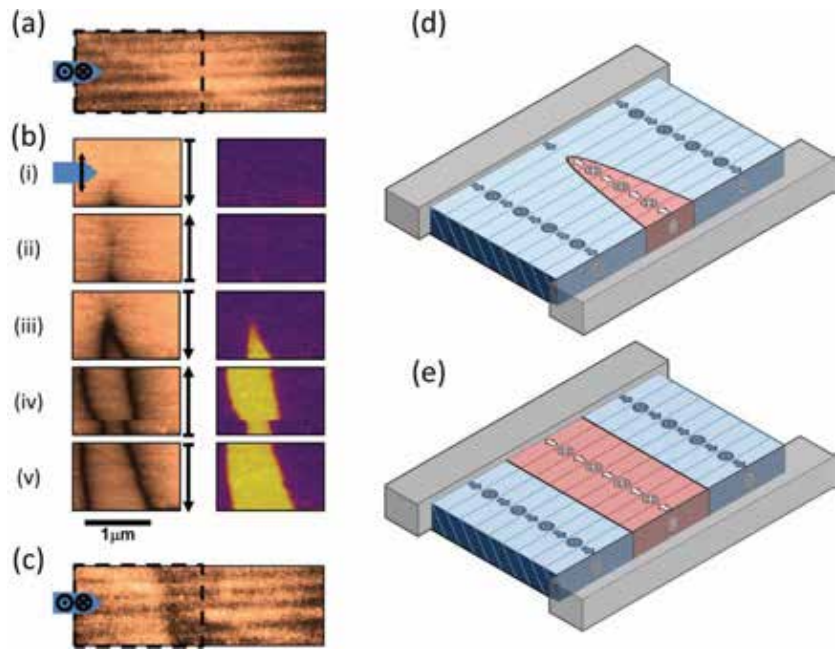

**Figure III.6.** (a) VPFM amplitude image of the initial domain structure of the $BaTiO_3$ lamella indicating the presence of a single superdomain. (b) LPFM amplitude (left panels) and phase (right panels) imaging of the area marked by a dotted line in (a) showing nucleation and growth of a new superdomain. The black arrows to the right of the amplitude panels indicate the slow scan direction during the PFM imaging (which could be used to deduce the time fl ow direction). In (i) and (ii), a small nucleated superdomain, not yet resolved in the phase signal, is revealed by the local reduction in LPFM amplitude signal. In (iii), the growth of a needle superdomain is apparent and the lateral movement of two distinct boundaries is observed in (iv) and (v). (c) VPFM amplitude image of the domain structure after backswitching where one of the superdomain boundaries is faintly visible. (d,e) A schematic illustration of the domain switching process visualized in (b). Reprinted with permission from [Sharma P, McQuaid R F P, McGilly L J, Gregg J M and Gruverman A 2013 *Adv. Mater.* **25** 1323-30]. Copyright 2013 Publisher Wiley-VCH Verlag GmbH & Co. KGaA [290].

### III.4.2. Lateral charge injection

As briefly mentioned in section III.4.1, the application of electric field can induce charge injection at lateral electrodes, that can clearly be visualized as injection and propagation of charged regions.[261] As shown in figure III.3, charge injection from the lateral electrode and propagation of the charges regions were observed in the surface potential images. As increasing the biasing time from 10 min to 1 h shown in figures III.7(a-c), the surface potential contrast at the nanowire becomes more faintly due to the surface charge mobility.[261] Further, after the negative bias was switched off, the dark contrast in the surface potential image [see figure III.7(e)] gradually becomes wider due to the surface charge mobility as well.

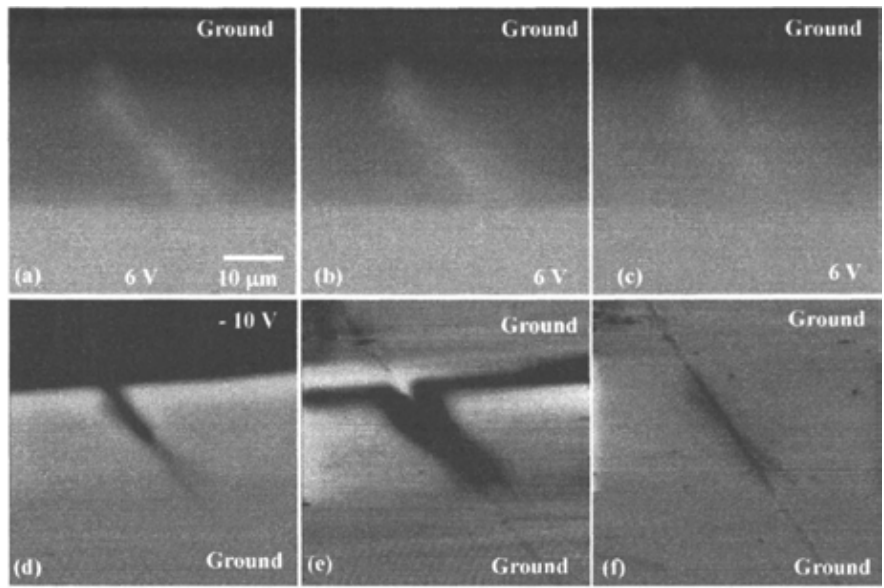

**Figure III.7.** Surface potential on the biased nanowire after (a) 10 min, (b) 20 min, and (c) 1 h scanning illustrating the smearing of potential contrast due to the mobile charge effect. (d) Surface potential at -10 V and (e) immediately after the bias is off illustrates the formation of charged negative halo. (f) After a week time, the halo disappears. The scale is 6 V [(a)-(c)], 10 V (d), and 1 V [(e) and (f)]. Reprinted with permission from [Kalinin S V, Shin J, Jesse S,

Geohegan D, Baddorf A P, Lilach Y, Moskovits M and Kolmakov A 2005 *J. Appl. Phys.* **98** 044503]. Copyright 2005, AIP Publishing LLC [261].

As a historical parallel, Shockley and Prim[292] reported space-charge limited emission for a planar electrode structures consisting of semiconductor layers. Strelcov et al.[293] reported that the charge injection can induce an metal-insulation transition because of removal of the Coulomb gap when charge injection increases the carrier concentration beyond a certain critical value. These charge injection were recently visualized by time-resolved KPFM (Tr-KPFM).[294-296] Figure III.8 presents voltage and time dependent surface potential behavior of which information can delivery charge dynamics in the planar electrode structures. The authors found that, while surface potential relaxation is correlated with surface charge motion at low biases and temperatures, ionic charge dynamics can be a dominant role in conductivity at high biases and temperatures.

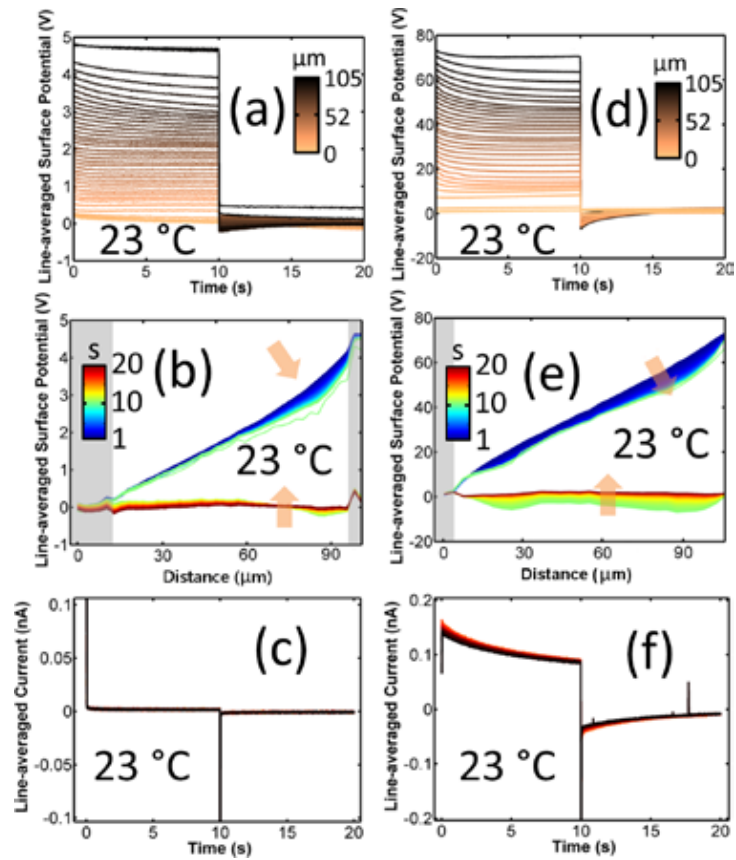

**Figure III.8.** (a-c) Line-averaged surface potential vs time and position, and interelectrode current vs time as measured at room temperature with 5Vappliedbetweenthe electrodes. (d-f) Same with 90Vappliedto electrodes; gray rectangles show positions of the lateral electrodes. Arrows and color bars in (b,e) indicate time sequence of the curves. Averaging was performed over spatial locations equidistant from the grounded electrode, which created a set of "lines", whose positions are indicated with the color bars in (a,d). Reprinted with permission from [Strelcov E, Jesse S, Huang Y, Teng Y, Kravchenko I I, Chu Y and Kalinin S V 2013 *ACS Nano* **7** 6806-15]. Copyright (2013) ACS Publications [295].

### III.4.3. Charge dynamics on ferroelectric surfaces

An interesting set of studies of surface potential behavior on ferroelectric surfaces as affected by lateral electric field and as a function of humidity was recently reported by Volinsky group.[297-299] The authors reported complete screening (disappearance of surface potential contrast) for strong humidity[297] and emergence of unscreened component for lower humidity. The authors also reported that application of strong electric bias across ferroelectric surface leads to reversible inversion of surface potential, as illustrated in figure III.9.[297, 298] This behavior can be attributed to partial removal of screening charged by sufficiently strong lateral electric field, leading to preponderance of Coulombic forces due to unscreened polarization charge.

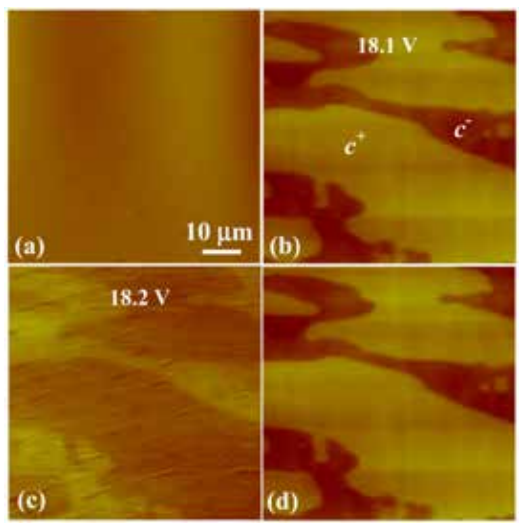

**Figure III.9.** (a) Topography image of the *c* domain. Surface potential map of the *c* domain at (b) 18.1 V and (c) 18.2 V showing surface potential inversion. (d) Complete recovery upon

switching the electric field off. Reprinted with permission from [He D Y, Qiao L J and Volinsky A A 2011 *J. Appl. Phys.* **110** 074104]. Copyright 2011, AIP Publishing LLC [297].

Strelcov et al.[294] reported charge injection through the lateral electrodes on the LiNbO$_3$ surface. They performed time-resolved KPFM at different polarizing biases. As shown in figure III.10(a), surface potential distribution for polarizing biases lower than 30 V is linear and there is only a very slight potential fluctuation on the line profile. However, as increasing the polarizing biases, surface potential curve grows larger and laterally extends from the biased electrode (see 50 and 70 V in figure III.10 (b,c). This difference between the highest surface potential and the polarizing bias at the electrode becomes larger. This behavior is attributed to the charge injection, which can be either electronic or ionic nature, from the biased electrode. Further, they directly showed the rearrangement of the screening charges by an application of the bias through the time dependent transient behavior.

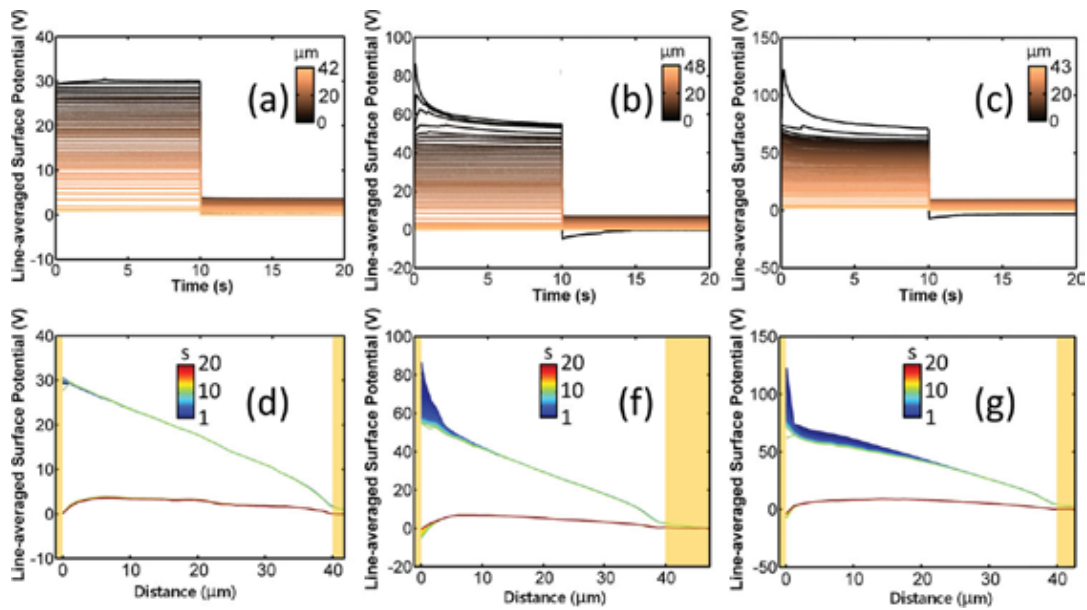

**Figure III.10.** Time evolution of the line-averaged surface potential and potential-distance profile as measured at the polarizing biases of (a, d) +30 V, (b, f) + 50V, (c, g) +70 V; electrodes positions are shown with golden rectangles; color bars in (a–c) indicate distance from the biased electrode; color bars in (d–g) indicates time; measurements performed with a flexible cantilever. Reprinted with permission from [Strelcov E, Ievlev A V, Jesse S, Kravchenko I I, Shur V Y and Kalinin S V 2014 *Adv. Mater.* **26** 958-63]. Copyright 2014 Publisher Wiley-VCH Verlag GmbH & Co. KGaA [294].

The authors further explored electrochemical polarization and relaxation. To quantify the electrochemical polarization/relaxation processes, they used a phenomenological model by fitting the surface potential evolution with an exponential decay function as follow:

$$f = A + B \times e^{-\frac{t}{\tau}} \quad (3)$$

where $\varphi$ is the surface potential, $A$ is the offset, $B$ is the pre-exponential factor, and $\tau$ is the mean lifetime. As shown in figure III.11(a), transient surface potential was observed because the bias application to the electrodes rearranges the screen charges through the charge injection. Further, as seen in figure III.11(b), the central domain has lower surface potential due to incomplete switching.

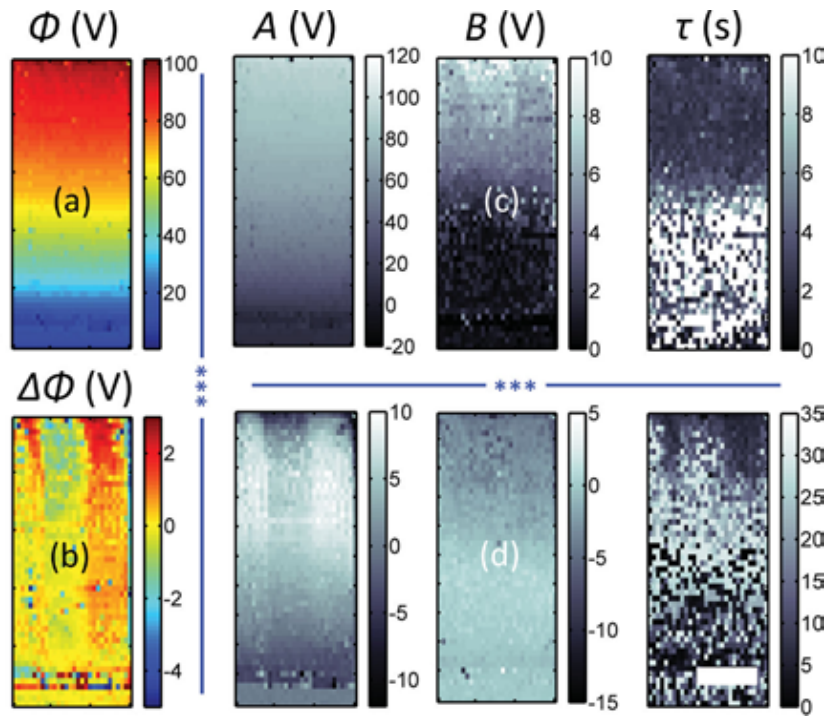

**Figure III.11.** Electrochemical-polarization by a 90 V bias pulse: (a) time-averaged surface potential distribution; (b) flattened time-averaged surface potential distribution showing ferroelectric domains. Fitting coefficients maps for: (c) electrochemical-polarization; (d)

electrochemical-relaxation. The lower part of the images was recorded on the grounded electrode; the top row was adjacent to the biased electrode. The scale bar is 10 μm. Reprinted with permission from [Strelcov E, Ievlev A V, Jesse S, Kravchenko I I, Shur V Y and Kalinin S V 2014 *Adv. Mater.* **26** 958-63]. Copyright 2014 Publisher Wiley-VCH Verlag GmbH & Co. KGaA [294].

**III.5. Variable temperature experiments**

The second aspect of domain and screening charge dynamics is their evolution with temperature. The temperature-dependent crystallographic structure and polarization in ferroelectrics are typically when known from scattering studies and can be with high degree of prediction be estimated from Ginzburg-Landau type theories. Hence, exploring kinetics and thermodynamics of surface properties (potential, piezoresponse, and topography) during the temperature changes provides direct insight into screening charge behavior.

The first studies of temperature-dependent ferroelectric properties were reported by Abplanalp by PFM[300] and Hamazaki et al.[301] by topographic imaging. In their original publication,[301] they explored temperature dependence of surface topography of materials such as BTO and $NaKC_4H_4O_6·4H_2O$, establishing excellent agrement between expected and experimentally measured corrugation values at the ferroelastic domain walls. Extensive imaging studies of temperature induced domain dynamics by PFM were reported by Allegrini group.[302, 303] They have reported domain formation and relaxation kinetics subjected to thermal cycles, behavior relevant to the local defect states. This early work was predominantly focused on the evolution of domain structures inside the material, however, quantitative studies of domain specific response as related to image formation mechanisms in SPM and screening mechanisms were of less interest.

The systematic study of domain specific properties of ferroelectric surfaces by KPFM and PFM were reported by Kalinin and Bonnell[274, 275, 306] as discussed below. In particular, here we summarize the observations of domain-specific potential evolution on heating and cooling through ferroelectric phase transition that led to observation of spurious potential increases above $T_C$, heating and cooling below phase transition that lead to observation of temperature induced potential inversion (TIPI), discuss these observation in the context of qualitative screening model, and discuss analysis of thermodynamics of screening process from this data.

### III.5.1. Surface potential evolution on heating

Polarization and charge dynamics on ferroelectric BTO (100) surfaces during on heating through ferroelectric phase transition were explored in Refs.[274, 304, 305] The temperature dependence of topographic structure and surface potential is illustrated in figure III.12. The surface topography illustrates the presence of corrugations due to ferroelastic domain walls between *a* (in plane) and *c* (out of plane) domains. While the relative number and orientation of domains does not change below $T_C$= 130°C, the surface corrugation angle, which is directly related to *c*/*a* ratio in the tetragonal unit cell, changes with temperature as shown in figure III.13.[306] Note the agreement between experimentally measured corrugation angle and the value calculated from the temperature dependence of *a*/*c* ratio in BTO, suggesting that domain structure is unclamped. The corrugation angle is established immediately after temperature change and remains constant under isothermal conditions.

The initial observations of domain potential contrast exhibit more complex dynamics, as illustrated in figure III.13(b).[304] In this case, a stepwise increase in temperature results in an *increase* of domain potential contrast, whereas isothermal annealing for ~30 min results in decrease of potential contrast. This behavior is somewhat unexpected, since polarization and hence surface polarization charge decreases with temperature, in agreement with both bulk data and measured corrugation angles.

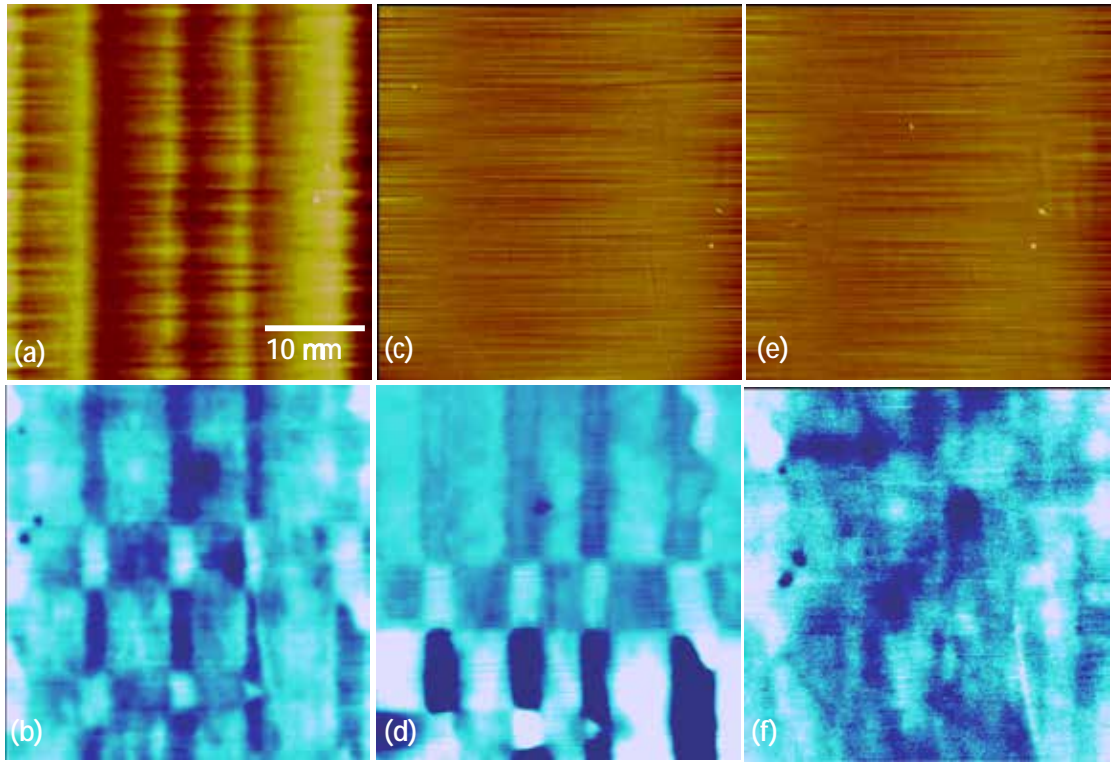

**Figure III.12.** (Top) Surface topography and (bottom) potential distribution at BTO (100) surface before ferroelectric phase transition at (a,b) 125°C, (c,d) 4 min after transition and (e,f) after 2.5 h annealing at 140 °C. Scale is (b) 0.1 V, (d) 0.5 V and (f) 0.05 V. Reprinted with permission from [Kalinin S V and Bonnell D A 2001 *Phys. Rev. B* **63** 125411]. Copyright (2001) by The American Physical Society [278].

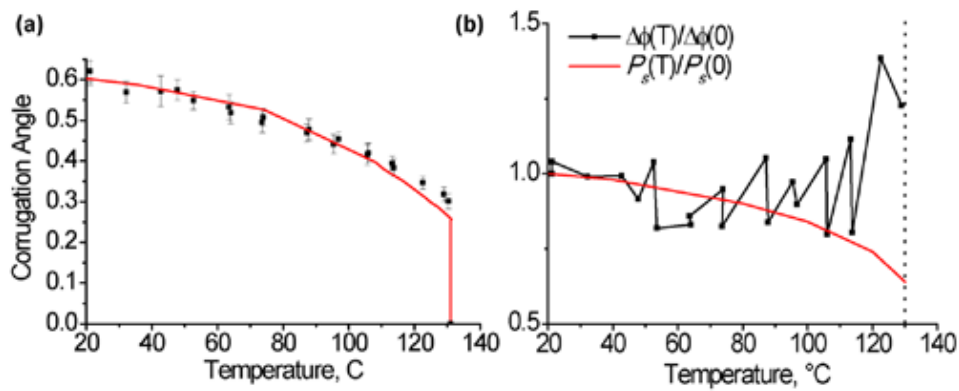

**Figure III.13.** (a) Temperature dependence of average corrugation angle on heating ( ) and cooling ( ) as compared to the calculated value (dashed line) and (b) domain potential contrast

in KPFM measurements below Curie temperature $T_C$. Panel (a) is reprinted with permission from [Kalinin S V and Bonnell D A 2000 *J. Appl. Phys.* **87** 3950]. Copyright 2000, AIP Publishing LLC [305]. Panel (b) is reprinted with permission from [Kalinin S V and Bonnell D A 2001 *Appl. Phys. Lett.* **78** 1116]. Copyright 2001, AIP Publishing LLC [304].

Even more intriguing phenomena are observed on ferroelectric transition during heating above $T_C$. In this case, ferroelectric polarization disappears as indicated by the absence of characteristic surface corrugations. However, the domain related potential features persist and potential amplitudes *increase* by almost 2 orders of magnitude. These potentials are metastable and rapidly decay with time, as illustrated in figure III.12 (image was acquired from bottom to top 4 min after the transition, total acquisition time - 11 min). After annealing for ~2 h, domain related potential contrast disappears.

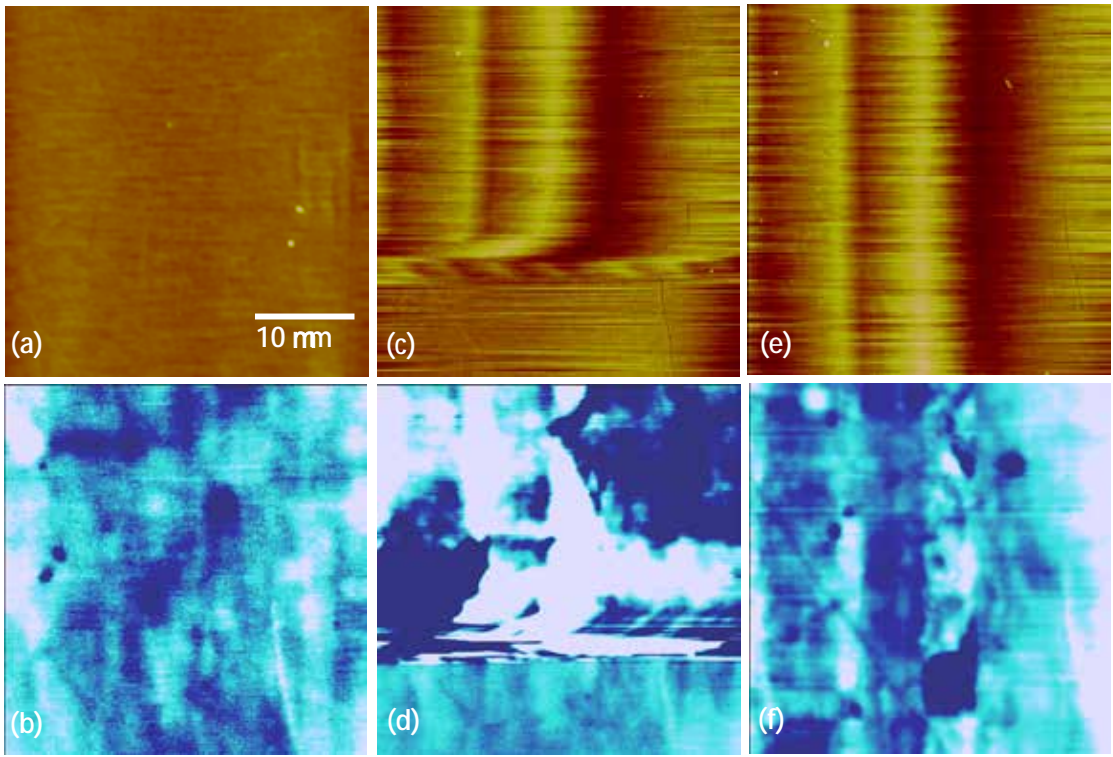

**Figure III.14.** (a,c,e) Surface topography and (b,d,f) surface potential distribution on BTO (100) surface above (a,b) $T_C$, (c,d) during the transition and (e,f) 1 h after the transition. Images are acquired from bottom to top. Scale is (a,c,e) 30 nm, (b) 0.05 V, (d,f) 0.1 V. Reprinted with permission from [Kalinin S V and Bonnell D A 2000 *J. Appl. Phys.* **87** 3950]. Copyright 2000, AIP Publishing LLC [305].

The sequence of events on the reverse transition is similar to those in the forward transition. Figure III.14(a-f) shows surface topography and surface potential distribution above the transition temperature, during the transition, and below the transition. It can be seen that during the transition the apparent topography is very volatile for ~30 s, then the new domain structure forms. New domains are oriented in the same direction as before the first transition; however, the size of the domains differs. At the transition surface potential exhibits large unstable potential amplitudes that may be attributed to depolarization currents associated with the formation of domain structure. After a relaxation period, the surface potential stabilizes and is again closely related to the new domain structure. Relaxation of newly formed potential features occurs much slower than on transition to the paraelectric phase.

The observation of slow potential dynamics on heating and spurious potentials above $T_C$ can be explained only assuming that the measured potential values are due to the interplay between (fast) polarization dynamics and (slow) screening charge dynamics. The sign of the surface potentials in this case is that of the screening charge. On increasing the temperature, polarization charge decreases, whereas screening charge relaxes only slowly with time. Hence, the increased contribution of screening charge to measured contrast results in increased surface potential. However, the system is far away of thermodynamics equilibrium and excess screening charge dissipates with time, leading to reduction of potential. Similar behavior is observed on increasing the temperature above $T_C$; however, the amount of polarization charge change in this case is considerably larger. Hence, the spurious surface potential features represent the effect of (now uncompensated) screening charges that can be observed directly by KPFM.

It is interesting to speculate what may be the role of these screening charges on other properties of ferroelectric surfaces. In particular, many groups reported the observation of weak surface ferroelectricity above bulk ferroelectric phase transitions. Given high polarizability of ferroelectric materials and potential for field-induced ferroelectric phase transition above $T_C$, the presence of the surface charges can result in surface ferroelectric phase.

## III.5.2. Temperature induced domain potential inversion

Further insight in polarization and screening charge dynamics can be obtained from variable temperature experiments below $T_C$, as reported in Ref.[306]. The evolution of surface potential and surface topography for *a-c* domain region on BTO (100) surface containing both *c+* and *c-* domains is illustrated in figure III.15(a-d). On increasing the temperature, the domain potential contrast increases, in agreement with earlier observations. Here, the relaxation of potential under isothermal conditions was explored, and the potential was found to decay with time to a stable and lower value. Typically, relaxation times of the order of 15 m – 1 h were observed, in agreement with other studies of ionic charge dynamics on oxide surfaces.

An unusual behavior is observed on decreasing the temperature, as illustrated in figure III.15(e-h). After a temperature decrease from 70°C to 50°C, the potential contrast between domains inverts, *i.e.* a positive *c* domain becomes negative. The potential difference between the domains decreases with time, passes through an isopotential point corresponding to zero domain potential contrast, and finally establishes an equilibrium value. This temperature induced domain potential inversion can be readily explained in the context of surface screening model. In this case, decrease of temperature results in the increase of polarization bound charge. For a certain amount of time, the Coulombic contribution of uncompensated polarization charge dominates the signal. Accumulation of screening charges inverts the contrast back to the case when polarity is dominated by the screening charges.

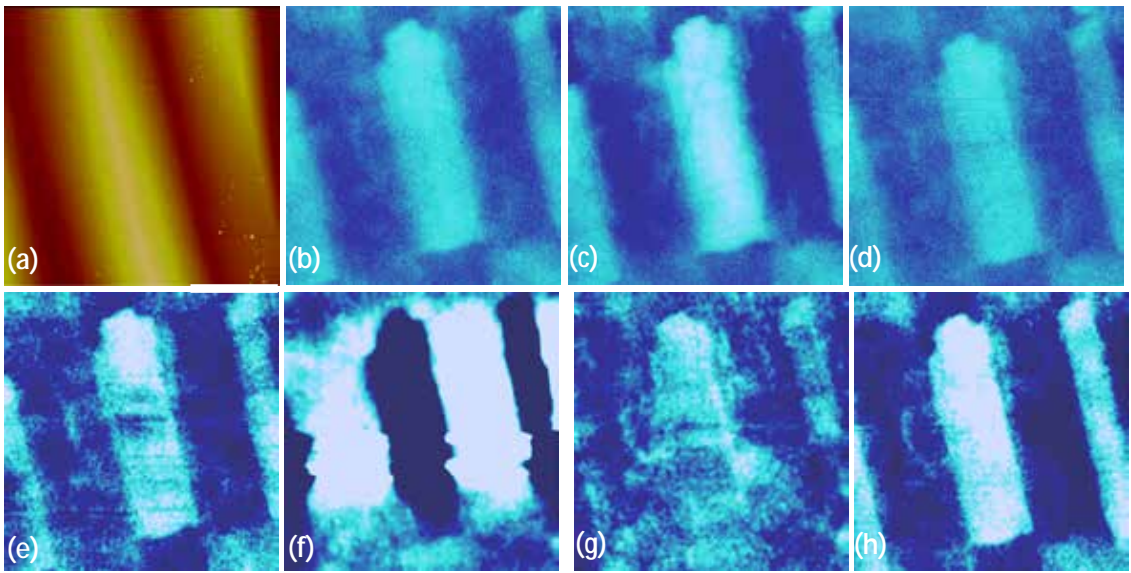

**Figure III.15.** (a) Surface topography and (b) surface potential of the ferroelectric domain structure on a BTO (100) surface at $T = 50$ °C. Surface potential (c) after heating from 50 °C to 70°C and (d) after annealing at 70 °C for 50 min. (e) Surface potential at T = 90 °C. Surface potential (f) during cooling from 90°C to 70°C, (g) at 70°C and (h) after annealing at 70°C for 50 min. Reprinted with permission from [Kalinin S V, Johnson C Y and Bonnell D A 2002 *J. Appl. Phys.* **91** 3816]. Copyright 2002, AIP Publishing LLC [306].

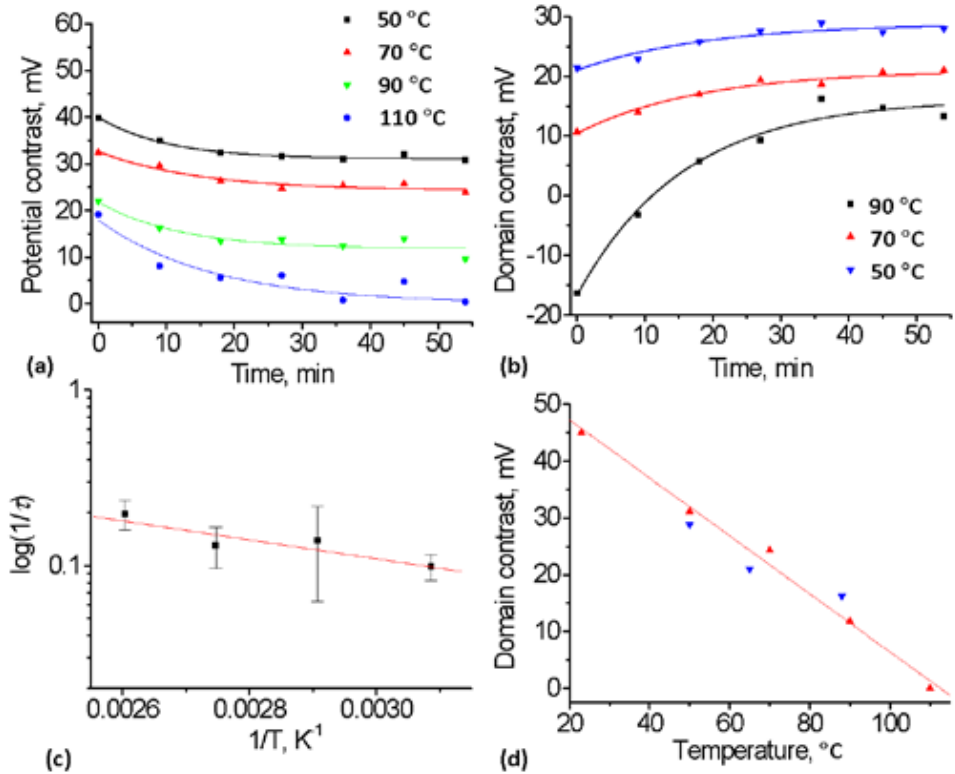

**Figure III.16.** Time dependence of domain potential contrast on (a) heating and (b) cooling. Solid lines are exponential fits. (c) Time constant for relaxation process on heating in Arrhenius coordinates and (d) temperature dependence of equilibrium domain potential contrast on heating (▲) and cooling (▼) and fit by equation (III.14) (solid line). Reprinted with permission from [Kalinin S V, Johnson C Y and Bonnell D A 2002 *J. Appl. Phys.* **91** 3816]. Copyright 2002, AIP Publishing LLC [306].

Experimentally measured kinetics of surface potential relaxation suggests that typical relaxation times are of the order of 10 min – 1 h, allowing the isothermal kinetics studies. The time dependence of domain potential contrast on heating and cooling is shown in figure III.16(a,b).[306] The kinetics of domain potential contrast, $\Delta\varphi$, was found to roughly follow the exponential law $\Delta\varphi = \Delta\varphi_0 + A\exp(-t/\tau)$, where $\tau$ is relaxation time and $A$ is a prefactor. The temperature dependence of the potential redistribution time is shown in figure III.16(c). The redistribution time is almost temperature independent, with associated relaxation energy of ~4 kJ/mole, suggesting that the kinetics of relaxation process is limited by the transport of chemical species to the surface. The characteristic redistribution time is ~ 20 min and is close to the relaxation time for domain potential contrast above $T_c$ (15 min). Notably, the SPM based studies have extremely limited bandwidth (~3-10 min for imaging experiments, 1-3 s for line imaging even excluding time for surface stabilization, typically of the order of 1-3 min). Hence, the presence of faster relaxation processes cannot be established from SPM data and measured dependence represents only the tails of relaxation process.

### III.5.3. Thermodynamics of screening

Fortuitously, the potential relaxation on heating and cooling is sufficiently fast so that thermodynamics equilibrium can be achieved in the experimentally accessible time scales. In particular, the relaxation both on heating and cooling to the same final temperature result in the same equilibrium value of domain potential contrast, $\Delta\varphi_0$. The temperature dependence of domain potential contrast in the temperature interval 30 °C < $T$ < 100 °C is linear

$$\Delta V_{dc} = 0.059 - 5.3 \times 10^{-4} T. \qquad (III.14)$$

Interestingly, the zero potential difference corresponding to temperature ~110 °C, well below the Curie temperature of BTO ($T_c$ = 130°C).

Experimentally observed dependence of surface potential on temperature suggests that equilibrium degree of screening depends on temperature. While at low temperatures screening is close to unity and measured potential is dominated by that of the dipole layer, on increasing the temperature the degree of screening decreases. At 110 °C, the contributions of the dipole layer and Coulombic forces from unscreened polarization charged are mutually compensated, corresponding to the absence of potential contrast. At higher temperatures, the sign of measured potential is that of polarization charges. Note that the temperature corresponding to the isopotential point depends on humidity. While no systematic KPFM *vs. T*, *p(H₂O)* studies were reported, isothermal variable humidity data by Volinsky group[297, 299] suggest that at room temperature isopotential point corresponds to 45 % relative humidity.

The observed dependence of potential contrast and degree of screening on temperature naturally leads to the question on whether thermodynamic parameters of the process can be obtained from these data.[306] This requires solution of two related problems, namely, (a) establishing the thermodynamics of screening process in terms of enthalpy and entropy of screening process and temperature-dependent ferroelectric properties of material, and (b) establishing the relationship between the effective potential measured by KPFM and degree of screening. Below we discuss both problems.

To describe the thermodynamics of screening surface, we assume that the surface of a ferroelectric material is characterized by a polarization charge density $\sigma = \mathbf{P} \cdot \mathbf{n}$, where $\mathbf{P}$ is the polarization vector and $\mathbf{n}$ is the unit normal to the surface. The free energy for screening process by external ionic charges can be written as:

$$E(\alpha,T) = E_{el}(\alpha,T) + \alpha \frac{P}{qN_a}\Delta H_{ads} - \alpha \frac{P}{qN_a} T \Delta S_{ads} + E_{dw} \quad \text{(III.15)}$$

where $q = 1.602 \times 10^{-19}$ C is electron charge, $P$ is spontaneous polarization, $N_a = 6.022 \times 10^{23}$ mol$^{-1}$ is Avogadro number, $\alpha$ is the degree of screening and $T$ is the temperature. Experimentally, the degree of screening is very close to unity, $\alpha \approx 1$. Therefore, it is convenient to introduce as the small parameter the fraction of the unscreened charge, $g = 1 - \alpha$. The enthalpy and entropy of adsorption are denoted $\Delta H_{ads}$ and $\Delta S_{ads}$, respectively. Experimentally, the domain wall area is observed to be constant during the measurement and hence the corresponding free energy, $E_{dw}$, is assumed independent on the degree of screening. The electrostatic contribution to the free energy, $E_{el}(g,T)$, can be derived as:

$$E_{el}(g,T) = \frac{P^2}{\varepsilon_0 + \sqrt{\varepsilon_x \varepsilon_z}} \left\{ g^2 L \frac{7\zeta(3)}{2\pi^3} + (1-g)h\frac{\varepsilon_0}{\varepsilon_2} + (1-g)^2 h \frac{\sqrt{\varepsilon_x \varepsilon_z}}{\varepsilon_2} \right\} \quad \text{(III.16)}$$

where $L$ is the domain size, $h$ is the screening layer width, $\varepsilon_2$ is the dielectric constant of the screening layer, $\varepsilon_x$ and $\varepsilon_z$ are the dielectric constants of the ferroelectric and $\varepsilon_0 = 8.854 \times 10^{-12}$ F/m is the dielectric constant of vacuum.

The temperature dependence of the equilibrium screening can be obtained from the condition of the minimum of free energy $\frac{\partial E(g,T)}{\partial g} = 0$. Since $E_{el}(g,T)$ is a quadratic function of $g$ this condition can be written as

$$\frac{\partial E_{el}}{\partial g} = -b_1(T)g + b_2(T), \qquad (III.17)$$

where $b_1$ and $b_2$ are material-dependent constants that can be estimated from the Ginzburg-Landau theory and parameters of domain structure. Calculated temperature dependencies of $b_1$ and $b_2$ for domain size $L = 10$ nm, $\varepsilon = 80$ (water) and $h = 0.1$ nm are shown on figure III.17. It is clearly seen that $b_1$ and $b_2$ are only weakly temperature dependent. The physical origin of this behavior is that the product $\varepsilon_0\varepsilon$ is only weakly temperature dependent. Consequently, $b_1$ and $b_2$ can be approximated by their room temperature values. At $T = 25°C$ for $h = 0.1$ nm $b_1 = 26.67$ J/m$^2$ and $b_2 = 0.02034$ J/m$^2$. Hence, equation (III.18) suggests that the degree of screening is a linear function of temperature.

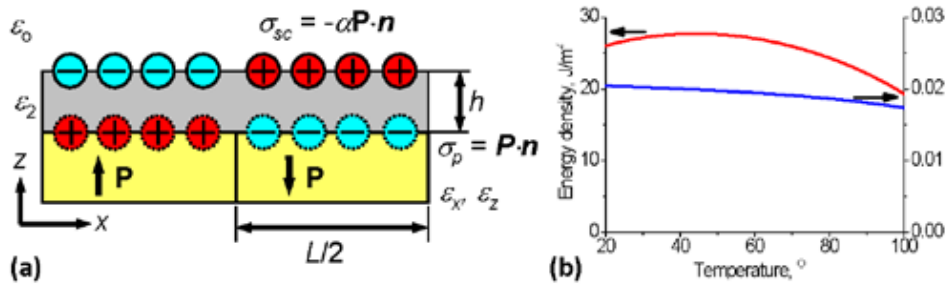

**Figure III.17.** (a) Charge distribution on the partially screened anisotropic ferroelectric surface and (b) temperature dependence of material constants. Reprinted with permission from [Kalinin S V, Johnson C Y and Bonnell D A 2002 *J. Appl. Phys.* **91** 3816]. Copyright 2002, AIP Publishing LLC [306].

From equation (III.15) the equilibrium degree of screening is defined by

$$g(T) = -\frac{P}{qN_a}\frac{\Delta H_{ads}}{b_1(T)} + \frac{b_2(T)}{b_1(T)} + T\frac{P}{qN_a}\frac{\Delta S_{ads}}{b_1(T)}, \qquad (III.18)$$

where $b_1(T)$ and $b_2(T)$ are temperature dependent coefficients defined by domain structure and material properties.

The second problem in the analysis of the KPFM data is the establishing the relationship between the measured effective potential and degree of screening. This analysis can be performed for a defined geometry of the SPM probe using image charge methods developed by Belaidi and others.[307] For a typical metal coated tip used in the KPFM measurements with $\theta = 17°$, $H \gg 10$ mm and tip-surface separation $z = 50$-$100$ nm equation (III.18) can be approximated as

$$\Delta V_{dc} = (V_1 - V_2) - 1.18 H (E_1 - E_2), \tag{III.19}$$

since the logarithmic term is only weakly dependent on the tip length. Under the experimental conditions (lift height 100 nm), the deviation between the true and measured domain potential difference does not exceed ~30 % and the uncertainties in the other parameters (tip shape model and materials properties) are expected to be comparable.

Using the representation of a partially screened ferroelectric surface as a superposition of completely unscreened and completely screened regions, the potential difference between domains of opposite polarity is

$$\Delta j_s = (1 - g) \frac{2h\sqrt{\varepsilon_x \varepsilon_z} P}{\varepsilon_2 (\varepsilon_0 + \sqrt{\varepsilon_x \varepsilon_z})}, \tag{III.20}$$

while the difference in the normal component of the electric field is

$$\Delta E_u = g \frac{P}{\varepsilon_0 + \sqrt{\varepsilon_x \varepsilon_z}}. \tag{III.21}$$

Hence, the measured potential difference between the domains is

$$\Delta V_{dc} = (1 - g) \frac{2h\sqrt{\varepsilon_x \varepsilon_z} P}{\varepsilon_2 (\varepsilon_0 + \sqrt{\varepsilon_x \varepsilon_z})} - bHg \frac{P}{\varepsilon_0 + \sqrt{\varepsilon_x \varepsilon_z}} \ln\left(\frac{H}{4d}\right)^{-1}, \tag{III.22}$$

*i.e.* domain potential contrast is a linear function of degree of screening.

When the force acting on the biased tip above a partially screened surface is written as

$$F = \frac{dC_z}{dz} (V_{tip} - V_{surf})^2 + C_z V_{tip} E_z \tag{III.23}$$

,where $E_z$ is the normal component of the electric field due to unscreened polarization charge, the combination of equations (III.22 and III.23) for a tip length of 10 mm yields the temperature dependence of equilibrium degree of screening

$$g = 1.627 \times 10^{-5} + 1.23 \times 10^{-6} T. \tag{III.24}$$

A comparison of equations (III.18 and III.24) allows the enthalpy, $\Delta H_{ads}$, and entropy, $\Delta S_{ads}$, of adsorption to be determined as $\Delta H_{ads}$ = 164.6 kJ/mole, $\Delta S_{ads}$ = -126.6 J/mole K. The enthalpy and entropy of adsorption thus obtained are within expected values in spite of the approximations inherent in this approach. Moreover, from equations (III.22 and III.24) the Coulombic contribution to the effective potential can be estimated as <10-20 % thus validating our previous conclusion that the surface is completely screened at room temperature. The nature of the screening charges cannot be determined from these experiments; however, these results are consistent with the well-known fact that water and hydroxyl groups, -OH, adsorb on oxide surfaces in air. The adsorbed water can provide the charge required to screen the polarization bound charge, since corresponding polarization charge densities are of order of 0.25 C/m$^2$ corresponding to 2.6×10$^{-6}$ mole/m$^2$. For a typical metal oxide surface with characteristic unit cell size of ~ 4 Å, this corresponds to the coverage of order of 0.25 ml. Dissociative adsorption of water as a dominant screening mechanism on BTO surface in air was verified using temperature programmed desorption experiments on poled BTO crystals. At the same perovskite (100) surfaces are free from the midgap surface states; therefore, the screening on BTO (100) surface cannot be attributed to the surface states filling.

## III.6. Environmental effects

The screening on ferroelectric surfaces can be expected to be strongly affected by the environment. First and foremost, the thermodynamics of screening process is directly controlled by the chemical potential of screening species. Furthermore, the properties of the screening layer will be strongly affected by the presence of *e.g.* wetting water layer, since associated high dielectric constant will strongly enhance dissociation of neutral species in charged ions. Here, we discuss the SPM based studies of environmental effects on screening as explored through the measured surface potentials.

As an example of such study, Kitamura group has explored surface potential of LiNbO$_3$ under different conditions, and observed strong dependence of KPFM contrast on environment, as could be expected for screening by ionic species.[281] The KPFM studies of BTO surface in ultrahigh vacuum are reported by Watanabe.[308, 309]. Interestingly, the authors report that the potential difference between *c+* and *c-* domains is ~100 mV, close to the value observed in ambient environments. The authors ascribe this behavior to the intrinsic shielding mechanisms and link this behavior to the formation of surface electron layers, in agreement with metallic-like conductivity of BTO surface reported by the same group.[139, 310] Notably, metallic conductance was recently discovered on the ferroelectric domain walls,[17] confirming this original explanation.

However, it should be noted that potential difference between domains of opposite polarity for intrinsic screening is expected to be of the order of band gap and hence of the order of 3 V for BTO.[311] The experimentally observed ~100 mV potential difference between the domains can be ascribed either to the finite resolution of KPFM, or the screening by surface adsorbates that persist on transfer the sample from ambient to UHV environment.

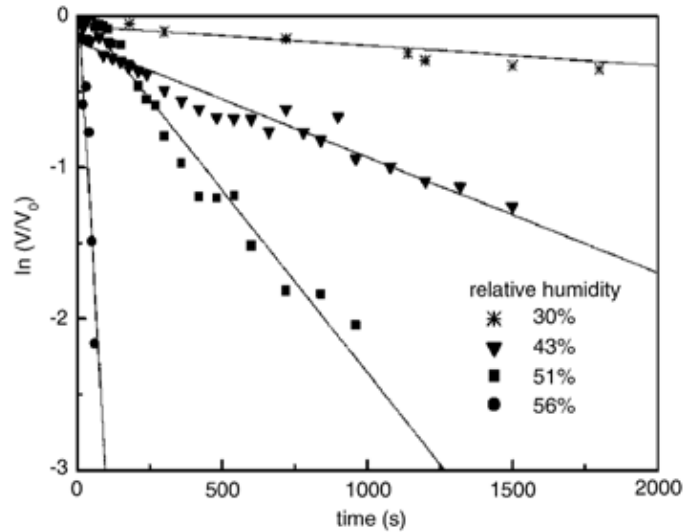

**Figure III.18.** The time dependence of the maximum potential $V$ normalized to the maximum potential $V_0$ of the first potential distribution after charging. Reprinted with permission from [Debska M 2005 *J. Electrostatics* **63** 1017]. Copyright (2005) with permission from Elsevier [312].

It is also possible that charge dynamics can be affected by local defects during and/or after switching and relative humidity.[284, 312] Since grain boundaries in polycrystalline films are relatively conductive, as mentioned before, the grain boundaries can affect screen charge migration on the film surfaces.[284] The charge dynamics explored macroscopically by induction probe shows faster relaxation under the high relative humidity in triglycine sulfate crystal (TGS) in figure III.18. Since the TGS crystal surface is dissolved by the water layer, the surface properties can be modified under the high humidity which can affect surface charge relaxation. This indicates the predominance of surface conduction under high relative humidity in TGS.[312]

Further systematic studies of environmental and humidity effects on domain-related surface potentials in ferroelectrics are reported by Volinsky group. The authors report complete disappearance of surface contrast for strong humidity. [313] Finally, it is well recognized that the thermodynamics of domain structure and hence screening can be affected not only by status of the top surface, but also behavior on the interfaces, especially for this films. An attempt to explore these phenomena in beveled PZT/Pt structures were reported by Lu et al.[314] The authors report on the presence of 240 nm transition layer visible both as a gradual decrease of the piezoresponse signal and a variation of the surface potential. The effect of this layer on retention behavior was further explored.

Several limitations of the KPFM and related force based SPM techniques for exploring screening phenomena on oxide surfaces should be mentioned. By definition, these techniques measure all electrostatic forces acting on the probe, and separation of dipole layer and uncompensated charge contributions requires numerical analysis of the tip-surface interactions.[244, 307, 315-317]. Such de-convolution has relatively low veracity, and encounter additional problems on topographically-non-uniform surfaces. The future progress can be achieved by probes that can separate electric field and surface potential effects. One such approach can be based on field effect probes. [318]

Another interesting direction can be based on using probes with defined and controlled charge state (as opposed to bias). This can be achieved using chemically functionalized probes,[319] similar to approach used in chemical force microscopy.[320, 321]

The theoretically expected role of screening charges on thermodynamics of domain formation suggests that tip-induced domain switching will be strongly dependent on environmental factors, most notably humidity. In recent years, humidity effects on tip-induced domain switching were explored by several groups.[322, 323] Shur et al.[323] has demonstrated the significant role of humidity on domain growth kinetics, behavior ascribed to the conductivity and charge mobility of screening depolarization charges. They have further analyzed the kinetics of the switching process is controlled by the interplay of the field of surface charges and depolarization and screening field. It should also be noted that similar phenomena were explored in the context of charge writing and tip induced electrochemistry.

### III.7. Interplay between screening and size effects in ferroelectric nanostructures

The interesting application of the screening charge dynamics to explore ferroelectric size effects was reported by Spanier et al.[324] Using single-crystalline nanowires[325] with controlled size, they used EFM to create the charged regions. The charge stability is directly related to the ferroelectric state of material, with charge stable in ferroelectric state when it is coupled to polarization, and unstable in ferroelectric state. Also note that while charge provides only indirect evidence on ferroelectric state, the non-contact measurements are much less destructive. For example, equivalent measurements by PFM would imply application of high electric fields to the material, and may affect the polarization distributions inside the material (especially close to $T_C$).

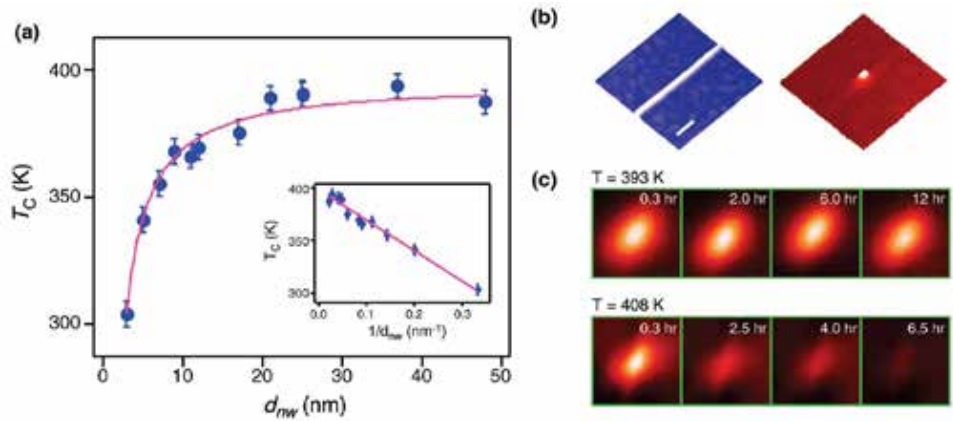

**Figure III.19.** (a) ferroelectric phase transition temperature, $T_C$, as a function of $d_{NW}$. The solid circles are the experimentally determined $T_C$. The magenta solid line is the result of the fit to the data using the $1/d_{NW}$ scaling relation. The inset plots $T_C$ as a function of $d_{NW}$ and illustrates the inverse-diameter dependence. (b) (left) A topographic image and (right) a spatial map of the electric polarization of an 11 nm diameter nanowire. (c) Time series of EFM images for a 25 nm diameter nanowire following the polarization writing below and above the phase transition temperature. The writing and reading conditions were identical to those in (b). Reprinted with permission from [Spanier J E, Kolpak A M, Urban J J, Grinberg I, Ouyang L, Yun W S, Rappe A M and Park H 2006 *Nano Lett.* 6 735-9]. Copyright (2006) ACS Publications [324].

As presented in figure III.19, the authors also reported that the Curie temperature is depressed as the nanowire diameter ($d_{NW}$) decreases, following a $1/d_{NW}$ scaling.[324] The diameter at which $T_C$ falls below room temperature is determined to be similar to 3 nm, and extrapolation of the data indicates that nanowires with $d_{NW}$ as small as 0.8 nm can support ferroelectricity at lower temperatures. The authors further analyzed the possible screening mechanisms, and provided evidence towards polarization stabilization by adsorption of charge ions. In particular, they report that ionic screening to be more efficient than that by conductive electrodes, in agreement with later studies by Gruverman and others.[326]

Groten et al.[227] reported thickness dependent pyroelectric effect and its correlation with the spontaneous polarization. The authors found that the induced temperature change decreases with increasing thickness of the substrate in the ferroelectric semicrystalline copolymer thin films. The authors further observed the temperature-induced change in the spontaneous polarization through the observation of the screening behavior. Later, Johann and Soergel[327] explored surface potential response to the induced temperature change in the lithium niobate. The authors found the linear relationship between the surface potential and the induced temperature change and, in particular, also directly probed relaxation of the surface potential under the 1 °C of temperature change.

**IV. Tip induced switching**

In general, when sufficient voltage is applied through the AFM tip in the ferroelectric materials, switching events can occur accompanied with charge injection. The injected charge often induces unexpected back switching due to the electric field induced by the injected charges. These injected charges are thermodynamically unstable because excess amounts of charges can be injected onto the ferroelectric surfaces during the bias application. As a corollary, the collection of the injected charges is feasible when a grounded tip is contacted with the charged ferroelectric surfaces. However, in fact, the charge injection can occur in not only ferroelectric and but also non-ferroelectric surfaces. Furthermore, it has been discovered that the switching events are possible when high pressure is applied to the ferroelectric surfaces, the behavior is originally attributed to flexoelectricity[328, 329] but that can also be explained by pressure induced charge dynamics. In this section, tip induced switching in both ferroelectric and non-ferroelectric surfaces are reviewed.

## IV.1. Charge injection in non-ferroelectric surfaces

When a voltage is applied to the sample surface through the AFM tip, charge injection can occur on the sample surface. In such a case, the injected charges can be trapped on the sample surface or in the bulk of the sample.[330] On the $SiO_2$ surfaces, the injected charges can be further detrapped by migration along the sample surface by Ohm's law and/or tunneling toward the underlying conducting layer if the film thickness is sufficiently thin. If the relaxation time for the detrapping of the injected charges is not significantly fast compared to the AFM imaging time, the injected changed can be imaged by either EFM or KPFM.

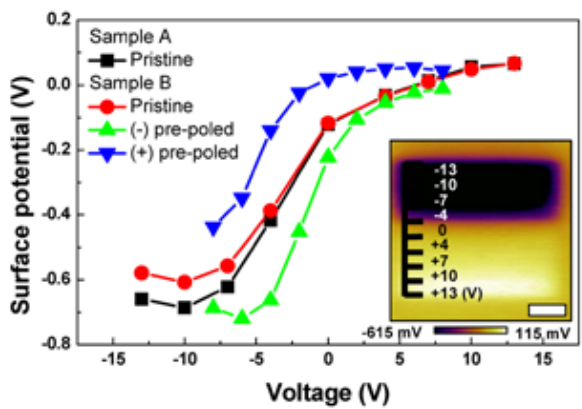

**Figure IV.1.** Tip induced surface potential as a function of poling bias on pristine (as-grown) surface (in $TiO_2$ samples *A* and *B* fabricated under the same conditions) and surface uniformly prepoled by ±8 V. Surface potential for each bias was averaged from each rectangular pattern. Inset shows the surface potential image of the area scanned with the applied biases from -13 to +13 V to the conductive probe for pristine $TiO_2$ thin film. The scale bar is 1.5 μm. Reprinted with permission from [Kim Y, Morozovska A N, Kumar A, Jesse S, Eliseev E A, Alibart F, Strukov D and Kalinin S V 2012 *ACS Nano* **6** 7026-33]. Copyright (2012) ACS Publications [331].

Shown in figure IV.1 is the example of the charge injection on the $TiO_2$ surface. The tip-induced surface potential was measured by KPFM after sequential biasing of the sample surface as a function of bias in a similar manner with figure IV.1.[331] As shown in figure IV.1, the surface potential is strongly dependent on the sign of the pre-biasing, *i.e.* pre-poling, and it is very analogous to the KPFM hysteresis loop. These changes in the surface potential are an expected effect of bulk charge injection as mentioned above.[332] As applying a negative (positive) bias to the AFM tip, negative (positive) charges are injected onto the sample surface. As a result, the negatively (positively) biased regions show relatively high (low) surface potential. However, since an initial surface potential depends on the pre-biasing state, hysteric behavior was observed in figure IV.1. The surface potential in the biased non-ferroelectric surfaces is mostly likely related to the charge injection.

**IV.2. Potential evolution during switching**

The charge dynamics and its corresponding surface potential under an application of a bias onto the ferroelectric surfaces can be somewhat different compared to the non-ferroelectric surfaces due to the existence of spontaneous polarization. Hence, of a paramount interest is the role of screening charges on the kinetics of tip (or, more generally, top electrode) induced switching processes. The polarization switching induced by micro-patterned top electrodes is of extreme interest in the context of electro-optical device fabrication.[333, 334] However, the wave of interest to these applications emerged in the context of PFM based studies of domain growth, and especially ferroelectric based data storage.[318, 335, 336] In fact, multiple authors reported ferroelectric bit sizes as small as several nanometers. J.-M. Triscone group reported that the minimum bit size can be determined by the radius of the AFM tip using PFM.[337] Using SNDM, Y. Cho group have demonstrated domains as small as 2.5 nm, corresponding to 10.1 Tb/in$^2$ storage density directly approaching molecular limits.[338] Further, N. Tayebi et al. demonstrate 2 nm (radius) of bit size, corresponding 40 Tbit/in$^2$, using single-walled carbon nanotube tip.[339]

The strong interest to tip-induced domain writing has spurred extensive theoretical effort for describing thermodynamics and kinetics of this process. Thermodynamics of domain formation in rigid dielectric approximation (*i.e.* polarization magnitude adopt bulk values within and outside the domain and changes jumpwise at the boundary) was explored in a series of works by Molotskii[340] and Morozovska,[341] with several special cases also reported by Durkan,[342] Emelyanov,[343] and Kalinin.[344] The dynamics in rigid approximation and assuming the viscous damping of domain walls was explored by Molotskii et al.[345] Finally, Chen et al. and Morozovska et al. explored the thermodynamics of switching assuming Ginzburg-Landau dynamics for polarization field, predicting intrinsic character of switching process in agreement with experimental data.[346, 347]

The notable aspect of this theoretical effort was the observation of the non-trivial role of electrostatic depolarization energy, and hence screening phenomena, on domain energy. In fact, Morozovska et al., demonstrated that the nucleation bias and activation energy for domain switching diverge with degree of screening, and, for unscreened surface, tip-induced polarization switching is thermodynamically impossible.[62] This observation suggests that while screening cannot be observed directly by PFM-based studies of domain dynamics, the indirect effects of screening on domain dynamics will be very significant. Furthermore, some information on charge dynamics during switching can be obtained by combination of PFM and KPFM studies, as described in the rest of this section.

The screen charge behavior during switching is complicated due to the charge compensation between polarization, screen, and injected charges.[277, 286, 348] Early work done by Chen et al. show that the surface charge trap is a dominant effect over the ferroelectric polarization when an external electric field is applied to the tip on PZT thin films.[286] It has also been demonstrated that the surface potential depends on pulse voltage and duration applied to the ferroelectric films.[286, 349] This work shows the injected charges during switching on the ferroelectric surfaces dominantly contributed to the resulted surface potential measured by KPFM. Indeed, in most of reports, the sign of the surface charge on the switched region was observed as an opposite to polarization charge due to the injected charges during switching.[350]. However, the dominant charge source can be dependent on the applying an electric field.[350] While polarization change can contribute dominantly to the surface potential under a low electric field, surface charge trap is a dominant effect under applying a high electric field. The contribution from surface charge trap is dependent on the materials. For instance, the $SrBi_2Ta_2O_9$ (SBT) thin films can easily trap surface charges than that of the PZT thin films. Also, the enhanced contribution from polarization charge can be observed after a grounded tip scan.

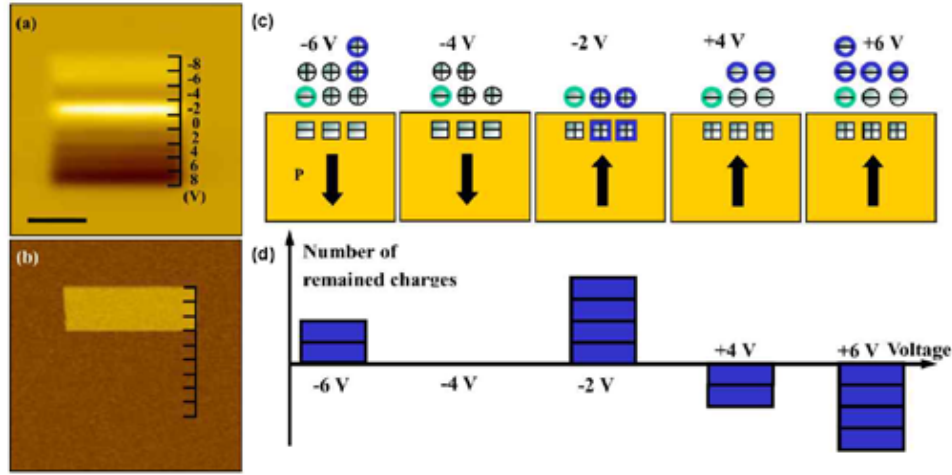

**Figure IV.2.** (a) KPFM surface potential distribution and (b) PFM phase image of the area scanned with the applied voltage biases from −8 to 8 V with a 2 V step (from top to bottom) to the bottom electrode. The black scale bar presents 2 µm. Schematic depiction of (c) polarization states and charge distributions on different applied biases and (d) the number of charges that remained after charge compensation. The rectangular and circular charges show, respectively, polarization and screen charges. The green circular charge represents the screen charge of the as-deposited state, and the blue charge represents the charge that remained after charge compensation. Reproduced with permission from [Kim Y, Bae C, Ryu K, Ko H, Kim Y K, Hong S and Shin H 2009 *Appl. Phys. Lett.* **94** 032907]. Copyright 2009, AIP Publishing LLC [277].

Finally, Kim et al. reported the origin of surface potential during switching by combination of PFM and KPFM.[277] Since PFM and KPFM allow exploring information on the polarization and the surface potential, respectively, and an amount of injected charges is dependent on the magnitude of applying voltage, the field dependent studies by combination of PFM and KPFM can provide full information on the charge sources during switching on the ferroelectric surfaces. The resultant absolute value and sign of surface potentials are defined by the interplay of polarization, injected charges, and as-deposited screening charges (see figure IV.2.). Notably, injected charges on the film surfaces can be removed by subsequent grounded tip scans which are correlated to the amounts of residual charges.[277, 351]

After the switching, screening charges are relaxed as a function of elapsed time. If polarization is stable after switching, the charge dynamics can be governed by Coulombic repulsion between screen charges.[284, 352, 353] Hence, the surface charges after applying a high electric field show faster relaxation due to larger Coulombic interaction between screen charges.[284, 353] In particular, in the polycrystalline films, the grain boundaries can affect relaxation of screen charge migration on the film surfaces since grain boundaries in polycrystalline films are relatively conductive.[283, 284]

## IV.3. Backswitching

An interesting aspect of the screening charge dynamics and its role of polarization switching in PFM is so called anomalous domain switching and formation of bubble domains. Experimentally, this phenomenon manifests as back-switching of polarization under the tip, *i.e.* formation of a small domain with polarization orientation against the field. This effect was first reported by Abplanalp et al.[354] and attributed to the higher-order polarization switching driven by Maxwell stress as shown in figure IV.3.

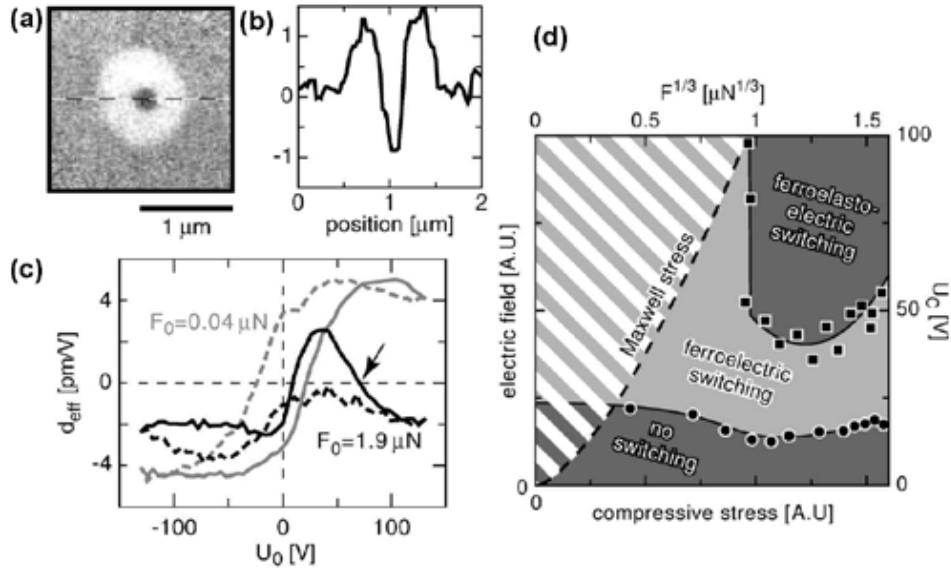

**Figure IV.3.** (a) Domain pattern created by a single voltage pulse applied under high mechanical force, imaged by PFM. The dark center at the place the tip was positioned during poling corresponds to an antiparallel alignment of the spontaneous polarization and poling field. (b) Piezoresponse measured along the horizontal line indicated in (a). (c) Local hysteresis loops measured under an applied force on a 4 μm thick BTO thin film. The solid lines correspond to increasing voltage and the dashed lines correspond to decreasing voltage. Under high mechanical load the response switches sign again (arrow), resulting in the domain pattern shown in (a). (d)

Dependence of the coercive voltage (points) and the critical voltage $U_C$ at which the piezoelectric response $d_{eff}$ is maximal (squares) is shown as a function of the third root of the total force present during domain formation. These experimental points separate areas of different switching mechanisms. The applied electric field results in the indicated Maxwell stress. The dashed area to the left was not accessible during these experiments because the Maxwell stress could not be complensated. Reprinted with permission from [Abplanalp M, Fousek J and Günter P 2001 *Phys. Rev. Lett.* **86** 5799]. Copyright (2001) by The American Physical Society [354].

This interpretation was challenged by Bühlmann et al,[355] who attributed this behavior to the effect of injected charges under the tip. This effect can be illustrated as follows (see figure IV.4). Application of the positive bias to the tip creates downward electric field in the tip-surface junction that results in the formation of downward oriented ferroelectric domain. However, this process is also associated with injection of positive charges on the sample surface that are generated in the tip surface junction and spread away from it under collective action of diffusion and migration transport. This charge spot can be easily detected by KPFM, as reported by several groups,[280, 351, 356] and the fact that measured potential is opposite to polarization charge suggests that the surface is over-screened.

On turning off the tip bias while still in contact with the surface, the tip is now effectively grounded. However, the presence of sluggish injected surface charges results in the onset of electric field oriented opposite to initial field. The action of this field can result in polarization back-switching, giving rise to the bubble domains.

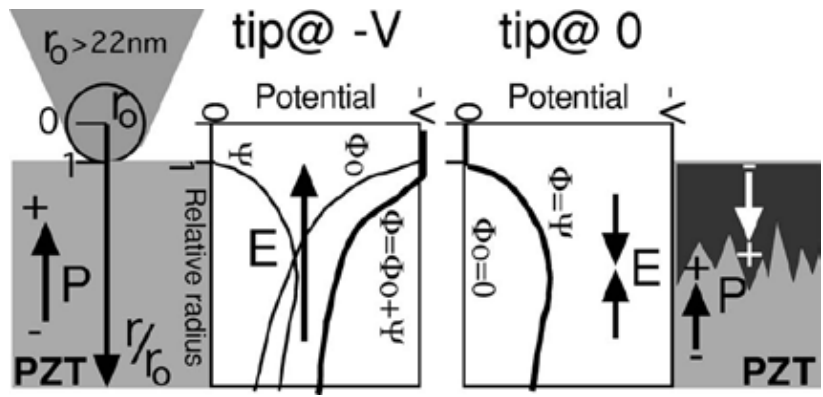

**Figure IV.4.** Schematic presentation of the anti-poling effect: (left) poling at high voltage with charge injection and (right) collection of surface charges by grounded tip and formation of opposite field in the upper part of the film, causing polarization switching. The zigzag shaped domain wall between the upper and lower film regions helps to reduce electrostatic energy of the head-to-head configuration. In the center, schematic representation of the contributing potentials, according the results of the spherical model. Reprinted with permission from [Bühlmann S, Colla E and Muralt P 2005 *Phys. Rev. B* **72** 214120]. Copyright (2005) by The American Physical Society [355].

Following this initial observation, the anomalous switching was explored by a number of authors, including Kim,[351, 356] Brugere,[357] Li,[358] Kan,[359] Kholkin,[280] Soergel,[360] and others. In particular, the injection charge assisted polarization reversal was not observed on epitaxial films with a thickness below 130 nm explored by Bühlmann et al.[355] However, Kim et al show the effect can occur even in 50 nm polycrystalline thin films due to the higher defect density.[356]

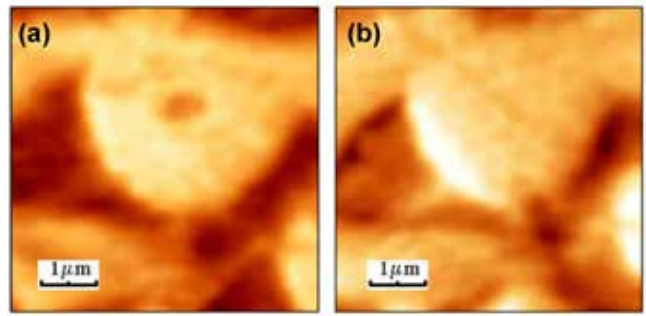

**Figure IV.5.** PFM images before (a) and after (b) illumination of an inverse domain with a band-gap UV light. Reprinted with permission from [Kholkin A L, Bdikin I K, Shvartsman V V and Pertsev N A 2007 Nanotechnology **18** 095502]. © IOP Publishing. Reproduced by permission of IOP Publishing. All right reserved (2007) [280].

Later, Kholkin et al. [280] reported that its origin is indeed correlated with trapped charges. They performed the similar experiments and observed the anomalous switching as well. To explore the origin of the observed anomalous switching, the anomalous switched region was monitored before and after illumination with a band-gap UV light. As presented in figure IV.5, the anomalous switched region was disappeared after the illumination with UV light. The illumination with a band-gap UV light to the ferroelectric can create free charge carriers of both signs, which neutralize the trapped charges and, accordingly, eliminate the internal field.

The role of injected charge and screening charge dynamics on polarization switching suggest that significant care must be made in the interpretation of the PFM switching experiments, including parameters such as minimal switchable domain size, and kinetics and thermodynamics of switching process. For example, universally observed nearly-logarithmic time dependence and almost linear voltage dependence of domain size reported by multiple groups[337, 361, 362] can be indicative of the kinetics of screening charge spreading, rather than intrinsic aspect of polarization dynamics and domain wall pinning.

**IV.4 Charge collection phenomena**

Contacting the surface with a grounded tip just after an application of a voltage bias could induce anomalous switching as discussed above. In addition, contacting or scanning the surface with a grounded tip can collect surface charges. In some cases, contacting the surface with a grounded tip can both induce anomalous switching and collect surface charges.

Kim et al reported the path dependent polarization reversal of the effect.[351] The grounded tip induces polarization reversal as well as surface charge transfer as shown in figure IV.6. Lilienblum and Soergel further have explored the effect through controlling of retracting of the tip, applying biases, and contact force.[360] They reported that this polarization reversal can happen by charge injection solely, but by ferroelastic switching. Kholkin group shows the inversed domain after illumination with UV light can be erased, which suggests that the effect is originated from the trapped charges (see Figure IV.5).[363]

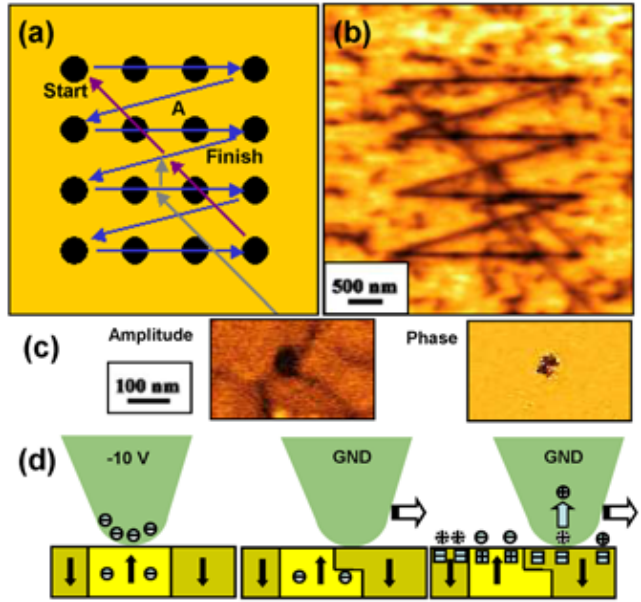

**Figure IV.6.** (a) Schematic of a trace of writing process for dot patterned domain. When the back-poling process finishes, the grounded tip is moved to the center of the box (gray line). Then, the grounded tip is moved to the left corner for writing process (violet line). The tip is moved to follow the trace of the blue line to write dots. Finally, the tip is moved to the center of the box (violet line). (b) Surface potential image acquired by KPFM. (c) PFM amplitude and phase images in the position *A* of (a) after the dot pattern. (d) Schematic of charge injection and transfer mechanisms: (left-hand) charge injection, (middle) polarization reversal, and (right-hand) charge transfer. The charge with dotted circle presents the transferred screen charge by a grounded tip. The bound charge near the ferroelectric surface is drawn as a rectangular shape. Reprinted with permission from [Kim Y, Kim J, Bühlmann S, Hong S, Kim Y K, Kim S-H and No K 2008 *Phys. Stat. sol. (RRL)* **2** 74]. Copyright 2008 Publisher Wiley-VCH Verlag GmbH & Co. KGaA [351]

As shown in figure IV.6(a), the black dots represent the locations of applied voltage for switching (see figure IV.6(c)) and solid lines represent trajectory of the grounded tip. As shown in figure IV.6(b), another mechanism produced when a grounded tip is in contact with the ferroelectric surface was identified and was referred to this phenomenon as screen charge transfer due to the electrostatic interaction. The dark lines, which are identical with the travel path of the grounded tip during pattern writing, can clearly be seen. Polarization can be excluded as a reason for the dark lines in KPFM based on the PFM results of figure IV.6(b). Rather, during pattern writing, the grounded tip removed the screen charges present on the ferroelectric surface.

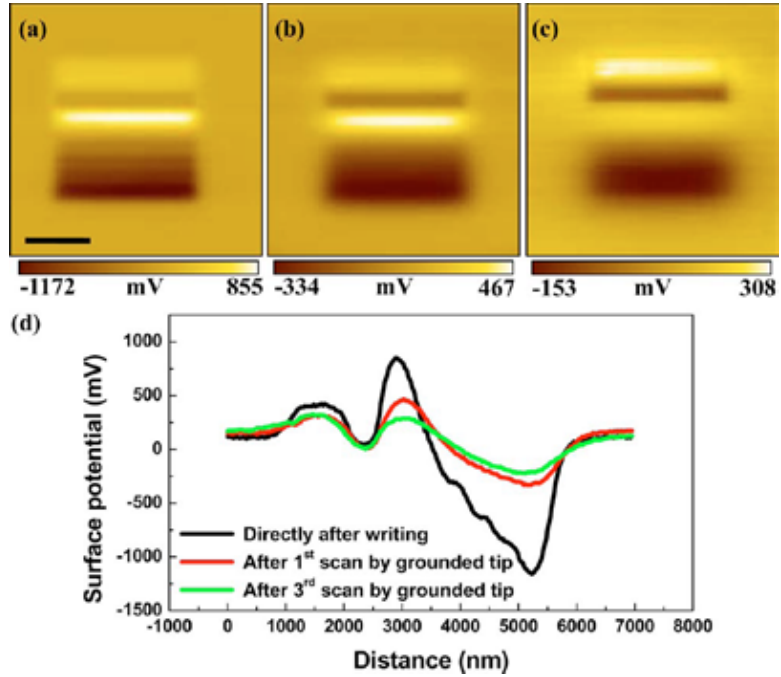

**Figure IV.7.** KPFM surface potential distributions after (a) the poling process and the (b) first and (c) third grounded tip scans. (d) Surface potential line profiles obtained from (a)-(c). Reproduced with permission from [Kim Y, Bae C, Ryu K, Ko H, Kim Y K, Hong S and Shin H 2009 *Appl. Phys. Lett.* **94** 032907]. Copyright 2009, AIP Publishing LLC [277].

Further, it was found that the amount of charge collection is dependent on the magnitude of applied voltage. The surface potential was resulted from competition between different sources of charges, *e.g.* polarization and injected charges, on the ferroelectric surfaces. Hence, the complex surface potential can be observed as shown in figure IV.7. Figures IV.7(a-c) show the surface potential evolution of the ferroelectric surfaces after contact mode scans by a grounded tip. When the electrically grounded tip was scanned by contact mode, the mobile screen charges on the ferroelectric surfaces will be swept to the grounded tip because of the electrical potential difference between the tip and the surface screen charge. Figure IV.7(a) shows the surface potential image acquired immediately after the poling process. Figure IV.7(b) and IV.7(c) present the surface potential images after the first and third scans by a grounded tip. As shown in figure IV.7, the surface potential contrast of biased regions changed, however the amount of the changed surface potential is strongly dependent on the initial surface potential value. Even though the surface potential behavior is rather complicated, this clearly presents that the larger surface potential, *i.e.* larger amount of surface charges, can be more readily removed compared to the smaller surface potential. That is, the charge collection is dependent on the amount of surface charges.

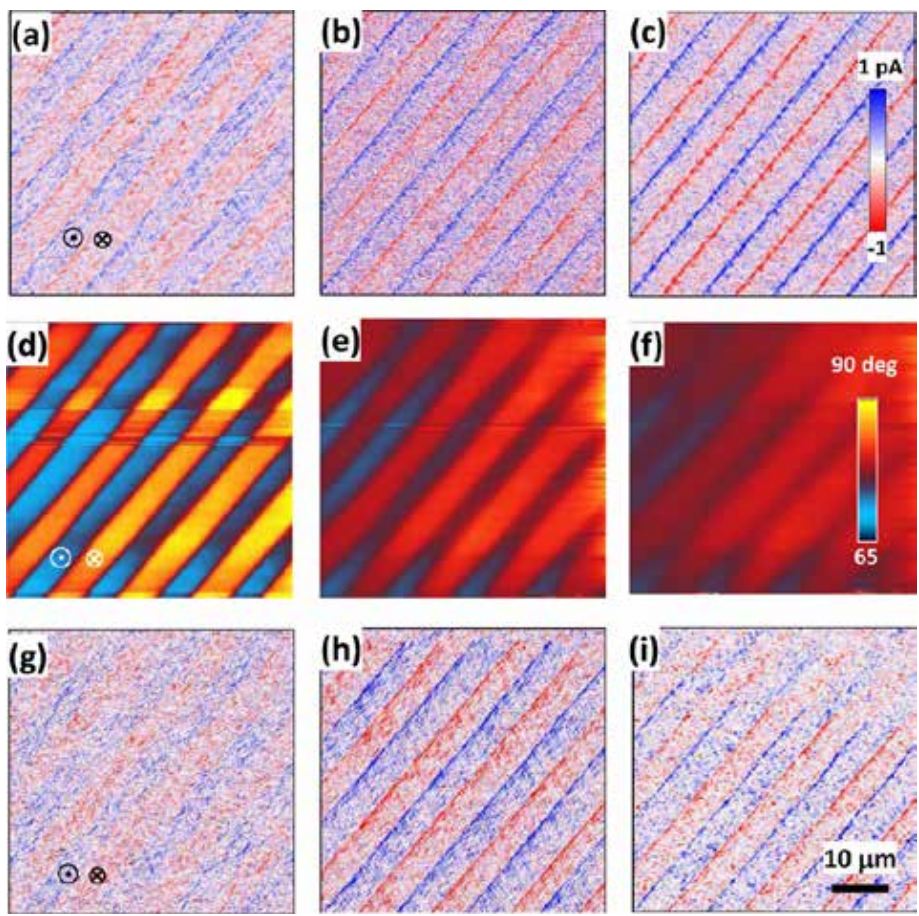

**Figure IV.8.** Influence of screening charges on the CGM signal patterns. (a–c) Three consecutive CGM SFMB scan images; (d–f) continuous EFM phase images at (d) 0–2 min, (e) 4–6 min, and (f) 12–14 min after the three CGM scans; (g–i) three consecutive CGM SFMB scan images after 20 min rescreening. Reprinted with permission from [Tong S, Jung I W, Choi Y Y, Hong S and Roelofs A 2016 *ACS Nano* **10** 2568-74]. Copyright (2016) ACS Publications [364].

Typically, the charge collection can be monitored by performing KPFM measurements before and after the grounded tip scans. Hong et al. reported that imaging mechanism of charge gradient microscopy (CGM) is also related to the charge collection.[364-367] They collected the current from CGM probe while scanning a ferroelectric surface. It is expected that the measured current dominantly originate from displacement current and charge flow from removal of screen charges. As shown in figure IV.8(f), EFM image show relatively small contrast, indicating fully screen state. In such a situation, the CGM imaging was performed for three times. As shown in figure IV.8(g-i), the domain signal was predominantly observed in the first CGM image, however the domain wall signal was observed in the third CGM image. The domain signal originates from the removal of surface screen charges.

Even though charge collection on the ferroelectric surfaces was discussed above, the similar charge collection is also possible even in non-ferroelectric surfaces.[368] In the recent report by Lee et al., charge injection was performed on the poly(methyl methacrylate) (PMMA) film surface by a similar manner. Even in the non-ferroelectric surfaces, unstable excess charges on the PMMA surface can readily move to the AFM tip, which is grounded during the scanning process. Obviously of interest is the interplay between charge collection and ferroelectric polarization, to be explored in the future.

## IV.5. Pressure effects

Figure IV.3 shows that the domain switching can be also related to the mechanically stimulus. Even though in this case it primarily originated from the internal electric field generating between a grounded tip and trapped charges, the multiple coupling mechanisms between polarization and pressure suggest potentially broader class of pressure induced phenomena on ferroelectric surfaces.

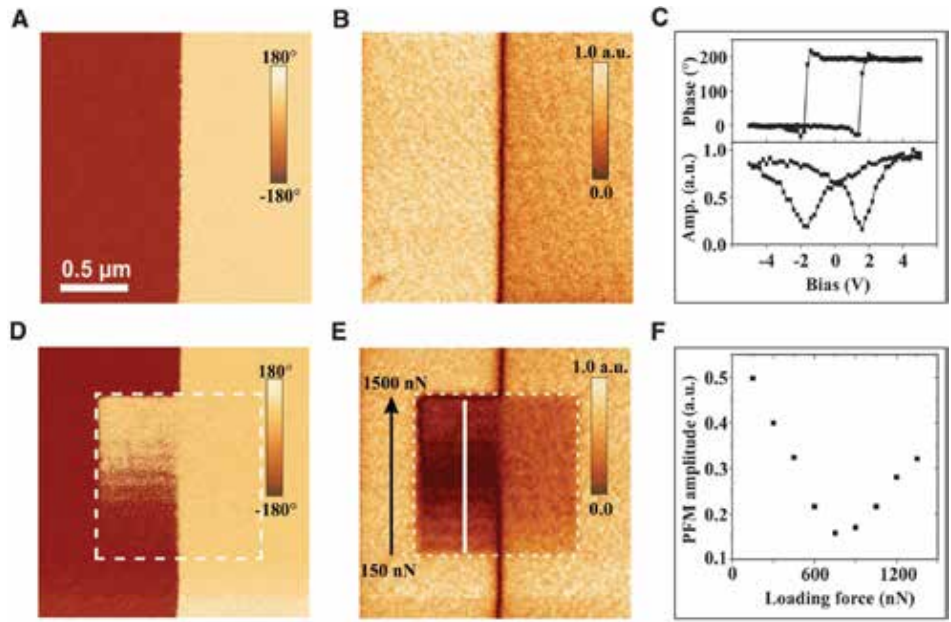

**Figure IV.9.** Mechanically induced reversal of ferroelectric polarization. (A and B) PFM phase (A) and amplitude (B) images of the bidomain pattern electrically written in the BaTiO$_3$ film. (C) Single-point PFM hysteresis loops of the BaTiO$_3$ film. (D and E) PFM phase (D) and amplitude (E) images of the same area after the 1 by 1 μm$^2$ area in the center (denoted by a dashed-line frame) has been scanned with the tip under an incrementally increasing loading force. The loading force was increasing in the bottom-up direction [denoted by a black arrow in (E)] from 150 to 1500 nN. (F) PFM amplitude as a function of the loading force obtained by cross-section analysis along the white vertical line in (E). Reprinted with permission from [Lu H, Bark C W. Esque de los Ojos D, Alcala J, Eom C B, Catalan G and Gruverman A 2012 *Science* **336** 59-61]. Reprinted with permission from AAAS (2012) [328].

Indeed, Lu et al. demonstrate that the strain gradient induced by the tip can mechanically switch the ferroelectric polarization.[328, 329] As presented in figure IV.9(d-f), the phase was reversed in the left half of the image and the amplitude decreases up to around 750 nN then increases again as increasing the loading force. On the basis of the phase and amplitude images, it can be concluded that the switching occurs at around 750 nN. The authors argued that this behavior originated from a flexoelectric effect. Although flexoelectricity is generally very weak compared to the piezoelectricity, strain gradient grow in inverse proportion to the relation length, indicating that flexoelectric effect can be very large at the nanoscale. In this context however it should be noted that pressure induced effects on screening charge can give rise to similar behavior, and the future will undoubtedly see further studies in this direction.

**V. Non-local interactions driven by screening charges**

As discussed throughout the review, ferroelectric polarization is unstable in the absence of screening by internal or external charge carriers. Correspondingly, screening processes will be inherently involved in any polarization reversal processes, and can significantly affect their thermodynamics and kinetics. While for system with metallic electrodes the screening charges are readily available and hence only the thermodynamics of the system is affected by the detailed structure of space charge-layers in ferroelectric, the situation can be completely different if the screening charges are not available or possess slow dynamics. In general case, the studies of such cases are hindered by the availability of the hierarchy of screening mechanisms. For example, rapid changes of polarization due to temperature change or switching in ambient environment will first cause fast, but energetically inefficient internal screening, and subsequent transition to slow but energetically favored ionic screening. However, in some cases the role of screening processes on domain dynamics can be clearly visualized via backswitching phenomena, or via non trivial non-local effects on domain dynamics stemming from the mass conservation laws of ionic species

**V.1. Spatiotemporal chaos during ferroelectric domain switching**

The chaotic phenomena during tip-induced domain switching were recently explored by Ievlev et al.[63, 369] Here, switching of the ferroelectric $LiNbO_3$ was studies as a function of temperature and humidity as shown in figure V.1. Application of bias pulse to the tip at predetermined locations results in polarization switching that was visualized via PFM imaging in a usual fashion. In the large regions of the V-T-d parameter space (where V is tip bias, T is temperature, and d is distance between domain centers) the domain size is uniform, indicative of the high quality of material. However, unusual effects were observed for small domain separations, including non-equal domain sizes and pronounced transient behavior, intermittency, and period tripling or the formation of non-periodic or long-periodic structures, as illustrated in figure V.1. Variation of point spacing and tip bias (for a fixed bias pulse duration of 250 ms) were used to construct the diagram of switching behaviors as shown in figure V.2(a). Here, application of biases below the nucleation threshold did not lead to domain formation (region I). Large biases resulted in continuous domain switching and the formation of "stripe" domain (region II). Application of moderate biases for large separations resulted in well-defined domain chains formed by uniform domains at each point of bias application, long explored for information storage.[197, 235, 362, 370] For very large separations, this corresponds to non-interacting domains in region IIIa. The uniform chains can also form for weak domain interaction, as shown in region IIIb. The boundary between the two was established based on the *transient* behavior in the chain, namely whether the first domain is larger than subsequent ones. Finally, intermittency and quasiperiodic switching was observed in region IV.

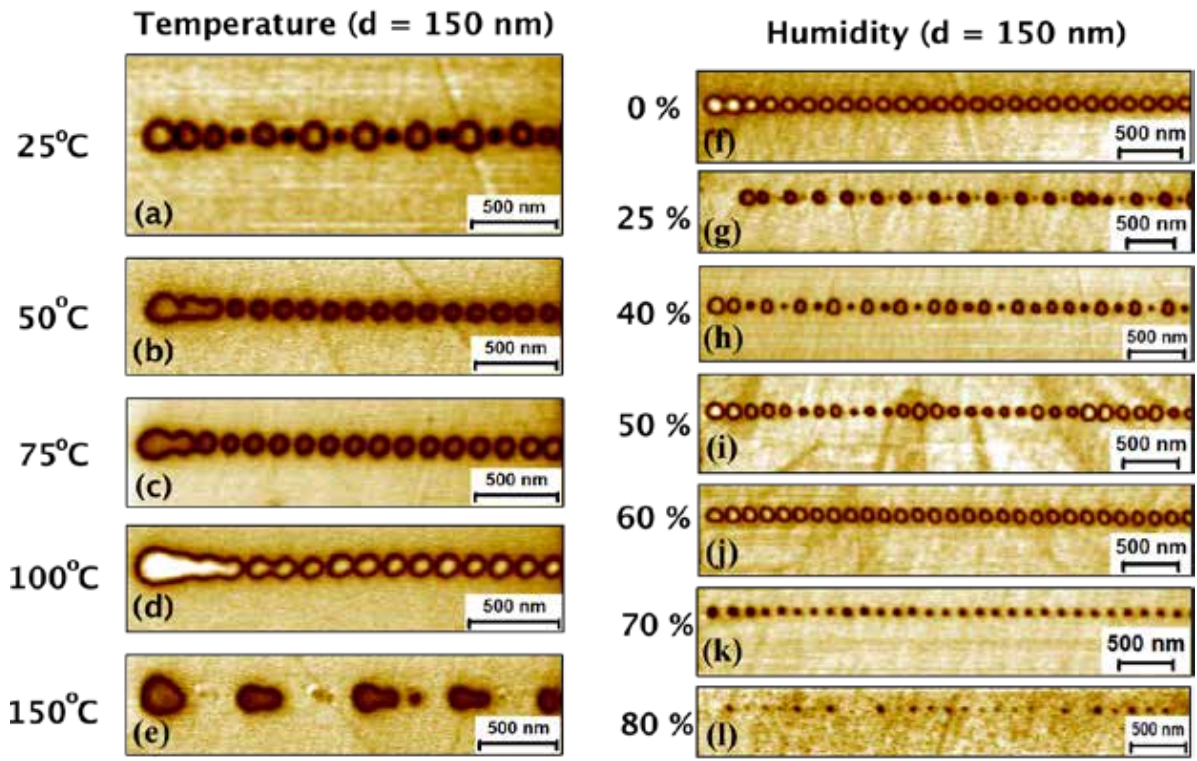

**Figure V.1**. (a-e) Evolution of the domain patterns for variable-temperature measurements. Note the rapid change of the switching character between 100 and 150 °C from period doubling (a) to uniform chains (b-d) to long-range periodicity (e). (f-l) Evolution of switching pattern as a function of relative humidity at room temperature. The data illustrates a transition between uniform switching (f), period doubling (g,h), long range periodicity (i) and back to uniform switching (j,k,l). Switching disappears at high humidity (decrease of domain sizes from j to l; no domain formed at 90%). Measurements in a-e are taken at 40% humidity (defined at RT),

measurements (f-l) are taken at RT. Reprinted with permission from [Ievlev A V, Jesse S, Morozovska A N, Strelcov E, Eliseev E A, Pershin Y V, Kumar A, Shur V Ya and Kalinin S V 2014 *Nat. Phys.* **10** 59] [63].

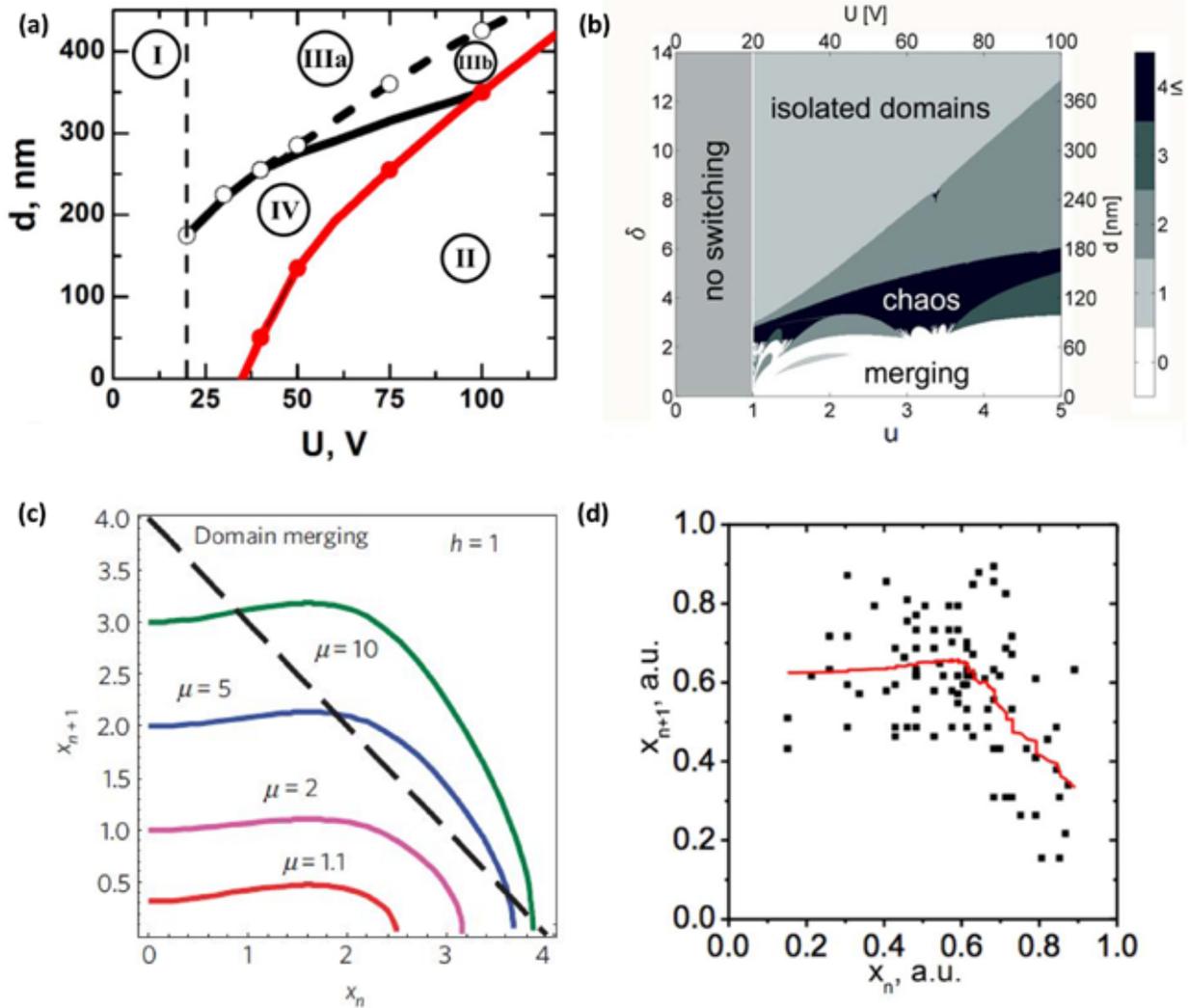

**Figure V.2.** (a) Phase diagram of switching behavior as a function of bias and domain spacing. Shown are regions of (I) no switching, (II) continuous switching, (IIIa) isolated uniform domains, (IIIb) uniform domains with transients, and (IV) quasiperiodic and chaotic behavior. (b) Phase diagram of switching behavior as a function of dimensionless voltage $u = (U/E_{cr}R_0)$ and domain spacing $\mathsf{d} = d/R_0$ calculated for $h=1$. Note the qualitative similarity between experimental diagram in figure (a) and theoretical figure (b). Also shown are axes obtained by rescaling to experimentally measurable parameters for $R_0 = 30$ nm, $E_{cr}R_0 = 20$ V. Using bulk

materials parameters $P_S$=0.75 C/m², $e_{33}$=30, $e_{11}$ = 84, the critical field for wall motion can be estimated as $E_{cr}$ = 0.67 V/nm and effective width of water layer is $H$=60 nm. Note the agreement between calculated patterns and those observed experimentally in Figure (a). (c) Dependence $x_{n+1}(x_n)$ calculated from equation (V.4) using equation (V.2) for and parameters $d = 4$, $l = 10$, and humidity $h = 10$ and different values of $m = u^{2/3}$ indicated near the curves. (d) Experimentally determined recursive relation $x_{n+1} = q(x_n)$ and smoothened dependence. Suppression of domain nucleation in the proximity of pre-existent domains is clearly seen ($x_n$ > 0.5), as well as a small upward trend for $x_n$ < 0.4. Reprinted with permission from [Ievlev A V, Jesse S, Morozovska A N, Strelcov E, Eliseev E A, Pershin Y V, Kumar A, Shur V Ya and Kalinin S V 2014 *Nat. Phys.* **10** 59] [63].

To explain the observed behaviors, the authors analyzed the ionic screening model assuming the conservation of the surface ionic species.[53, 60] In this case, polarization switching in the selected region results in the formation of the excess uncompensated charge, and can be represented as:

$$([+P] – OH^-) + H_2O + 2e^- = ([-P] – H^+) + 2OH^- \quad (V.1)$$

Here, ([+P] – OH⁻) is the positive polarization charge bound with the screening hydroxyl group (see Refs.[244, 306] for discussion of equilibrium degree of screening), and ([-P] – H⁺) is the negative polarization charge bound with a screening proton. The formed hydroxyls diffuse laterally across the sample surface, creating a "halo" surrounding the switched area. This "halo" will in turn suppress the polarization switching in the adjacent regions via the classical Le Chatelier effect.

This analysis was confirmed by subsequent studies of switching behaviors as a function of temperature and humidity. On increasing the temperature for parameters corresponding to intermittent switching at room temperature, transition to the regular writing occurred. A series of remarkable transitions was observed on increasing the humidity. For (nominally) zero humidity, the writing yields a series of the equal sized domains. For higher humidity, the onset of period doubling is observed at 25% RH. At higher humidity, quasiperiodic structures are observed for 35% and long-range periodicity (up to 8-10 domains) is observed for 50 % RH. For 60% humidity the domains again become uniform. For further increase of humidity, the domain size starts to reduce, transient behavior becomes less pronounced, and for (nominally) 90% humidity and above domains no longer form at this bias. This behavior is also confirmed by locally measured hysteresis loops. These observations can be immediately rationalized by considering the T and humidity effect on the mobility of screening charges. Namely, for low humidity the thickness of the adsorbed water layer on the surface is minimal, and hence the charge halo around the domain is highly localized and does not affect the formation of adjacent domains. For intermediate humidity values, the charges are localized in the vicinity of the formed domain, and intermittent dynamics are observed. Finally, for high humidity the charge spreads rapidly across the sample surface, obviating this effect. At the same time, the water layer effectively screens the surface, precluding polarization switching.

The authors have further proceed to quantify this behavior by treating the domain sequences as a classical time series for which transition to complex dynamic behaviors occur are well-known.[371, 372] By analyzing the structure of the depolarizing field created by the charge halo under some non-restrictive assumptions on its density profile, the authors have demonstrated the non-monotonic effect of existing domain on the next domain in the chain, and were further able to reduce this dynamics to a classical logistic map , $z_{n+1} = a z_n (1 - z_n)$, properties of which as a function of a master parameter $a$ are extremely well studied.[372] This analysis allowed to reconstruct the theoretical domain behavior diagram in dimensionless units which was very similar to experimental observations. Furthermore, matching the corresponding material-related parameters yielded the realistic estimates for the Peierls field and domain wall energies. Finally, the direct observation of recursive relationship was demonstrated.

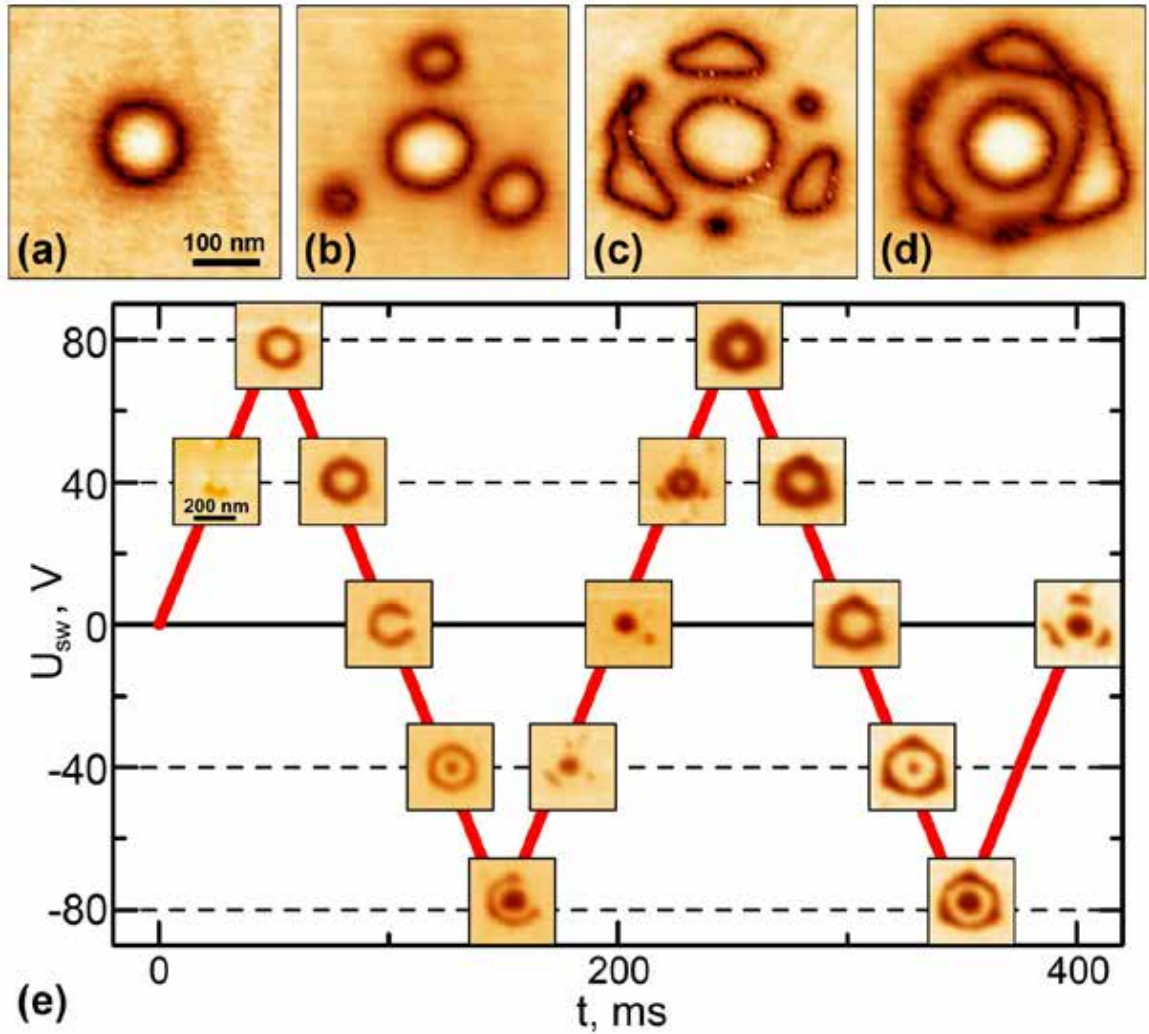

**Figure V.3.** (a)-(d) Shapes of the domains explored after switching by sequence of bipolar triangular pulses at 75% of relative humidity. $U_{sw} = 80$ V; (a) 1, (b) 2, (c) 3 and (d) 5 cycles. (e) Evolution of the domain structure as a result of tip-induced switching by sequence of two bipolar triangular pulses with amplitude $U_{sw} = 80$ V. Shown in the signals of amplitude (a)-(d) and combined piezoresponse signal (e). Reprinted with permission from [Ievlev A V, Morozovska A N, Eliseev E A, Shur V Ya and Kalinin S V 2014 *Nat. Commun.* **5** 4545] [369].

The spectroscopic measurements demonstrated that normal polarization reversal was observed after application of the electric field with direction opposite to $P_s$ immediately after application of bias. Application of the nominally non-switching rectangular pulses also leads to polarization reversal, i.e. switching against applied electric field. However, in this case switching is observed at the falling edge of the switching pulse, rather than during application of the field as is the case for normal switching. PFM visualization after abnormal switching showed formation of the ring-shaped domains with radius above 120 nm, significantly exceeding the size of the domains formed as a result of normal switching. The authors postulated that the abnormal switching is induced by the redistribution of the screening charges on the surface and top layer of ferroelectric. On application of tip bias, applied electric filed along $P_s$ is screened by surface and bulk charges of the opposite sign. After switching the bias off, these accumulated sluggish charges produce electric field that induces polarization reversal nominally against the field. The ring shape of the domain can then be explained by back switching of polarization, as is directly confirmed by observed time dependencies of the piezoresponse signals.

**V.2. Domain shape instabilities during switching**

Similar effects were observed in the shape of individual domains formed during bipolar switching, as shown in figure V.3. Here, switching process gave rise to highly non-trivial domain shapes at the different stages of triangular wave. Interestingly, application of the electric field with the same direction as spontaneous polarization ($U_{sw} > 0$ V) led to formation of the ring-shaped domains. Application of the electric field with opposite direction led to normal switching with formation of circular domains. The subsequent cycling leads to the loss of stability of the ring domains, with the formation of well-defined satellites

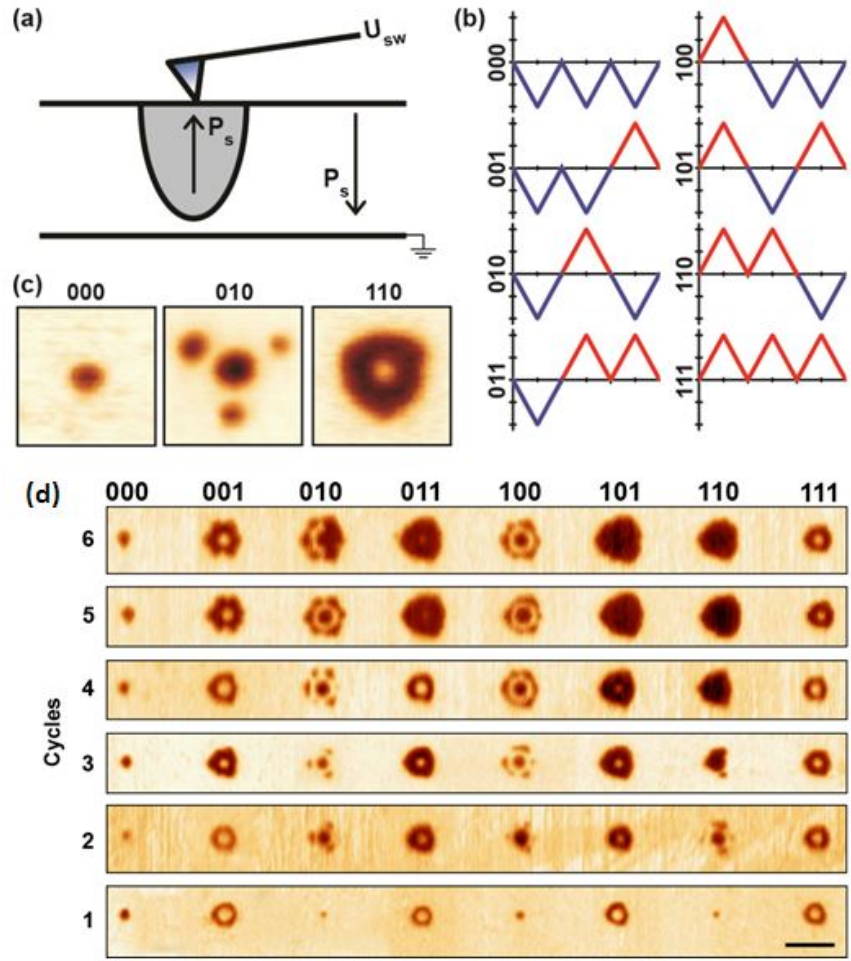

**Figure V.4.** Switching by sequences of positive and negative triangular pulses. (a) Scheme of switching procedure. (b) Used switching combinations and corresponding codes. (c) Examples of the domains formed after switching by different sequences (3 switching cycles). (d) Domain shapes as a function of sequence and number of switching cycles. Scale bar is 500 nm. Reprinted with permission from [Ievlev A V, Morozovska A N, Eliseev E A, Shur V Ya and Kalinin S V 2014 *Nat. Commun.* **5** 4545] [369].

These observations further enabled information coding in a single domain shape, as shown in figure V.4. Each negative pulse is labeled as "0", positive as "1". Combinations of 3 consequent pulses have been used. This enables 8 possible unique switching combinations from 000 to 111 (figure V.4(b)). Complex investigation of the tip-induced switching as a function of used sequence and number of switching cycles demonstrated unexpectedly wide variety of the domain morphologies (figure V.4(d)). In particular, one switching cycle showed formation of the domains of two types: circular and ring-like. Domain shape was defined by last pulse in the sequence only. Sequences ending by negative pulse (000, 010, 100, 110) led to formation small circular domains while sequences ending by positive pulse (001, 011, 101, 111) – ring-like domains. However, increasing of the number of switching cycles revealed dependence of the resulted domain shape on the structure of whole sequence. Switching by 3-5 cycles allowed differentiation of 6-7 different domain morphologies. Subsequently, the specially trained neural network was able to recognize the 3 element sequence with 80% probability, illustrating the information encoding in the domain shape.

**V.3. Modeling of screening effects impact on the tip-induced polarization switching**

For most SPM experiments in humid atmosphere surface screening charges dynamics appear under the tip-induced polarization reversal can be attributed to existing of the water meniscus between tip and the sample surface. Results of the numerical modeling of spatial distributions of the tip-induced electric potential and field $E_{tip}$ under the sample surface are shown in figure V.5 for different height $h_m$ of the meniscus.

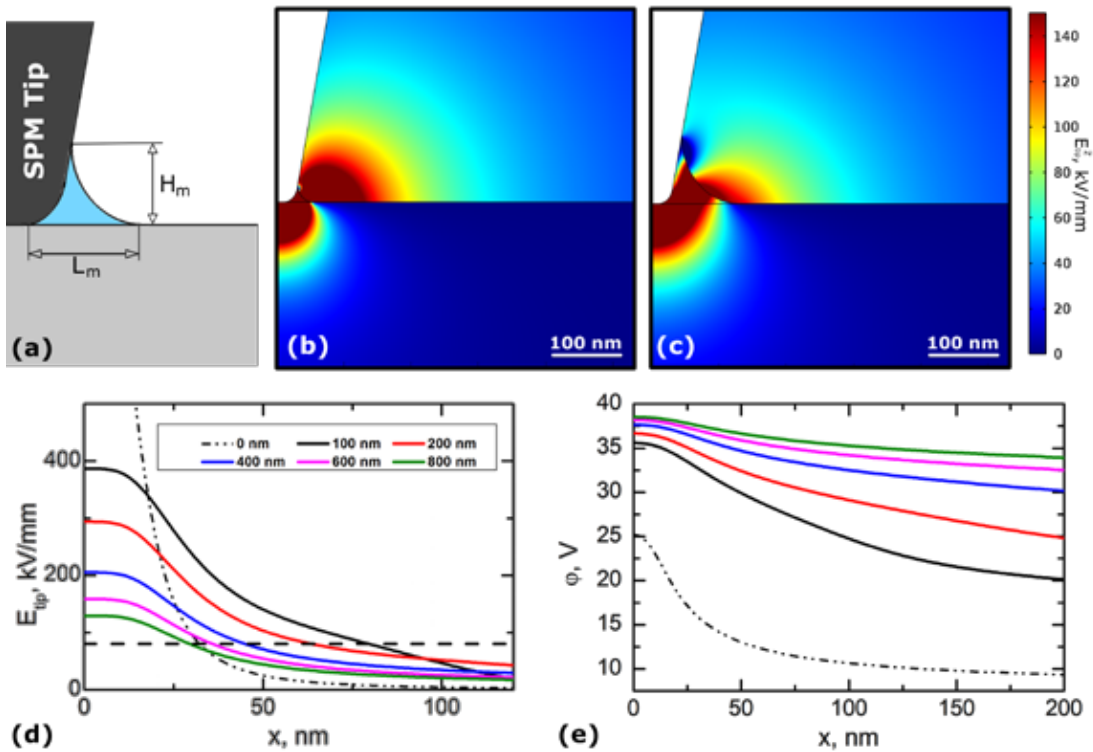

**Figure V.5** Results of numerical simulations of z-component the electric field produced by biased SPM tip in presence of water meniscus. (a) Scheme of the model; (b), (c) 2D maps of the spatial distribution of $E_{tip}^Z$ for (b) $h_m = 25$ nm and (c) $h_m = 100$ nm; (d), (e) distributions of $E_{tip}^Z$ and potential along polar direction at 10 nm under the surface for different $h_m$. Reprinted with permission from [Ievlev A V, Morozovska A N, Shur V Ya and Kalinin S V 2014 *Appl. Phys. Lett.* **104** 092908]. Copyright 2014, AIP Publishing LLC [64].

As can be seen from figures V.5(d-e), the appearance of the meniscus leads to decreasing of the electric field in the area under the tip and it spatial delocalization. Using an effective point charge model[362, 373-375] for the description of SPM tip field, simulated distributions of $E_{tip}^z(r)$ can be fitted by the following expression:

$$E_{tip}^z(r) = \frac{Q^*}{2\pi\varepsilon_0\left(\sqrt{\varepsilon_c\varepsilon_a}+1\right)}\sqrt{\frac{\varepsilon_a}{\varepsilon_c}}\frac{d^*}{\left((d^*)^2+r^2\right)^{3/2}} \quad (V.2)$$

At that the values of the effective tip radius $d^*$ and effective charge $Q^*$ are fitting parameters extracted from either experiment or numerical modeling. Typical dependences of the $Q^*$ and $d^*$ on the meniscus height $h_m$ are shown in figure V.6. Particularly, effective tip radius $d^*$ changes for the small heights $h_m £ 100$ nm and saturates for larger heights. $Q^*$ has short part of the growth at $h_m £ 100$ nm and then asymptotically decreases with $h_m$ increase. Extracted values of the effective tip radius and charge allows to describe the growth of the isolated domain with presence of the water meniscus using the model[62].

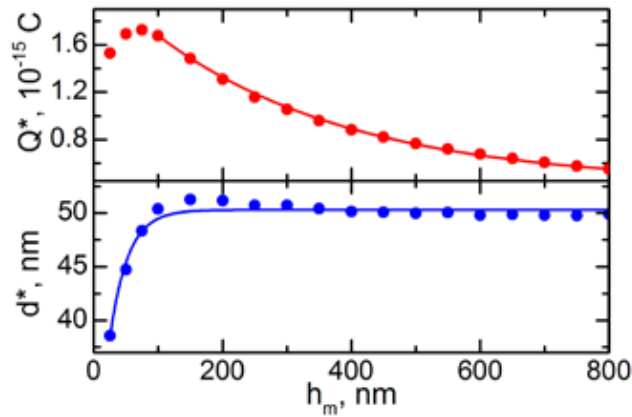

**Figure V.6.** Values of the effective charge $Q^*$ and effective tip radius $d^*$ vs. meniscus height in the point charge model. Reprinted with permission from [Ievlev A V, Morozovska A N, Shur V

Ya and Kalinin S V 2014 *Appl. Phys. Lett.* **104** 092908]. Copyright 2014, AIP Publishing LLC [64].

Namely, to fit the point data shown in figure V.6 the following trial functions have been selected:

$$d^*(h_m) = d_\infty \left(1 - \frac{d_\infty - d_0}{d_\infty} e^{-\frac{h_m}{h_q}}\right), \qquad Q^*(h_m) = Q_0 \left(c\frac{d_\infty}{d_0} + \left(1 - c\frac{d_\infty}{d_0}\right)e^{-\frac{h_m}{h_q}}\right), \qquad (V.3)$$

Here $d^*(0) = d_0$ and $d^*(\infty) = d_\infty$, at that $h_m = 0$ corresponds to the dry atmosphere and in this case tip has effective radius $d_0$. Effective tip charge $Q_0 \approx U_{sw} d_0$ is proportional to switching bias $U_{sw}$ applied to the tip and its radius. The limit $h_m = \infty$ corresponds to the tip under the water (liquid PFM). In this case the tip has effective radius $d_\infty$ which is higher than the tip radius in dry air due to delocalization of the field. Effective charge is still proportional to the tip radius $d_\infty$ and bias $U_{sw}$, but reduced by factor due to higher permittivity of the water $\varepsilon_w = 80$. $h_q$ and $h_d$ are characteristic distances for the effective tip radius and effective charge vs. meniscus size dependences correspondingly.

To relate the meniscus height $h_m$ with the value of relative humidity H as $h_m(H) = h_0 \exp\left(-\frac{H_{cr}}{H}\right)$, where the parameters $h_0$ and $H_{cr}$ are defined from the fitting of experimental data. Finally, domain radius as a function of relative humidity $H$ and bias $U_{sw}$ was be obtained:[64]

$$r(U_{sw}, H) \approx r_0 + \beta d^*(H) \sqrt{\left(\frac{Q^*(U_{sw}, H)}{U_{cr} d^*(H) G}\right)^\alpha - 1}, \qquad (V.4)$$

where is a power factor $2/3 < \alpha < 2$, dimensionless parameter $\beta$ reflects the tip form-factor and has the order of unity, $U_{cr}$ has sense of a critical voltage for a "dry" ferroelectric surface without water meniscus ($H = 0$). The effective tip parameters $d^*$ and $Q^*$ are given by equation (V.3), $G\sim 1$ is a factor reflecting the tip geometry.

Note that equations (V.2-4) consider the influence of the top water layer on polarization reversal due to redistribution of the electric field produced by the tip only. But the influence doesn't limited by this phenomenon. In addition, the presence of the top water layer changes all screening conditions by redistribution of the charge carriers across the layer. This phenomenon can be used for explanation of the inconsistence between these results and those in refs.[322, 323] In the both papers growth of the micron-sized domains was observed. At such distances redistribution of the electric filed caused by water meniscus is not so pronounced and cannot lead to essential change of the domain kinetics. However external screening caused by charge carriers in the adsorbed surface layer supports the switching far from the tip and leads to formation of the large domains.

Finally, a self-consistent theoretical approach capable to describe the features of the anisotropic nanodomain formation induced by a strongly inhomogeneous electric field of charged SPM tip on non-polar cuts of ferroelectrics has been proposed.[376] Here it was shown that a threshold field is an anisotropic function that is specified from the polar properties and lattice pinning anisotropy of a given ferroelectric in a self-consistent way. The proposed method for the calculation of the anisotropic threshold field is not material-specific, thus the field should be anisotropic in all ferroelectrics with the spontaneous polarization anisotropy along the main crystallographic directions. The most evident examples are uniaxial ferroelectrics, layered ferroelectric perovskites and low symmetry incommensurate ferroelectrics. Obtained results quantitatively describe difference in several times in nanodomain length experimentally observed on X- and Y-cuts of $LiNbO_3$[377] and can give insight into the anisotropic dynamics of nanoscale polarization reversal in strongly inhomogeneous electric fields. The threshold fields along different crystallographic directions can be significantly different due to crystal anisotropy of the inter-atomic relief and energy barriers. 3D-atomic structure of $LiNbO_3$ crystallographic cuts are shown in figures V.7(a-c) using the coordinates from Boysen and Altorfer.[378]

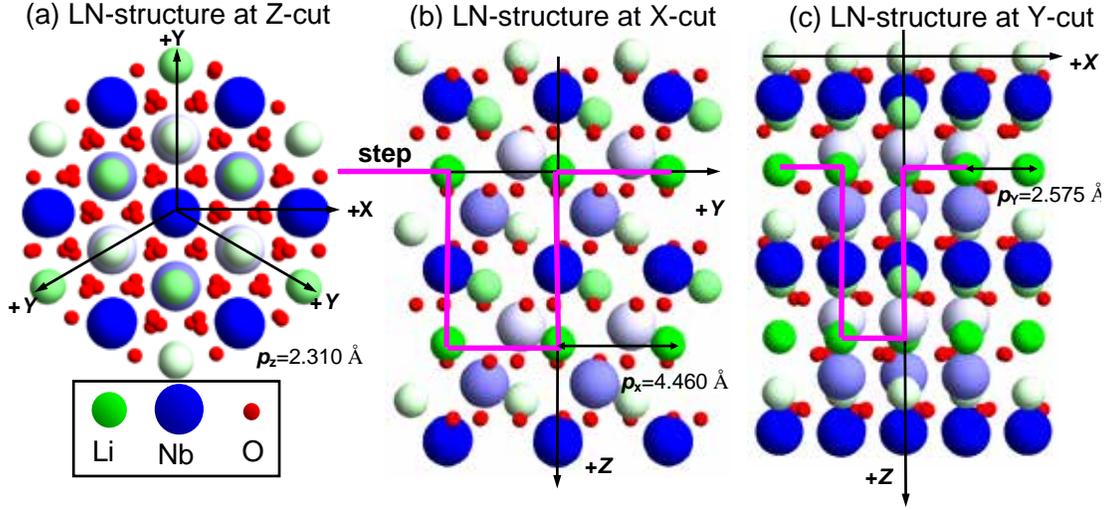

**Figure V.7.** Atomic structure of LiNbO$_3$ Z-cut (XY plane) (a) X-cut (ZY plane) (b) and Y-cut (ZX-plane) (c). Big blue balls are Nb atoms, smaller green balls are Li atoms and the smallest red balls are O atoms. Reprinted with permission from [Morozovska A N, Ievlev A V, Obukhovskii V V, Fomichov Y, Varenyk O V, Shur V Ya, Kalinin S V and Eliseev E A 2016 *Phys. Rev. B* **93** 165439]. Copyright (2016) by The American Physical Society [376].

A suggested step-like path of the domain wall motion in the polar direction Z on the non-polar X- and Y-cuts is shown by an elementary step in figures. V.7(b-c). The step-like path is defined as the elementary Li-Li distances in the directions perpendicular to the polar Z-axes. Then using rhombohedral lattice parameters of LiNbO$_3$, $a = 5.15$ Å, $c = 13.86$ Å, angle $\alpha_c = 55°53'$[379, 380], the minimal distances $p_{[abc]}$ between the equilibrium positions of uncharged domain wall planes at different crystallographic cuts, are $p_{[100]} = \sqrt{3}a/2 \approx 4.460$ Å for X-cut, $p_{[010]} = a/2 \approx 2.575$ Å for Y-cut and $p_{[001]} = c/6 \approx 2.310$ Å for Z-cut.

The anisotropic threshold field accounts for the anisotropy of the minimal distance between the equilibrium positions of the uncharged domain wall in the Ishibashi formula[381] in the following way:

$$E_{th}^{[abc]} = -e^4(\pi/2)^{7/2}\alpha P_S(w/p_{[abc]})^3 \exp(-\pi^2 w/p_{[abc]}), \quad (V.5)$$

Here the half-width of the domain wall $w$ is normalized on the minimal distance $p_{[abc]}$ between the equilibrium positions of the uncharged domain wall plane propagating in the crystallographic direction [abc]. Hereinafter we associate [001] with a Z-cut, [010] with Y-cut and [001] with X-cut.

Figure V.8 illustrates the anisotropic threshold field calculated using equation (V.4) for LiNbO$_3$ parameters a, $P_S$ and different domain wall half-width $w$, since the latter can be strongly affected by depolarization field and depends on the wall bound charge (e.g. incline angle with respect to the polar direction). As one can see, the value of $E_{th}$ differs on the one or even several orders of magnitude for different direction of the domain wall motion. In addition it strongly decreases with $p_{[abc]}$ increase and vary in the range ($10^{-3} - 10^{+2}$) kV/mm. $E_{th}$ monotonically and rapidly decreases with $w$ increase more than 1 Å for any period $p_{[abc]}$. Note, that smaller $w$ values are unlikely physical. At fixed $w>1$ Å the highest fields correspond to the smallest period $p_{[abc]}$, i.e. $E_{th}^Z < E_{th}^Y < E_{th}^X$ since $p_Z < p_Y < p_X$. This exactly means that the threshold field is the smallest for Z-cut, intermediate for Y-cut and the highest for the X-cut of the crystal.

Equation (V.5) allows us to estimate the ratio of the threshold and activation fields in different directions for known distances $p_{[abc]}$ [figure V.8(b)]. The ratio of the threshold fields $E_{th}^X/E_{th}^Y$ changes from 1.5 to 5, the ratios $E_{th}^Y/E_{th}^Z$ and $E_{th}^X/E_{th}^Z$ are within the range of 1.5 to 100 for realistic values of the domain wall width. The ratio of the activation fields $E_a^X/E_a^Y$ changes from 1.5 to 10, the ratios $E_a^Y/E_a^Z$ and $E_a^X/E_a^Z$ are within the range of 0 to 3.

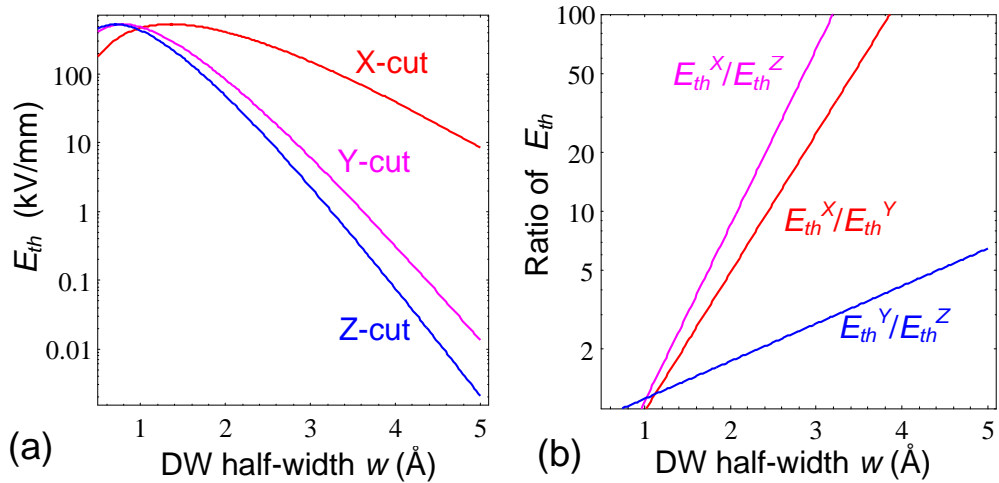

**Figure V.8**. (a) Threshold fields dependence on the domain wall half-width $w$ calculated within Suzuki-Ishibashi model for LiNbO$_3$ parameters a = - 1.95´10$^9$ m/F, $P_S$ = 0.735 C/m$^2$. (b) The threshold fields' ratio vs. the domain wall half-width *w*. Reprinted with permission from [Morozovska A N, Ievlev A V, Obukhovskii V V, Fomichov Y, Varenyk O V, Shur V Ya, Kalinin S V and Eliseev E A 2016 *Phys. Rev. B* **93** 165439]. Copyright (2016) by The American Physical Society [376].

Note, that DFT calculations of the $E_{th}$ values for three crystallographic directions can be very helpful for both the anisotropic model verification and comparison with experiment.[58]

**V.4. Screening effects during in-plane domain switching**

Application of the bis to the tip in contact with the surface with p[urely in-plane polarization can give rise to in-plane domain switching due to radial part of tip field. Such switching necessarily leads to formation of charged domain walls, screened by bulk and surface charges. Hence, observations of such domains and their time evolution provides insight into screening mehcnisms. Alikin et al[377] measured experimentally the domain shape and sizes on the non-polar X- and Y-cuts of CLN. Corresponding domain length and width at the non-polar surfaces of the CLN are shown by symbols with error bars in figures. V.9(a) and (b). Solid curves are the best fitting within the afore-presented anisotropic model[376]. Using the fitting parameters we calculated that the threshold field $E_{th}$ for X-cut is about 420 kV/mm and about 250 kV/mm for Y-cut. Note that due to the scattering of the width for X- and especially Y-cut the experimental data for domain width is much less reliable than the data for length. So that it has a little sense to conclude about the method validity from the domain width fitting. However a reasonable agreement with experiment by changing the fitting parameters for both domain length and width is present.

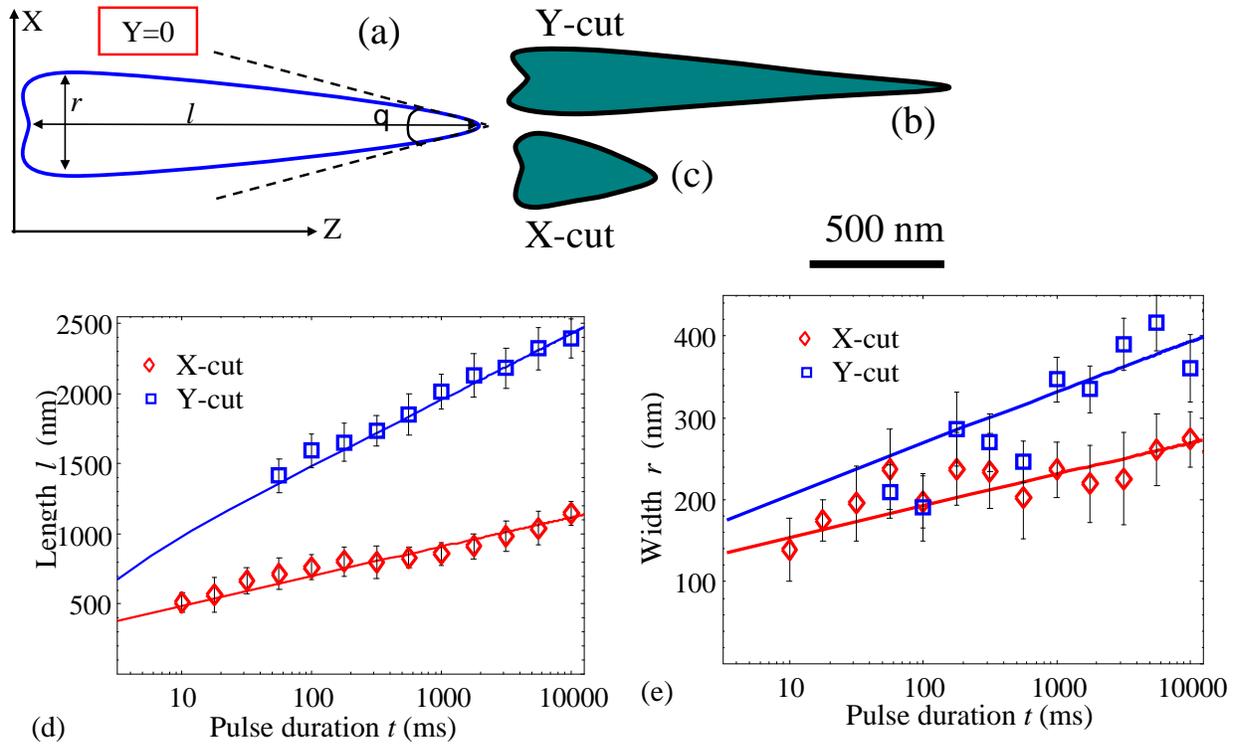

**Figure V.9.** (a) Sketch of the domain shape in the XZ plane (Y=0) induced by AFM probe on congruent LiNbO$_3$ (CLN) non-polar cuts. (b-c) Experimentally observed domain shape from Alikin et al on (b) Y- and (c) X-cuts of 20-mm-thick CLN. Dependencies of (a) domain length and (b) domain width vs. the switching pulse duration on X- and Y-cuts of LiNbO$_3$. Symbols with error bars are experimental data from Alikin et al for X- and Y- cuts of 20-mm-thick CLN placed in dry nitrogen, solid curves are the fitting. Reprinted with permission from [Morozovska A N, Ievlev A V, Obukhovskii V V, Fomichov Y, Varenyk O V, Shur V Ya, Kalinin S V and Eliseev E A 2016 *Phys. Rev. B* **93** 165439]. Copyright (2016) by The American Physical Society [376].

Finally, we discuss the question about the difference in domain depth for the cases when the writing electric field acts on different LiNbO$_3$ cuts. As it was reported earlier by Molotskii et al[382] for the case of nanodomain formation on polar Z-cut their depth (called length because of the radial symmetry of domain cross-section) can reach micron distances due to the breakdown effect. Alikin et al concluded from a selective etching that the domain depth on the Y-cut is rather small in comparison with the one on the X-cut. Moreover, "Y-cut domains" most likely remained nanosized in Y-direction, while "X-cut domain" can be much deeper, but not needle-like as "Z-cut domains". Anisotropic approach can explain these facts, because it account for the anisotropy of lattice barriers and depolarization effects at the charged domain walls. In particular, the longest needle-like shape of Z-cut domains is conditioned by the smallest threshold field $E_{th}(p_Z)$ and domain breakdown in Z-direction takes place. The smallest depth Y-cut domain in X-direction originated from inequality $E_{th}(p_X) \gg E_{th}(p_Y) > E_{th}(p_Z)$, since the smaller is the threshold field the bigger is the domain size. These speculations can be quantified using the $E_{th}(p_{[abc]})$ ratios for different crystallographic cuts presented in figure V.9 and corresponding minimal distance $p_Z \approx 2.310$ Å, $p_Y \approx 2.575$ Å, $p_X \approx 4.469$ Å.

The authors have argued that observed domain-domain interactions and complex shapes may potentially be used for a variety of information technology applications. However, the most significant aspect of these observations is the direct visualization of the role that slow screening charge dynamics can play in ferroelectric switching. While classical observations of domain wall motion cannot distinguish the possible mechanisms – e.g. pinning in the material vs. screening charge dynamics, the observation of long range interactions, back switching, and complex domain shapes clearly illustrate the role of screening processes in domain switching.

## VI. Chemical and light effects

As discussed above, the surface state of ferroelectrics is very sensitive to external environment. Correspondingly, it can be affected by chemistry and instant photo carriers via photovoltaic effects. Here, we discuss the chemical switching, light effects, and internal screening photodeposition on the ferroelectric surfaces.

## VI.1. Chemical switching

Reversible chemical switching was demonstrated in PbTiO$_3$ thin film by using x-ray scattering measurements.[24] In their work, they reported reversal chemical switching assisted by amount of oxygen content. This chemically induced changes in ferroelectric domain structures was observed using SPM.

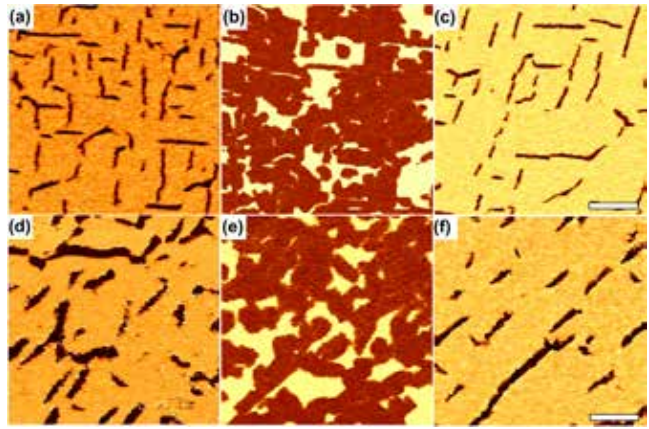

**Figure VI.1.** PFM phase images of different states: [(a) and (d)] as-deposited, [(b) and (e)] oxygen plasma treated for 10 min, and (c) and (f) after subsequent annealing at 350 °C for 30 min, for BFO thin films of [(a)-(c)] 30 nm and [(d)-(f)] 60 nm thickness. The scale bars of (c) and (f) correspond to 400 nm and 300 nm, respectively. Reprinted with permission from [Kim Y, Vrejoiu I, Hesse D and Alexe M 2010 *Appl. Phys. Lett.* **96** 202902]. Copyright 2010, AIP Publishing LLC [383].

Reversible chemical switching was also reported in $BiFeO_3$ (BFO) thin films using PFM.[383] The oxygen plasma and vacuum thermal annealing were used to chemically control the polarization switching as similar to atmospheres with high and low oxygen partial pressures, respectively, in ref. [134] As shown in figure VI.1, reversible chemical switching was well achieved by surface chemical reactions by the oxygen plasma and subsequent vacuum thermal annealing. They further directly observed that the later domain growth during the plasma treatment depends on the duration of the treatment.

The chemical switching of the ferroelectric materials is strongly correlated with chemistry of ferroelectric surfaces.[384] Bonnell group reported that chemical species such as methanol and $CO_2$ adsorb on $BaTiO_3$ (BTO) surfaces at specific defect sties, here oxygen vacancy sites.[157] Further, they presented adsorption and reaction of chemical species on the ferroelectric surfaces are strongly dependent on the ferroelectric polarization.[156, 385, 386]

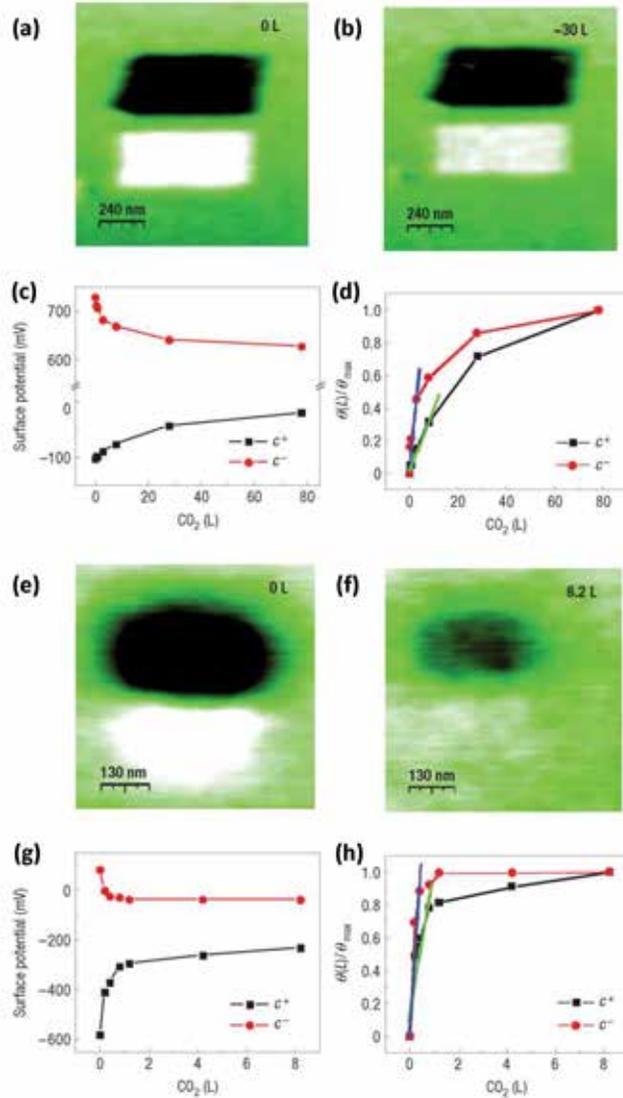

**Figure VI.2.** Influence of $CO_2$ adsorption on the surface potential of BTO (001) and a PZT thin film. (a) Surface potential maps of $c^+$ and $c^-$ domains on BTO (001), which were poled by an external electrical field applied by a conductive AFM tip scanning over the surface. Ferroelectric domains were poled positively (negatively) in the dark (bright) area and showed negative (positive) surface potential due to the opposite compensating charges. The areas surrounding $c^+$ and $c^-$ domains represent in-plane a domains. (b) Surface potential maps after exposure to 30 L of $CO_2$. (c) Average surface potential versus $CO_2$ dose on BTO (001). (d) $\theta(L)/\theta_{max}$ versus $CO_2$ dose. $S$ is proportional to the slope of the line. (e–h) Corresponding data for $CO_2$ adsorption on the polycrystalline PZT thin film. Reprinted with permission from [Li D, Zhao M H, Garra J, Kolpak A M, Rappe A M, Bonnell D A and Vohs J M 2008 *Nat. Mater.* **7** 473-7] [186].

Later, Bonnell group directly observed effect of ferroelectric polarization on the physisorption energies as shown in figure VI.2. Figure VI.2 shows surface potential as a function of $CO_2$ dose and corresponding coverage of $CO_2$ on BTO and PZT thin films. They found that the gas-phase molecule such $CO_2$ first physisorbs to the ferroelectric oxide surfaces then it diffuse until it chemisorbes at the oxygen vacancy sites. The amount of chemisorption is dependent on the polarization direction, $CO_2$ dose, as well as materials.

**VI.2. Light effects**

A wealth of information on the role of screening phenomena on domain dynamics can be obtained using studies under strong light irradiation. The original exploration of this area dates back to 60ies to early work of Fridkin, Sturman and others, area generally referred as photoferroelectrics or ferroelectric semiconductors.[311, 387-390] The central premise of these studies was that irradiation of ferroelectric surface with sub-band gap light results in generation of carriers in the near-surface layers. The interaction of photogenerated carriers and polarization fields give rise to a broad set of phenomena ranging from formation of unusual domain structures,[388] conductivity,[389] significant changes in coercive fields for polarization switching,[391] and elastic[390] and electromechanical responses.[392] The theory of these phenomena including formation of spatially-inhomogeneous polarization patterns and dynamic solutions (moving solitons, etc) was considered by a number of authors.[393]

However, these early observations were significantly hindered by the lack of high-resolution probes of domain structures. More recently, Gruverman et al.[394] reported that UV illumination of the ferroelectric thin films results in a voltage shift by an interaction between photo induced and polarization charges. In particular, they directly observed domain pinning induced by the illumination at the nanoscale using PFM.[394] In a similar study, Shao et al. has explored to directly observe photo induced carrier screening on the ferroelectric surfaces.[395] As presented in figure VI.3, surface potential is rapidly reduced by UV illumination and recovers after the switch-off of the UV illumination. These relaxation phenomena are due to the charging and discharging of photo induced carriers.

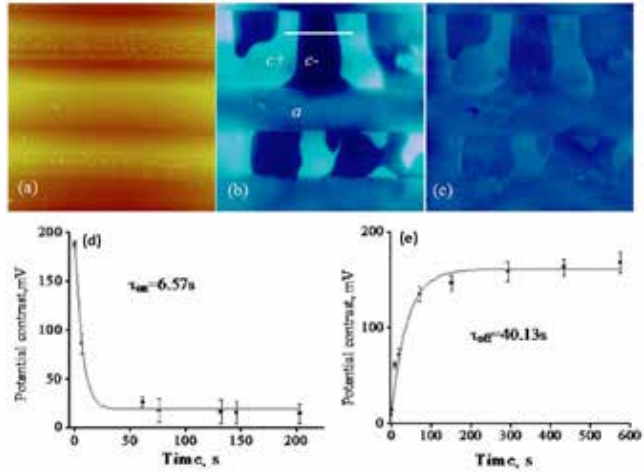

**Figure VI.3.** (a) AFM topography of a BTO (001) surface that shows three corrugations due to *a-c* domain walls. *z* scale is 200 nm. (b) The surface potential image of the same area as shown in (a) shows (c) domains with curved domain walls. *z* scale is 0.25 V (c) The surface potential image when the UV light is on. *z* scale is 0.25 V. Time dependence of surface potential contrast between $c^+$ and $c^-$ domains immediately after UV light is switched (a) on and (b) off. Reprinted with permission from [Shao R, Nikiforov M P and Bonnell D A 2006 *Appl. Phys. Lett.* **89** 112904]. Copyright 2006, AIP Publishing LLC [395].

Finally, recent years have seen a dramatic growth of interest to photovoltaic phenomena in ferroelectric materials following the works by Yang,[396] Choi,[397] and Alexe.[398, 399] Alexe et al.[398] reported unusual photovoltaic effect in BiFeO$_3$ single crystals. As shown in Figure VI.4, the local and macroscopic open circuit voltages from the photocurrent-voltage characteristics are well matched to each other. However, if one can calculate the current density, one can find that the current density of the local measurements is significantly higher than that of the macroscopic measurements. This is due to the fact that the AFM tip is very efficient in collecting the photo induced carriers.

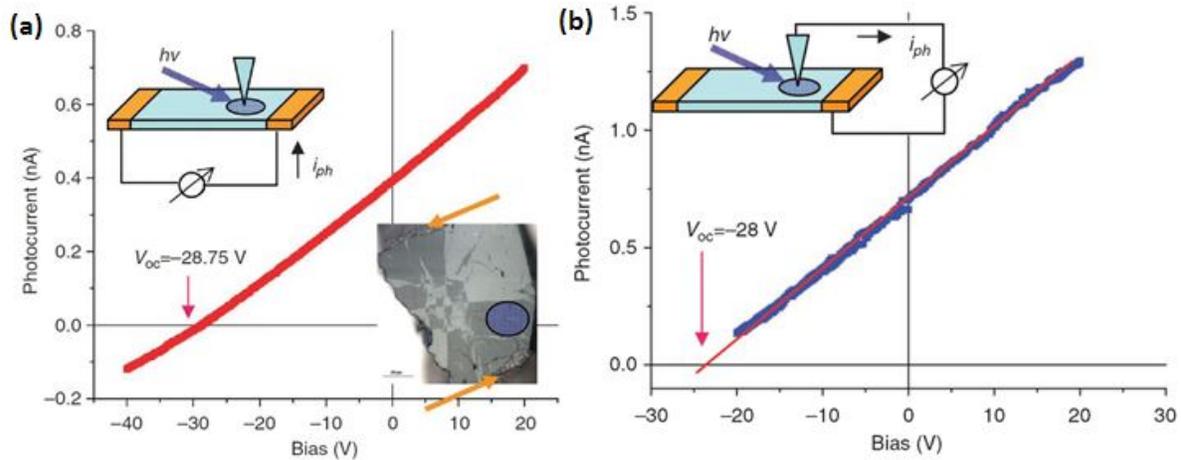

**Figure VI.4.** Photocurrent–voltage characteristics. (a) Measured macroscopically across the entire crystal by illumination in the region marked in the lower right inset, and (b) measured by probing with the AFM tip in the middle of the illuminated area. The upper left inset in (a) schematically shows the measurement setup and the connection to the macroscopic measurement; the lower right inset in (a) shows the crystal with silver electrodes marked by arrows and approximately the illuminated area by the laser spot. The inset in (b) shows the measurement setup for the local measurements by the AFM tip. Reprinted with permission from [Alexe M and Hesse D 2011 *Nat. Commun.* **2** 256] [398].

Yang et al.[396] studied ferroelectric domain wall dependent photovoltaic effects in the BFO thin films. They found that current-voltage characteristics in two different electrode geometries (one is perpendicular and the other is parallel to the domain walls) reveal significantly different photovoltaic behaviors (see figure VI.5). Because the illuminated domain wall enables a more efficient separation of the excitons, a net voltage observed across the sample is resulted from the combination of the domain walls and excess charge carries by illumination.

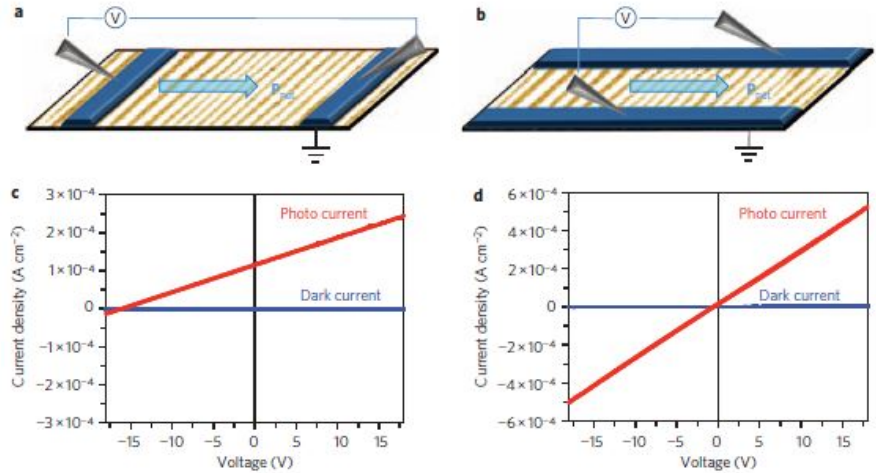

**Figure VI.5**. Light and dark I–V measurements. (a,b) Schematics of the perpendicular (DW ⊥) (a) and the parallel (DW ∥) (b) device geometries. (c,d) Corresponding I–V measurements of the DW ⊥ (c) and DW ∥ (d) devices, respectively. Reprinted with permission from [Yang S Y, Seidel J, Byrnes S J, Shafer P, Yang C-H, Rossell M D, Yu P, Chu Y-H, Scott J F, Ager J W, Martin L W and Ramesh R 2010 *Nat. Nanotech.* **5** 143-7] [396].

## VI.3. Domain dependent photodeposition

The SPM studies of the ferroelectric surfaces are directly sensitive to the electrostatic fields above the surface (KPFM and EFM) or bulk material properties in the near-surface layer (KPFM). As such, they are more sensitive to the external screening behavior, whereas the structure of potential and field distributions inside material is generally inaccessible by SPM and can be explored by depth-resolved electron microscopy and scattering methods. However, certain insight into the internal screening on ferroelectric surfaces can be obtained from the observation of photovoltaic, chemical and photoelectrochemical properties of ferroelectrics. In these cases, light irradiation of ferroelectric results in generation of electron-hole pairs, and their subsequent dynamics can be explored though SPM of photochemical reactivity.

A notable example of such behavior is domain specific photochemical deposition of metal and oxides on ferroelectrics as originally reported and extensively explored by Rohrer and Giocondi.[400, 401] As presented in figure VI.8, the authors reported that illumination of BTO surface in $AgNO_3$ solutions lead to preferential silver deposition on positive domains, whereas negative domains remained uncovered by silver. This behavior can be readily explained assuming that the positive polarization charge results in downward band bending. Under illumination, the photogenerated electron-hole pair is split by the near-surface field, and electron moves towards surface. In the absence of electrochemically active species, electron accumulation will result in the onset of flat band conditions, similarly to open solar cell. However, in the presence of reducible cations chemical reaction and deposition of metallic silver is possible. Notably, positive carriers are extracted in negative domains and likely result in water oxidation, completing mass and electron balance.

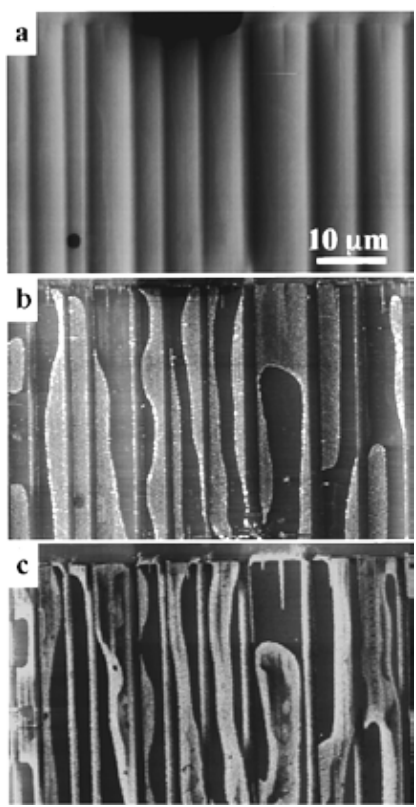

**Figure VI.6.** Topographic AFM images of the surface of a BaTiO$_3$ single crystal. (a) Before the reactions. (b) The same area of the surface after illumination in an aqueous AgNO$_3$ solution. The white contrast corresponds to silver. (c) The same area of the surface after it was cleaned and illuminated in an aqueous lead acetate solution. The white contrast corresponds to lead containing deposits. The ranges of the vertical black-to-white contrast in topography (a−c) are 80, 100, and 110 nm, respectively. Reprinted with permission from [Giocondi J L and Rohrer G S 2001 *J. Phys. Chem. B* **105** 8275-7]. Copyright (2001) ACS Publications [400].

The evidence for this mechanism can be obtained from observations of photochemical reactivity in poorly conducting materials such as $LiNbO_3$ (LNO). As shown in Figure VI.7, metal deposition on the periodically poled lithium niobate (PPLN) surface was observed only in the vicinity of ferroelectric domain walls, *i.e.* compensation is possible only on the length scale of carrier diffusion in the materials.[402]

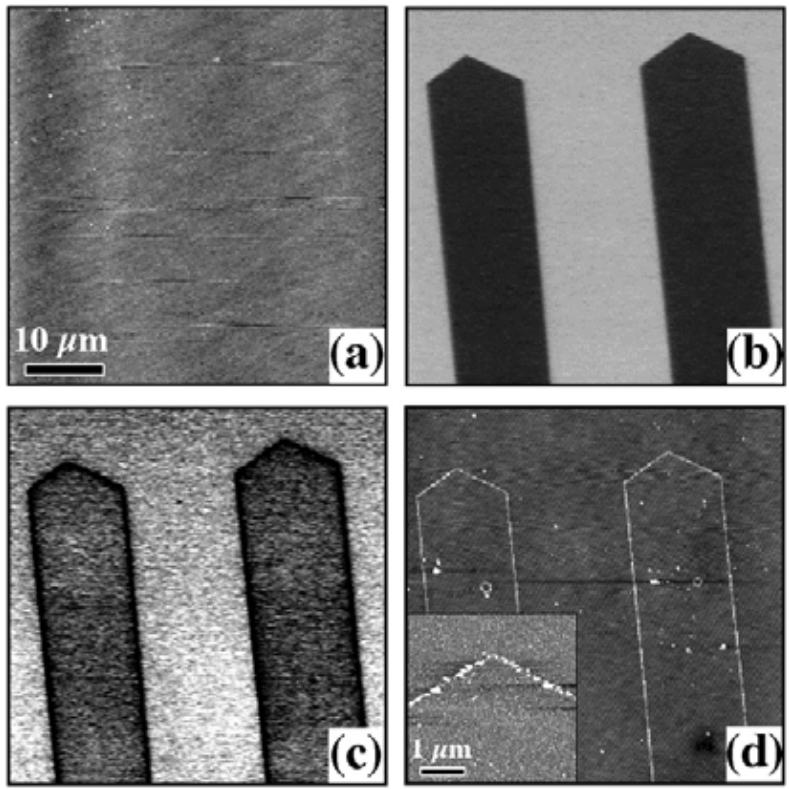

**Figure VI.7.** (a) Topographic image of the PPLN sample before deposition; ((b), (c)) corresponding PFM phase and amplitude images, respectively, of the PPLN sample; (d) topographic image of the same sample after deposition (inset: higher-resolution topographic image illustrating the structure of the lines as formed by silver particles). Reprinted with permission from [Hanson J N, Rodriguez B J, Nemanich R J and Gruverman A 2006 Nanotechnology **17** 4946-9]. © IOP Publishing. Reproduced by permission of IOP Publishing. All right reserved (2006) [402].

In addition to the photochemical deposition, there are many tries for understanding and probing of screening phenomena on ferroelectric surface as discussed in Section VI.2.[280, 281, 395, 403, 404] Beyond fundamental probing of screening phenomena on the ferroelectric surfaces, these studies open the pathway for polarization controlled creation of metallic and other nanostructures. In this ferroelectric lithography as pioneered by Bonnell group,[405] the charged SPM tip is used to create domain pattern and subsequent metal photodeposition creates nanostructure. The latter can then be studied by *e.g.* contacting to outside worlds, chemical modification, etc., or potentially transferred to different substrate. Notably, the metal photodeposition is not associated with significant changes in ferroelectric surfaces and hence can be performed multiple times, effectively providing a master. This approach was recently extensively explored by Gruverman,[394] Eng.[406] Kitamura[407] and others.[405] As an example, Eng et al., prepared Pt nanostructures through creation of domain patterns and subsequent photochemical reduction.[406] Further, as shown in figure VI.8, Kalinin et al., controlled local ferroelectric domains by application of an electric field through the conductive AFM tip then the subsequent photo reduction enabled deposition of Ag nanoparticles only on $c^+$ domains in the pattern.

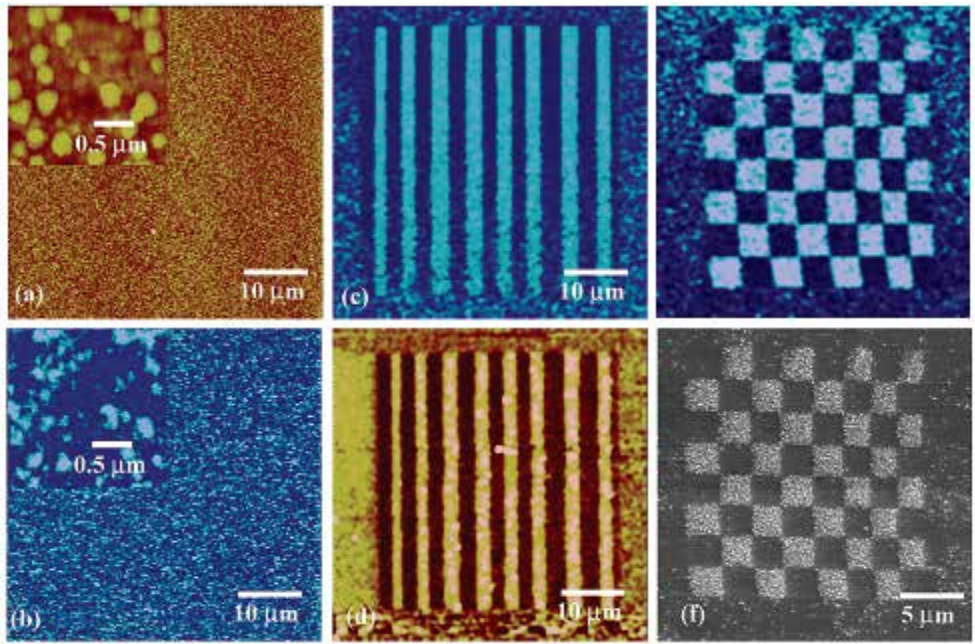

**Figure VI.8.** Surface topography (a) and piezoresponse image (b) of PZT thin film. The inset shows that the PFM contrast is not random but is due to the small (~50-100 nm) ferroelectric

domains associated with grains. PFM image (c) of lines patterned with alternating +10 and -10 Vdc. Surface topography (d) after deposition of Ag nanoparticles. Note one-to-one correspondence between tip-induced polarization distribution and metal deposition pattern. The features consist of closely packed metal nanoparticles of 3-10 nm. Piezoresponse image of checkerboard domain structure fabricated using in-house lithographic system (e) and SEM image of corresponding silver photodeposition pattern (f). Reprinted with permission from [Kalinin S V, Bonnell D A, Alvarez T, Lei X, Hu Z, Ferris J H, Zhang Q and Dunn S 2002 *Nano Lett.* **2** 589-93]. Copyright (2002) ACS Publications [405].

**VII. Conclusion and outlook**

Ferroelectric surfaces are the inherent part of ferroelectric materials, and via modern scanning probe imaging techniques provide the window into the bulk physics and domain dynamics. However, the presence of the normal polarization component at surfaces necessitates polarization screening that can be realized via surface band bending or dissociative chemisorption of reactive species. Consequently, screening charge dynamics becomes non-trivial component of the static and dynamic properties of ferroelectric surfaces and near surface regions. Polarization cannot be switched unless screening charge redistribute – hence transport of screening species can become rate-limiting step of switching phenomena. The conservation of screening charge species can give rise to nontrivial static and dynamic phenomena, as illustrated by bistability of domain wall motion and formation of chaotic structures. However, it also controls switching that leads to classical domain geometries, and hence should be considered along with bulk wall motion and pinning mechanisms. Surface screening significantly affects the basic structure of the ferroelectric domain wall-surface junctions, and hence can be highly relevant for domain wall conduction mechanisms as observed by AFM.[13, 15, 18, 21, 22, 408-410] Finally, in thin films the coupling between the ferroelectricity and surface electrochemistry can give rise to the continuum of coupled ferroelectric-electrochemical states, reminiscent of relaxors.

Scanning probe microscopy techniques have provided valuable insight into structure and properties of ferroelectric surfaces via dual observables of surface potential and electromechanical response. Combined with variable temperature and time dependent studies, these observations provide the definitive evidence towards surface ionic screening. However, observations such as metal photodeposition suggest that there are also changes in the surface electronic structures. These considerations necessitate local structure- and functionality sensitive probing via electron microscopy and scattering techniques, as e.g. demonstrated recently by Evans group for switching effect on local crystallography.[411]. In the future, direct correlation between probe and X-ray imaging tools in the form of multimodal imaging will allow to enable direct data mining [412] of structure property relationships of ferroelectric surfaces.

Finally, understanding of ferroelectric surfaces necessitates the development and mesoscopic theoretical models that capture and combine ferroelectric and (electro)chemical degrees of freedom, and can be applied for other polar oxide surfaces.

**Acknowledgement**

Research was supported (S.V.K. and D.F.) by the U.S. Department of Energy, Office of Science, Basic Energy Sciences, Materials Sciences and Engineering Division. This research was conducted at the Center for Nanophase Materials Sciences, which is a DOE Office of Science User Facility. This work was partially supported (Y.K.) by Basic Science Research program through the National Research Foundation of Korea funded by the Ministry of Science, ICT & Future Planning (NRF-2014R1A4A1008474).

# References

[1] Tagantsev A K, Cross L E and Fousek J 2010 *Domains in Ferroic Crystals and Thin Films* (New York: Springer)
[2] Scott J F 2007 *Science* **315** 954-9
[3] Setter N, Damjanovic D, Eng L, Fox G, Gevorgian S, Hong S, Kingon A, Kohlstedt H, Park N Y, Stephenson G B, Stolitchnov I, Tagantsev A K, Taylor D V, Yamada T and Streiffer S 2006 *J. Appl. Phys.* **100** 051606
[4] Vugmeister B E 2006 *Phys. Rev. B* **73** 174117.
[5] Glinchuk M D and Stephanovich V A 1999 *J. Appl. Phys.* **85** 1722-6
[6] Glinchuk M D and Stephanovich V A 1998 *J. Phys.-Condens. Matter* **10** 11081-94
[7] Westphal V, Kleemann W and Glinchuk M D 1992 *Phys. Rev. Lett.* **68** 847-50
[8] Spaldin N A and Fiebig M 2005 *Science* **309** 391-2
[9] Fiebig M 2005 *J. Phys. D-Appl. Phys.* **38** R123-R52
[10] Fiebig M, Lottermoser T, Frohlich D, Goltsev A V and Pisarev R V 2002 *Nature* **419** 818-20
[11] Wang J, Neaton J B, Zheng H, Nagarajan V, Ogale S B, Liu B, Viehland D, Vaithyanathan V, Schlom D G, Waghmare U V, Spaldin N A, Rabe K M, Wuttig M and Ramesh R 2003 *Science* **299** 1719-22
[12] Cheong S W and Mostovoy M 2007 *Nat. Mater.* **6** 13-20
[13] Seidel J, Martin L W, He Q, Zhan Q, Chu Y H, Rother A, Hawkridge M E, Maksymovych P, Yu P, Gajek M, Balke N, Kalinin S V, Gemming S, Wang F, Catalan G, Scott J F, Spaldin N A, Orenstein J and Ramesh R 2009 *Nat. Mater.* **8** 229-34
[14] Daraktchiev M, Catalan G and Scott J F 2010 *Phys. Rev. B* **81** 224118
[15] Vasudevan R K, Wu W D, Guest J R, Baddorf A P, Morozovska A N, Eliseev E A, Balke N, Nagarajan V, Maksymovych P and Kalinin S V 2013 *Adv. Funct. Mater.* **23** 2592-616
[16] Balke N, Winchester B, Ren W, Chu Y H, Morozovska A N, Eliseev E A, Huijben M, Vasudevan R K, Maksymovych P, Britson J, Jesse S, Kornev I, Ramesh R, Bellaiche L, Chen L Q and Kalinin S V 2012 *Nat. Phys.* **8** 81-8
[17] Maksymovych P, Morozovska A N, Yu P, Eliseev E A, Chu Y H, Ramesh R, Baddorf A P and Kalinin S V 2012 *Nano Lett.* **12** 209-13
[18] Eliseev E A, Morozovska A N, Svechnikov G S, Maksymovych P and Kalinin S V 2012 *Phys. Rev. B* **85** 045312
[19] Morozovska A N, Vasudevan R K, Maksymovych P, Kalinin S V and Eliseev E A 2012 *Phys. Rev. B* **86** 085315
[20] Maksymovych P, Seidel J, Chu Y H, Wu P P, Baddorf A P, Chen L Q, Kalinin S V and Ramesh R 2011 *Nano Lett.* **11** 1906-12
[21] Farokhipoor S and Noheda B 2011 *Phys. Rev. Lett.* **107** 127601
[22] Guyonnet J, Gaponenko I, Gariglio S and Paruch P 2011 *Adv. Mater.* **23** 5377-82
[23] Stachiotti M G and Sepliarsky M 2011 *Phys. Rev. Lett.* **106** 137601
[24] Fong D D, Kolpak A M, Eastman J A, Streiffer S K, Fuoss P H, Stephenson G B, Thompson C, Kim D M, Choi K J, Eom C B, Grinberg I and Rappe A M 2006 *Phys. Rev. Lett.* **96** 127601
[25] Fister T T, Fong D D, Eastman J A, Iddir H, Zapol P, Fuoss P H, Balasubramanian M, Gordon R A, Balasubramaniam K R and Salvador P A 2011 *Phys. Rev. Lett.* **106** 037401
[26] Highland M J, Fister T T, Fong D D, Fuoss P H, Thompson C, Eastman J A, Streiffer S K and Stephenson G B 2011 *Phys. Rev. Lett.* **107** 187602
[27] Streiffer S K, Eastman J A, Fong D D, Thompson C, Munkholm A, Murty M V R, Auciello O, Bai G R and Stephenson G B 2002 *Phys. Rev. Lett.* **89** 067601
[28] Jia C L, Urban K W, Alexe M, Hesse D and Vrejoiu I 2011 *Science* **331** 1420-3

[29] Jia C L, Nagarajan V, He J Q, Houben L, Zhao T, Ramesh R, Urban K and Waser R 2007 *Nat. Mater.* **6** 64-9
[30] Jia C L, Mi S B, Urban K, Vrejoiu I, Alexe M and Hesse D 2008 *Nat. Mater.* **7** 57-61
[31] Borisevich A Y, Chang H J, Huijben M, Oxley M P, Okamoto S, Niranjan M K, Burton J D, Tsymbal E Y, Chu Y H, Yu P, Ramesh R, Kalinin S V and Pennycook S J 2010 *Phys. Rev. Lett.* **105** 087204
[32] Borisevich A Y, Morozovska A N, Kim Y M, Leonard D, Oxley M P, Biegalski M D, Eliseev E A and Kalinin S V 2012 *Phys. Rev. Lett.* **109** 065702
[33] Borisevich A Y, Lupini A R, He J, Eliseev E A, Morozovska A N, Svechnikov G S, Yu P, Chu Y H, Ramesh R, Pantelides S T, Kalinin S V and Pennycook S J 2012 *Phys. Rev. B* **86** 140102
[34] Chang H J, Kalinin S V, Morozovska A N, Huijben M, Chu Y H, Yu P, Ramesh R, Eliseev E A, Svechnikov G S, Pennycook S J and Borisevich A Y 2011 *Adv. Mater.* **23** 2474-9
[35] Kingsmith R D and Vanderbilt D 1993 *Phys. Rev. B* **47** 1651-4
[36] Vanderbilt D and Kingsmith R D 1993 *Phys. Rev. B* **48** 4442-55
[37] Resta R, Posternak M and Baldereschi A 1993 *Phys. Rev. Lett.* **70** 1010-3
[38] Dawber M, Rabe K M and Scott J F 2005 *Rev. Mod. Phys.* **77** 1083-130
[39] Rosenman G, Shur D, Krasik Y E and Dunaevsky A 2000 *J. Appl. Phys.* **88** 6109-61
[40] Puchkarev V F and Mesyats G A 1995 *J. Appl. Phys.* **78** 5633-7
[41] Riege H 1994 *Nucl. Instrum. Methods Phys. Res. Sect. A-Accel. Spectrom. Dect. Assoc. Equip.* **340** 80-9
[42] Rosenman G and Rez I 1993 *J. Appl. Phys.* **73** 1904-8
[43] Klopfer M, Wolowiec T, Satchouk V, Alivov Y and Molloi S 2012 *Nucl. Instrum. Methods Phys. Res. Sect. A-Accel. Spectrom. Dect. Assoc. Equip.* **689** 47-51
[44] Naranjo B, Gimzewski J K and Putterman S 2005 *Nature* **434** 1115-7
[45] Thompson C, Fong D D, Wang R V, Jiang F, Streiffer S K, Latifi K, Eastman J A, Fuoss P H and Stephenson G B 2008 *Appl. Phys. Lett.* **93** 182901
[46] Ohtomo A, Muller D A, Grazul J L and Hwang H Y 2002 *Nature* **419** 378-80
[47] Mannhart J and Schlom D G 2010 *Science* **327** 1607-11
[48] Zhong Z C, Xu P X and Kelly P J 2010 *Phys. Rev. B* **82** 165127
[49] Kalabukhov A, Gunnarsson R, Borjesson J, Olsson E, Claeson T and Winkler D 2007 *Phys. Rev. B* **75** 121404
[50] Cheng G L, Siles P F, Bi F, Cen C, Bogorin D F, Bark C W, Folkman C M, Park J W, Eom C B, Medeiros-Ribeiro G and Levy J 2011 *Nat. Nanotech.* **6** 343-7
[51] Bi F, Bogorin D F, Cen C, Bark C W, Park J W, Eom C B and Levy J 2010 *Appl. Phys. Lett.* **97** 173110
[52] Bark C W, Sharma P, Wang Y, Baek S H, Lee S, Ryu S, Folkman C M, Paudel T R, Kumar A, Kalinin S V, Sokolov A, Tsymbal E Y, Rzchowski M S, Gruverman A and Eom C B 2012 *Nano Lett.* **12** 1765-71
[53] Kumar A, Arruda T M, Kim Y, Ivanov I N, Jesse S, Bark C W, Bristowe N C, Artacho E, Littlewood P B, Eom C B and Kalinin S V 2012 *ACS Nano* **6** 3841-52
[54] Kittel C 1995 *Introduction to solid state physics* (Wiley)
[55] Catalan G, Scott J F, Schilling A and Gregg J M 2007 *J. Phys.-Condens. Matter* **19** 132201
[56] Dawber M, Chandra P, Littlewood P B and Scott J F 2003 *J. Phys.-Condens. Matter* **15** L393-L8
[57] Chisholm M F, Luo W D, Oxley M P, Pantelides S T and Lee H N 2010 *Phys. Rev. Lett.* **105** 197602
[58] Aravind V R, Morozovska A N, Bhattacharyya S, Lee D, Jesse S, Grinberg I, Li Y L, Choudhury S, Wu P, Seal K, Rappe A M, Svechnikov S V, Eliseev E A, Phillpot S R, Chen L Q, Gopalan V and Kalinin S V 2010 *Phys. Rev. B* **82** 024111
[59] Bocher L, Gloter A, Crassous A, Garcia V, March K, Zobelli A, Valencia S, Enouz-Vedrenne S, Moya X, Marthur N D, Deranlot C, Fusil S, Bouzehouane K, Bibes M, Barthelemy A, Colliex C and Stephan O 2012 *Nano Lett.* **12** 376-82

[60] Kalinin S V and Bonnell D A 2004 *Nano Lett.* **4** 555-60
[61] Jesse S, Baddorf A P and Kalinin S V 2006 *Appl. Phys. Lett.* **88** 062908
[62] Morozovska A N, Svechnikov S V, Eliseev E A, Jesse S, Rodriguez B J and Kalinin S V 2007 *J. Appl. Phys.* **102** 114108
[63] Ievlev A V, Jesse S, Morozovska A N, Strelcov E, Eliseev E A, Pershin Y V, Kumar A, Shur V Y and Kalinin S V 2014 *Nat. Phys.* **10** 59-66
[64] Ievlev A V, Morozovska A N, Shur V Y and Kalinin S V 2014 *Appl. Phys. Lett.* **104** 092908
[65] Bardeen J 1947 *Phys. Rev.* **71** 717-27
[66] Kreher K 1980 *Kristall und Technik* **15** 1392
[67] Pintilie L, Stancu V, Trupina L and Pintilie I 2010 *Phys. Rev. B* **82** 085319
[68] Morozovska A N, Eliseev E A, Svechnikov S V, Krutov A D, Shur V Y, Borisevich A Y, Maksymovych P and Kalinin S V 2010 *Phys. Rev. B* **81** 205308
[69] Itskovsky M A 1974 *Fiz. Tverd. Tela (Leningrad)* **16** 2065
[70] Tagantsev A K, Cross L E and Fousek J 2010 *Domains in ferroic crystals and thin film* (New York: Springer)
[71] Eliseev E A, Kalinin S V and Morozovska A N 2015 *J. Appl. Phys.* **117** 034102
[72] Bratkovsky A M and Levanyuk A P 2009 *J. Comput. Theor. Nanos.* **6** 465-89
[73] Bratkovsky A M and Levanyuk A P 2011 *Phys. Rev. B.* **84** 045401
[74] Chenskii E and Tarasenko V 1982 *Zh. Eksp. Teor. Fiz.* **83** 1089-99
[75] Jia C L, Urban K W, Alexe M, Hesse D and Vrejoiu I 2011 *Science* **331** 1420-3
[76] Tang Y L, Zhu Y L, Ma X L, Borisevich A Y, Morozovska A N, Eliseev E A, Wang W Y, Wang Y J, Xu Y B, Zhang Z D and Pennycook S J 2015 *Science* **348** 547-51
[77] Darinskii B M, Lazarev A P and Sidorkin A S 1989 *Fizika tverdogo tela* 31 287-9.
[78] Eliseev E A, Morozovska A N, Kalinin S V, Li Y L, Shen J, Glinchuk M D, Chen L Q and Gopalan V 2009 *J. Appl. Phys.* **106** 084102
[79] Shur V Y, Gruverman A, Kuminov V and Tonkachyova N 1990 *Ferroelectrics* **111** 197-206
[80] Eliseev E A, Morozovska A N, Svechnikov G S, Rumyantsev E L, Shishkin E I, Shur V Y and Kalinin S V 2008 *Phys. Rev. B* **78** 245409
[81] Schmickler W 1996 *Interfacial Electrochemistry* (Oxford University Press, Oxford) ch 4
[82] Timoshenko S P and Goodier J N 1970 Theory of Elasticity (New York: McGraw-Hill)
[83] Landau L D and Lifshitz E M 1986 *Course of Theoretical Physics: Vol. 7 Theory of Elasticity* (Butterworth-Heinemann, Oxford) **3** 109
[84] Glinchuk M D, Eliseev E A, Stephanovich V A and Farhi R 2003 *J. Appl. Phys.* **93** 1150-9
[85] Glinchuk M D, Eliseev E A, Deineka A, Jastrabik L, Suchaneck G, Sandner T, Gerlach G and Hrabovsky M 2001 *Integr. Ferroelectr.* **38** 745-54
[86] Woo C H and Zheng Y 2008 *Appl. Phys. A-Mater.* **91** 59-63
[87] Damon Y and Cho Y 2006 *Jpn. J. Appl. Phys. 2* **45** L1304-L6
[88] Daimon Y and Cho Y 2007 *Appl. Phys. Lett.* **90** 192906
[89] Kalinin S V and Bonnell D A 2002 *Electrically Based Microstructural Characterization III,* ed. Gerhardt R A *et al.* pp 101-6
[90] Tagantsev A K and Gerra G 2006 *J. Appl. Phys.* **100** 051607
[91] Catalan G, Bea H, Fusil S, Bibes M, Paruch P, Barthelemy A and Scott J F 2008 *Phys. Rev. Lett.* **100** 027602
[92] This behavior can be illustrated by considering a simple system consisting of the finite strip with width b with constant surface charge placed at the distance H from the electrode. In the case the full field distribution can be obtained in the evident form. The field behaves as at the film surface near the edge of the strip and vanishes with b increase.
[93] Sawa A 2008 *Mater. Today* **11** 28-36
[94] Magyari-Kope B, Tendulkar M, Park S G, Lee H D and Nishi Y 2011 *Nanotechnology* **22** 254029
[95] Shang D S, Shi L, Sun J R and Shen B G 2011 *Nanotechnology* **22** 254008

[96]   Kim Y, Kelly S J, Morozovska A, Rahani E K, Strelcov E, Eliseev E, Jesse S, Biegalski M D, Balke N, Benedek N, Strukov D, Aarts J, Hwang I, Oh S, Choi J S, Choi T, Park B H, Shenoy V B, Maksymovych P and Kalinin S V 2013 *Nano Lett.* **13** 4068-74
[97]   Waser R and Aono M 2007 *Nat. Mater.* **6** 833-40
[98]   Szot K, Rogala M, Speier W, Klusek Z, Besmehn A and Waser R 2011 *Nanotechnology* **22** 254001
[99]   Morozovska A N, Eliseev E A and Kalinin S V 2010 *Appl. Phys. Lett.* **96** 222906
[100]  Morozovska A N, Eliseev E A, Balke N and Kalinin S V 2010 *J. Appl. Phys.* **108** 053712
[101]  Balke N, Jesse S, Morozovska A N, Eliseev E, Chung D W, Kim Y, Adamczyk L, Garcia R E, Dudney N and Kalinin S V 2010 *Nat. Nanotech.* **5** 749-54
[102]  Morozovska A N, Eliseev E A, Varenyk O V, Kim Y, Strelcov E, Tselev A, Morozovsky N V and Kalinin S V 2014 *J. Appl. Phys.* **116** 066808
[103]  McLachlan M A, McComb D W, Ryan M P, Morozovska A N, Eliseev E A, Payzant E A, Jesse S, Seal K, Baddorf A P and Kalinin S V 2011 *Adv. Funct. Mater.* **21** 941-7
[104]  Morozovska A N, Eliseev E A, Svechnikov G S and Kalinin S V 2011 *Phys. Rev. B* **84** 045402
[105]  Morozovska A N, Eliseev E A and Kalinin S V 2012 *J. Appl. Phys.* **111** 014114
[106]  Morozovska A N, Eliseev E A, Bravina S L, Ciucci F, Svechnikov G S, Chen L Q and Kalinin S V 2012 *J. Appl. Phys.* **111** 014107
[107]  Chua L 1971 *IEEE Trans. Circuit Theory* **18** 507-19
[108]  Chua L O and Sung Mo K 1976 *P. IEEE* **64** 209-23
[109]  Mathur N D 2008 *Nature* **455** E13
[110]  Gil Y, Umurhan O M and Riess I 2007 *Solid State Ion.* **178** 1-12
[111]  Gil Y, Umurhan O M and Riess I 2008 *J. Appl. Phys.* **104** 084504
[112]  Allen S M and Cahn J W 1979 *Acta Metall.* **27** 1085-95
[113]  Zhang X C, Shyy W and Sastry A M 2007 *J. Electrochem. Soc.* **154** A910-A6
[114]  Tang M, Huang H Y, Meethong N, Kao Y H, Carter W C and Chiang Y M 2009 *Chem. Mater.* **21** 1557-71
[115]  Tang M, Carter W C, Belak J F and Chiang Y-M 2010 *Electrochim. Acta* **56** 969-76
[116]  Morozovska A N, Glinchuk M D, 2016 *J. Appl. Phys.* **119**, 094109
[117]  Morozovska A N, Eliseev E A and Kalinin S V (unpublished)
[118]  Highland M J, Fister T T, Fong D D, Fuoss P H, Thompson C, Eastman J A, Streiffer S K and Stephenson G B 2011 *Phys. Rev. Lett.* **107** 187602
[119]  Stephenson G B and Highland M J 2011 *Phys. Rev. B* **84** 064107
[120]  Morozovska A N, Eliseev E A, Genenko Y A, Vorotiahin I S, Silibin M V, Cao Y, Kim Y, Glinchuk M D and Kalinin S V 2016 *Phys. Rev. B* **94** 174101
[121]  Waser R, Böttger U and Tiedke S 2004 *Polar Oxides: Properties, Characterization, and Imaging* (Wiley) ch 8
[122]  Fong D D and Thompson C 2006 *Annu. Rev. Mater. Res.* **36** 431-65
[123]  Shin J, Nascimento V B, Borisevich A Y, Plummer E W, Kalinin S V and Baddorf A P 2008 *Phys. Rev. B* **77** 245437
[124]  Shin J, Nascimento V B, Geneste G, Rundgren J, Plummer E W, Dkhil B, Kalinin S V and Baddorf A P 2009 *Nano Lett.* **9** 3720-5
[125]  Wang J L, Gaillard F, Pancotti A, Gautier B, Niu G, Vilquin B, Pillard V, Rodrigues G L M P and Barrett N 2012 *J. Phys. Chem. C* **116** 21802-9
[126]  Stephenson G B, Fong D D, Murty M V R, Streiffer S K, Eastman J A, Auciello O, Fuoss P H, Munkholm A, Aanerud M E M and Thompson C 2003 *Physica B* **336** 81-9
[127]  Rusanov A I 2005 *Surf. Sci. Rep.* **58** 111-239
[128]  Cammarata R C 2009 *Solid State Phys.* **61** 1-75
[129]  Munkholm A, Streiffer S K, Murty M V R, Eastman J A, Thompson C, Auciello O, Thompson L, Moore J F and Stephenson G B 2002 *Phys. Rev. Lett.* **88** 016101

[130] Bousquet E, Dawber M, Stucki N, Lichtensteiger C, Hermet P, Gariglio S, Triscone J M and Ghosez P 2008 *Nature* **452** 732-6
[131] Fong D D, Stephenson G B, Streiffer S K, Eastman J A, Auciello O, Fuoss P H and Thompson C 2004 *Science* **304** 1650-3
[132] Stephenson G B and Elder K R 2006 *J. Appl. Phys.* **100** 051601
[133] Fong D D, Cionca C, Yacoby Y, Stephenson G B, Eastman J A, Fuoss P H, Streiffer S K, Thompson C, Clarke R, Pindak R and Stern E A 2005 *Phys. Rev. B.* **71** 144112
[134] Wang R V, Fong D D, Jiang F, Highland M J, Fuoss P H, Thompson C, Kolpak A M, Eastman J A, Streiffer S K, Rappe A M and Stephenson G B 2009 *Phys. Rev. Lett.* **102** 047601
[135] Highland M J, Fister T T, Richard M I, Fong D D, Fuoss P H, Thompson C, Eastman J A, Streiffer S K and Stephenson G B 2010 *Phys. Rev. Lett.* **105** 167601
[136] Alahmed Z and Fu H X 2007 *Phys. Rev. B* **76** 224101
[137] Apostol N G, Stoflea L E, Lungu G A, Tache C A, Popescu D G, Pintilie L and Teodorescu C M 2013 *Mater. Sci. Eng. B-Adv.* **178** 1317-22
[138] Stoflea L E, Apostol N G, Trupina L and Teodorescu C M 2014 *J. Mater. Chem. A* **2** 14386-92
[139] Watanabe Y, Okano M and Masuda A 2001 *Phys. Rev. Lett.* **86** 332-5
[140] He J, Stephenson G B and Nakhmanson S M 2012 *J. Appl. Phys.* **112** 054112
[141] Hong S, Nakhmanson S M and Fong D D 2016 *Rep. Prog. Phys.* **79** 076501
[142] Liviu Cristian Tănase N G A, Laura Elena Abramiuc, Cristian Alexandru Tache, Luminița Hrib, Lucian Trupină, Lucian Pintilie and Cristian Mihail Teodorescu 2016 *Sci. Rep.* **6** 35301
[143] Nassreddine S, Morfin F, Niu G, Vilquin B, Gaillard F and Piccolo L 2014 *Surf. Interface Anal.* **46** 721-5
[144] Kakekhani A and Ismail-Beigi S 2016 *J. Mater. Chem. A* **4** 5235-46
[145] Kakekhani A and Ismail-Beigi S 2016 *Phys. Chem. Chem. Phys.* **18** 19676-95
[146] Kolpak A M, Grinberg I and Rappe A M 2007 *Phys. Rev. Lett.* **98** 166101
[147] Kim S, Schoenberg M R and Rappe A M 2011 *Phys. Rev. Lett.* **107** 076102
[148] Yasunobu Inoue I Y, and Kazunori Sat 1984 *J. Phys. Chem.* **88** 1148-51
[149] Yun Y, Pilet N, Schwarz U D and Altman E I 2009 *Surf. Sci.* **603** 3145-54
[150] Garrity K, Kakekhani A, Kolpak A and Ismail-Beigi S 2013 *Phys. Rev. B* **88** 045401
[151] Kakekhani A and Ismail-Beigi S 2015 *ACS Catal.* **5** 4537-45
[152] Khan M A, Nadeem M A and Idrissn H 2016 *Surf. Sci. Rep.* **71** 1-31
[153] Yun Y, Kampschulte L, Li M, Liao D and Altman E I 2007 *J. Phys. Chem. C* **111** 13951-6
[154] Yun Y and Altman E I 2007 *J. Am. Chem. Soc.* **129** 15684-9
[155] Kakekhani A, Ismail-Beigi S and Altman E I 2016 *Surf. Sci.* **650** 302-16
[156] Garra J, Vohs J M and Bonnell D A 2009 *Surf. Sci.* **603** 1106-14
[157] Garra J, Vohs J M and Bonnell D A 2009 *J. Vac. Sci. Technol. A* **27** L13-L7
[158] Popescu D G, Husanu M A, Trupina L, Hrib L, Pintilie L, Barinov A, Lizzit S, Lacovig P and Teodorescu C M 2015 *Phys. Chem. Chem. Phys.* **17** 509-20
[159] Husanu M A, Popescu D G, Tache C A, Apostol N G, Barinov A, Lizzit S, Lacovig P and Teodorescu C M 2015 *Appl. Surf. Sci.* **352** 73-81
[160] Rault J E, Mentes T O, Locatelli A and Barrett N 2014 *Sci. Rep.* **4** 6792
[161] Thomas A G, Muryn C A, Hardman P J, Dürr H A, Owen I W, Thornton G, Quinn F M, Rosenberg R A, Love P J and Rehn V 1994 *Surf. Sci.* **307-309** 355-9
[162] Li D B and Bonnell D A 2008 *Annu. Rev. Mater. Res.* **38** 351-68
[163] Nelson C T, Gao P, Jokisaari J R, Heikes C, Adamo C, Melville A, Baek S H, Folkman C M, Winchester B, Gu Y J, Liu Y M, Zhang K, Wang E G, Li J Y, Chen L Q, Eom C B, Schlom D G and Pan X Q 2011 *Science* **334** 968-71
[164] Cherifi S, Hertel R, Fusil S, Bea H, Bouzehouane K, Allibe J, Bibes M and Barthelemy A 2010 *Phys. Status Solidi-R* **4** 22-4
[165] Wang Y G, Dec J and Kleemann W 1998 *J. Appl. Phys.* **84** 6795-9

[166] Gruverman A L, Hatano J and Tokumoto H 1997 *Jpn. J. Appl. Phys. Part 1 - Regul. Pap. Short Notes Rev. Pap.* **36** 2207-11
[167] Takashige M, Hamazaki S, Fukurai N and Shimizu F 1997 *J. Phys. Soc. Jpn.* **66** 1848-9
[168] Takashige M, Hamazaki S I, Fukurai N, Shimizu F and Kojima S 1997 *Ferroelectrics* **203** 221-5
[169] Takashige M, Hamazaki S I, Fukurai N, Shimizu F and Kojima S 1996 *Jpn. J. Appl. Phys. Part 1 - Regul. Pap. Short Notes Rev. Pap.* **35** 5181-4
[170] Balakumar S, Xu J B, Ma J X, Ganesamoorthy S and Wilson I H 1997 *Jpn. J. Appl. Phys. Part 1 - Regul. Pap. Short Notes Rev. Pap.* **36** 5566-9
[171] Pang G K H and Baba-Kishi K Z 1998 *J. Phys. D-Appl. Phys.* **31** 2846-53
[172] Eng L M, Friedrich M, Fousek J and Gunter P 1996 *Ferroelectrics* **186** 49-52
[173] Eng L M, Friedrich M, Fousek J and Gunter P 1996 *J. Vac. Sci. Technol. B* **14** 1191-6
[174] Bluhm H, Wiesendanger R and Meyer K P 1996 *J. Vac. Sci. Technol. B B* **14** 1180-3
[175] Shilo D, Ravichandran G and Bhattacharya K 2004 *Nat. Mater.* **3** 453-7
[176] Bluhm H, Schwarz U D and Wiesendanger R 1998 *Phys. Rev. B* **57** 161-9
[177] Correia A, Massanell J, Garcia N, Levanyuk A P, Zlatkin A and Przeslawski J 1996 *Appl. Phys. Lett.* **68** 2796-8
[178] Tolstikhina A L, Belugina N V and Shikin S A 2000 *Ultramicroscopy* **82** 149-52
[179] Bluhm H, Schwarz U D and Meyer K P 1995 *Appl. Phys. A-Mater. Sci. Process.* **61** 525-33
[180] Labardi M, Allegrini M, Fuso F, Leccabue F, Watts B, Ascoli C and Frediani C 1995 *Integr. Ferroelectr.* **8** 143-50
[181] Qi Y B, Park J Y, Hendriksen B L M, Ogletree D F and Salmeron M 2008 *Phys. Rev. B* **77** 184105
[182] Park J Y, Qi Y B, Ogletree D F, Thiel P A and Salmeron M 2007 *Phys. Rev. B* **76** 064108
[183] Park J Y, Ogletree D F, Thiel P A and Salmeron M 2006 *Science* **313** 186
[184] Rabe U, Kopycinska M, Hirsekorn S, Saldana J M, Schneider G A and Arnold W 2002 *J. Phys. D-Appl. Phys.* **35** 2621-35
[185] Liu X X, Heiderhoff R, Abicht H P and Balk L J 2002 *J. Phys. D-Appl. Phys.* **35** 74-87
[186] Li D B, Zhao M H, Garra J, Kolpak A M, Rappe A M, Bonnell D A and Vohs J M 2008 *Nat. Mater.* **7** 473-7
[187] Yin Q R, Zeng H R, Yu H F and Li G R 2006 *J. Mater. Sci.* **41** 259-70
[188] Yu H F, Zeng H R, Ding A L, Li G R, Luo H S and Yin Q R 2006 *Integr. Ferroelectr.* **78** 309-17
[189] Yu H F, Zeng H R, Ding A L, Li G R, Luo H S and Yin Q R 2005 *Integr. Ferroelectr.* **73** 165-71
[190] Yu H F, Zeng H R, Ma X D, Chu R Q, Li G R, Luo H S and Yin Q R 2005 *Phys. Status Solidi A-Appl. Res.* **202** R10-R2
[191] Jesse S and Kalinin S V 2011 *J. Phys. D-Appl. Phys.* **44** 464006
[192] Kalinin S V, Karapetian E and Kachanov M 2004 *Phys. Rev. B* **70** 184101
[193] Karapetian E, Kachanov M and Kalinin S V 2005 *Philos. Mag.* **85** 1017-51
[194] Hong J W, Park S I and Khim Z G 1999 *Rev. Sci. Instrum.* **70** 1735-9
[195] Takata K, Kushida K and Torii K 1995 *Jpn. J. Appl. Phys. Part 1 - Regul. Pap. Short Notes Rev. Pap.* **34** 2890-4
[196] Franke K and Weihnacht M 1995 *Ferroelectr. Lett. Sect.* **19** 25-33
[197] Franke K, Besold J, Haessler W and Seegebarth C 1994 *Surf. Sci.* **302** L283-L8
[198] Nonnenmacher M, Oboyle M P and Wickramasinghe H K 1991 *Appl. Phys. Lett.* **58** 2921-3
[199] S. Sadewasser and T. Glatzel E 2011 *Kelvin Probe Force Microscopy* (New York: Springer Science)
[200] Kitamura S and Iwatsuki M 1998 *Appl. Phys. Lett.* **72** 3154-6
[201] Zerweck U, Loppacher C, Otto T, Grafstrom S and Eng L M 2005 *Phys. Rev. B* **71** 125424
[202] Loppacher C, Zerweck U, Teich S, Beyreuther E, Otto T, Grafstrom S and Eng L M 2005 *Nanotechnology* **16** S1-S6
[203] Sommerhalter C, Matthes T W, Glatzel T, Jager-Waldau A and Lux-Steiner M C 1999 *Appl. Phys. Lett.* **75** 286-8

[204] Ziegler D and Stemmer A 2011 *Nanotechnology* **22** 075501
[205] Liscio A, Palermo V and Samori P 2010 *Accounts Chem. Res.* **43** 541-50
[206] Kalinin S V and Bonnell D A 2002 *Phys. Rev. B* **65** 125408
[207] Huey B D, Ramanujan C, Bobji M, Blendell J, White G, Szoszkiewicz R and Kulik A 2004 *J. Electroceram.* **13** 287-91
[208] Harnagea C, Alexe M, Hesse D and Pignolet A 2003 *Appl. Phys. Lett.* **83** 338-40
[209] Seal K, Jesse S, Rodriguez B J, Baddorf A P and Kalinin S V 2007 *Appl. Phys. Lett.* **91** 232904
[210] Jesse S, Baddorf A P and Kalinin S V 2006 *Nanotechnology* **17** 1615-28
[211] Hong S, Shin H, Woo J and No K 2002 *Appl. Phys. Lett.* **80** 1453-5
[212] Hong S, Woo J, Shin H, Jeon J U, Pak Y E, Colla E L, Setter N, Kim E and No K 2001 *J. Appl. Phys.* **89** 1377-86
[213] Gruverman A, Auciello O and Tokumoto H 1998 *Annu. Rev. Mater. Sci.* **28** 101-23
[214] Gruverman A, Auciello O, Ramesh R and Tokumoto H 1997 *Nanotechnology* **8** A38-A43
[215] Eng L M, Bammerlin M, Loppacher C, Guggisberg M, Bennewitz R, Luthi R, Meyer E, Huser T, Heinzelmann H and Guntherodt H J 1999 *Ferroelectrics* **222** 153-62
[216] Bonnell D A, Kalinin S V, Kholkin A L and Gruverman A 2009 *MRS Bulletin* **34** 648-57
[217] Gruverman A and Kholkin A 2006 *Rep. Prog. Phys.* **69** 2443-74
[218] Gruverman A and Kalinin S V 2006 *J. Mater. Sci.* **41** 107-16
[219] Kalinin S V, Rar A and Jesse S 2006 *IEEE T. Ultrason. Ferr.* **53** 2226-52
[220] Balke N, Bdikin I, Kalinin S V and Kholkin A L 2009 *J. Am. Ceram. Soc.* **92** 1629-47
[221] Kalinin S V, Morozovska A N, Chen L Q and Rodriguez B J 2010 *Rep. Prog. Phys.* **73** 056502
[222] Kalinin S V, Setter N and Kholkin A L 2009 *MRS Bulletin* **34** 634-42
[223] Soergel E 2011 *J. Phys. D-Appl. Phys.* **44** 464003
[224] Kalinin S V, Jesse S, Rodriguez B J, Seal K, Baddorf A P, Zhao T, Chu Y H, Ramesh R, Eliseev E A, Morozovska A N, Mirman B and Karapetian E 2007 *Jpn. J. Appl. Phys. Part 1 - Regul. Pap. Short Notes Rev. Pap.* **46** 5674-85
[225] Cunningham S, Larkin I A and Davis J H 1998 *Appl. Phys. Lett.* **73** 123-5
[226] Terris B D, Stern J E, Rugar D and Mamin H J 1989 *Phys. Rev. Lett.* **63** 2669-72
[227] Groten J, Zirkl M, Jakopic G, Leitner A and Stadlober B 2010 *Phys. Rev. B* **82** 054112
[228] Canet-Ferrer J, Martin-Carron L, Martinez-Pastor J, Valdes J L, Pena A, Carvajal J J and Diaz F 2007 *J. Microsc.-Oxford* **226** 133-9
[229] Levy J, Hubert C and Trivelli A 2000 *J. Chem. Phys.* **112** 7848-55
[230] Eng L M and Guntherodt H J 2000 *Ferroelectrics* **236** 35
[231] Sakai A, Sasaki N, Tamate T and Ninomiya T 2003 *Ferroelectrics* **284** 189-93
[232] Odagawa H and Cho Y 2000 *Surf. Sci.* **463** L621-L5
[233] Odagawa H and Cho Y 2000 *Jpn. J. Appl. Phys. Part 1 - Regul. Pap. Short Notes Rev. Pap.* **39** 5719-22
[234] Cho Y S 2011 *J. Mater. Res.* **26** 2007-16
[235] Tanaka K, Kurihashi Y, Uda T, Daimon Y, Odagawa N, Hirose R, Hiranaga Y and Cho Y 2008 *Jpn. J. Appl. Phys.* **47** 3311-25
[236] Sugimura H, Ishida Y, Hayashi K, Takai O and Nakagiri N 2002 *Appl. Phys. Lett.* **80** 1459-61
[237] Fujihira M 1999 *Annu. Rev. Mater. Sci.* **29** 353-80
[238] Melitz W, Shen J, Kummel A C and Lee S 2011 *Surf. Sci. Rep.* **66** 1-27
[239] Baumgart C, Helm M and Schmidt H 2009 *Phys. Rev. B* **80** 085305
[240] Jacobs H O, Knapp H F and Stemmer A 1999 *Rev. Sci. Instrum.* **70** 1756-60
[241] Charrier D S H, Kemerink M, Smalbrugge B E, de Vries T and Janssen R A J 2008 *ACS Nano* **2** 622-6
[242] Sadewasser S, Glatzel T, Shikler R, Rosenwaks Y and Lux-Steiner M C 2003 *Appl. Surf. Sci.* **210** 32-6
[243] Blumel A, Plank H, Klug A, Fisslthaler E, Sezen M, Grogger W and List E J W 2010 *Rev. Sci. Instrum.* **81** 056107

[244] Kalinin S V and Bonnell D A 2001 *Phys. Rev. B* **63** 125411
[245] Kalinin S V and Bonnell D A 2000 *Phys. Rev. B* **62** 10419-30
[246] Wu Y and Shannon M A 2006 *Rev. Sci. Instrum.* **77** 043711
[247] Efimov A and Cohen S R 2000 *J. Vac. Sci. Technol. A-Vac. Surf. Films* **18** 1051-5
[248] Strassburg E, Boag A and Rosenwaks Y 2005 *Rev. Sci. Instrum.* **76** 083705
[249] Jesse S, Guo S, Kumar A, Rodriguez B J, Proksch R and Kalinin S V 2010 *Nanotechnology* **21** 405703
[250] Sadewasser S, Jelinek P, Fang C K, Custance O, Yamada Y, Sugimoto Y, Abe M and Morita S 2009 *Phys. Rev. Lett.* **103** 266103
[251] Bocquet F, Nony L, Loppacher C and Glatzel T 2008 *Phys. Rev. B* **78** 035410
[252] Barth C, Foster A S, Henry C R and Shluger A L 2011 *Adv. Mater.* **23** 477-501
[253] Okamoto K, Sugawara Y and Morita S 2003 *Jpn. J. Appl. Phys. Part 1 - Regul. Pap. Short Notes Rev. Pap.* **42** 7163-8
[254] Lee W, Lee M, Kim Y B and Prinz F B 2009 *Nanotechnology* **20** 445706
[255] Glatzel T, Marron D F, Schedel-Niedrig T, Sadewasser S and Lux-Steiner M C 2002 *Appl. Phys. Lett.* **81** 2017-9
[256] Visoly-Fisher I, Cohen S R, Gartsman K, Ruzin A and Cahen D 2006 *Adv. Funct. Mater.* **16** 649-60
[257] Coffey D C and Ginger D S 2006 *Nat. Mater.* **5** 735-40
[258] Palermo V, Palma M and Samori P 2006 *Adv. Mater.* **18** 145-64
[259] Ellison D J, Lee B, Podzorov V and Frisbie C D 2011 *Adv. Mater.* **23** 502-7
[260] Hoppe H, Glatzel T, Niggemann M, Hinsch A, Lux-Steiner M C and Sariciftci N S 2005 *Nano Lett.* **5** 269-74
[261] Kalinin S V, Shin J, Jesse S, Geohegan D, Baddorf A P, Lilach Y, Moskovits M and Kolmakov A 2005 *J. Appl. Phys.* **98** 044503
[262] Schujman S B, Vajtai R, Biswas S, Dewhirst B, Schowalter L J and Ajayan P 2002 *Appl. Phys. Lett.* **81** 541-3
[263] Woodside M T and McEuen P L 2002 *Science* **296** 1098-101
[264] McCormick K L, Woodside M T, Huang M, Wu M S, McEuen P L, Duruoz C and Harris J S 1999 *Phys. Rev. B* **59** 4654-7
[265] Miyato Y, Kobayashi K, Matsushige K and Yamada H 2005 *Jpn. J. Appl. Phys. Part 1 - Regul. Pap. Short Notes Rev. Pap.* **44** 1633-6
[266] Huey B D and Bonnell D A 2000 *Appl. Phys. Lett.* **76** 1012-4
[267] Bonnell D A, Huey B and Carroll D 1995 *Solid State Ion.* **75** 35-42
[268] Kalinin S V, Suchomel M R, Davies P K and Bonnell D A 2002 *J. Am. Ceram. Soc.* **85** 3011-7
[269] Tanimoto M and Vatel O 1996 *J. Vac. Sci. Technol. B* **14** 1547-51
[270] Vatel O and Tanimoto M 1995 *J. Appl. Phys.* **77** 2358-62
[271] Chung S Y, Kim I D and Kang S J L 2004 *Nat. Mater.* **3** 774-8
[272] Schneider G A, Felten F and McMeeking R M 2003 *Acta Mater.* **51** 2235-41
[273] Kalinin S V and Bonnell D A 1999 *Z. Metallkde.* **90** 983-9
[274] Kalinin S V and Bonnell D A 2000 *Ferroelectric Thin Films VIII,* ed R W Schwartz *et al.* pp 327-32
[275] Shvebelman M M, Agronin A G, Urenski R P, Rosenwaks Y and Rosenman G I 2002 *Nano Lett.* **2** 455-8
[276] Kramer B 2001 *Advances in Solid State Physics 41* (Berlin: Springer-Verlag Berlin) pp 287-98
[277] Kim Y, Bae C, Ryu K, Ko H, Kim Y K, Hong S and Shin H 2009 *Appl. Phys. Lett.* **94** 032907
[278] Kalinin S V and Bonnell D A 2001 *Phys. Rev. B* **63** 125411
[279] Kalinin S V, Bonnell D A, Freitag M and Johnson A T 2002 *Appl. Phys. Lett.* **81** 754-6
[280] Kholkin A L, Bdikin I K, Shvartsman V V and Pertsev N A 2007 *Nanotechnology* **18** 095502
[281] Liu X Y, Kitamura K and Terabe K 2006 *Appl. Phys. Lett.* **89** 132905
[282] Choi S, Heo J, Kim D and Chung I S 2004 *Thin Solid Films* **464** 277-81

[283] Kim Y, Buhlmann S, Kim J, Park M, No K, Kim Y K and Hong S 2007 *Appl. Phys. Lett.* **91** 052906
[284] Kim Y, Park M, Buhlmann S, Hong S, Kim Y K, Ko H, Kim J and No K 2010 *J. Appl. Phys.* **107** 054103
[285] Chen X Q, Yamada H, Horiuchi T and Matsushige K 1999 *Jpn. J. Appl. Phys. Part 1 - Regul. Pap. Short Notes Rev. Pap.* **38** 3932-5
[286] Chen X Q, Yamada H, Horiuchi T, Matsushige K, Watanabe S, Kawai M and Weiss P S 1999 *J. Vac. Sci. Technol. B* **17** 1930-4
[287] Iwata M, Katsuraya K, Suzuki I, Maeda M, Yasuda N and Ishibashi Y 2005 *Mater. Sci. Eng. B-Solid State Mater. Adv. Technol.* **120** 88-90
[288] Balke N, Gajek M, Tagantsev A K, Martin L W, Chu Y H, Ramesh R and Kalinin S V 2010 *Adv. Funct. Mater.* **20** 3466-75
[289] Zou X, You L, Chen W G, Ding H, Wu D, Wu T, Chen L and Wang J 2012 *ACS Nano* **6** 8997-9004
[290] Sharma P, McQuaid R G P, McGilly L J, Gregg J M and Gruverman A 2013 *Adv. Mater.* **25** 1323-30
[291] McQuaid R G P, McGilly L J, Sharma P, Gruverman A and Gregg J M 2011 *Nat. Commun.* **2** 404
[292] Prim W S a R C 1953 *Phys. Rev.* **90** 753-8
[293] Strelcov E, Tselev A, Ivanov I, Budai J D, Zhang J, Tischler J Z, Kravchenko I, Kalinin S V and Kolmakov A 2012 *Nano Lett.* **12** 6198-205
[294] Strelcov E, Ievlev A V, Jesse S, Kravchenko I I, Shur V Y and Kalinin S V 2014 *Adv. Mater.* **26** 958-63
[295] Strelcov E, Jesse S, Huang Y L, Teng Y C, Kravchenko I I, Chu Y H and Kalinin S V 2013 *ACS Nano* **7** 6806-15
[296] Ding J L, Strelcov E, Kalinin S V and Bassiri-Gharb N 2015 *Nano Lett.* **15** 3669-76
[297] He D Y, Qiao L J and Volinsky A A 2011 *J. Appl. Phys.* **110** 074104
[298] He D Y, Qiao L J, Volinsky A A, Bai Y and Guo L Q 2011 *Phys. Rev. B* **84** 024101
[299] He D Y, Qiao L J, Volinsky A A, Bai Y, Wu M and Chu W Y 2011 *Appl. Phys. Lett.* **98** 062905
[300] Abplanalp M 2001 *PhD thesis* Swiss Federal Institute of Technology, Switzerland
[301] Hamazaki S I, Takahashi Y, Shimizu F and Takashige M 2001 *Ferroelectrics* **251** 101-8
[302] Likodimos V, Labardi M, Orlik X K, Pardi L, Allegrini M, Emonin S and Marti O 2001 *Phys. Rev. B* **63** 064104
[303] Likodimos V, Labardi M and Allegrini M 2002 *Phys. Rev. B* **66** 024104
[304] Kalinin S V and Bonnell D A 2001 *Appl. Phys. Lett.* **78** 1116-8
[305] Kalinin S V and Bonnell D A 2000 *J. Appl. Phys.* **87** 3950-7
[306] Kalinin S V, Johnson C Y and Bonnell D A 2002 *J. Appl. Phys.* **91** 3816-23
[307] Belaidi S, Girard P and Leveque G 1997 *J. Appl. Phys.* **81** 1023-30
[308] Kaku S, Eto H, Nakamura K and Watanabe Y 2009 *ISAF: 2009 18th IEEE International Symposium on the Applications of Ferroelectrics* (New York: IEEE) pp 29-32
[309] Watanabe Y 2011 *Ferroelectrics* **419** 28-32
[310] Watanabe Y, Kaku S, Matsumoto D, Urakami Y and Cheong S W 2009 *Ferroelectrics* **379** 381-91
[311] Fridkin V M 1980 *Ferroelectric semiconductors* (New York: Consultants Bureau)
[312] Debska M 2005 *J Electrostat* **63** 1017-23
[313] He D Y, Qiao L J, Volinsky A A, Bai Y, Wu M and Chu W Y 2011 *Appl. Phys. Lett.* **98** 062905
[314] Lu X M, Schlaphof F, Grafstrom S, Loppacher C, Eng L M, Suchaneck G and Gerlach G 2002 *Appl. Phys. Lett.* **81** 3215-7
[315] Fumagalli L, Gramse G, Esteban-Ferrer D, Edwards M A and Gomila G 2010 *Appl. Phys. Lett.* **96** 183107
[316] Gramse G, Casuso I, Toset J, Fumagalli L and Gomila G 2009 *Nanotechnology* **20** 395702

[317] Fumagalli L, Ferrari G, Sampietro M and Gomila G 2007 *Appl. Phys. Lett.* **91** 243110
[318] Ko H, Ryu K, Park H, Park C, Jeon D, Kim Y K, Jung J, Min D K, Kim Y, Lee H N, Park Y, Shin H and Hong S 2011 *Nano Lett.* **11** 1428-33
[319] Nicklaus M, Pignolet A, Harnagea C and Ruediger A 2011 *Appl. Phys. Lett.* **98** 162901
[320] Noy A, Vezenov D V and Lieber C M 1997 *Annu. Rev. Mater. Sci.* **27** 381-421
[321] Noy A, Frisbie C D, Rozsnyai L F, Wrighton M S and Lieber C M 1995 *J. Am. Chem. Soc.* **117** 7943-51
[322] Dahan D, Molotskii M, Rosenman G and Rosenwaks Y 2006 *Appl. Phys. Lett.* **89** 152902
[323] Shur V Y, Ievlev A V, Nikolaeva E V, Shishkin E I and Neradovskiy M M 2011 *J. Appl. Phys.* **110** 052017
[324] Spanier J E, Kolpak A M, Urban J J, Grinberg I, Lian O Y, Yun W S, Rappe A M and Park H 2006 *Nano Lett.* **6** 735-9
[325] Urban J J, Spanier J E, Lian O Y, Yun W S and Park H 2003 *Adv. Mater.* **15** 423-6
[326] Gruverman A and Tanaka M 2001 *J. Appl. Phys.* **89** 1836-43
[327] Johann F and Soergel E 2009 *Appl. Phys. Lett.* **95** 232906
[328] Lu H, Bark C W, de los Ojos D E, Alcala J, Eom C B, Catalan G and Gruverman A 2012 *Science* **336** 59-61
[329] Lu H, Kim D J, Bark C W, Ryu S, Eom C B, Tsymbal E Y and Gruverman A 2012 *Nano Lett.* **12** 6289-92
[330] Buh G H, Chung H J and Kuk Y 2001 *Appl. Phys. Lett.* **79** 2010-2
[331] Kim Y, Morozovska A N, Kumar A, Jesse S, Eliseev E A, Alibart F, Strukov D and Kalinin S V 2012 *ACS Nano* **6** 7026-33
[332] Schirmeisen A, Taskiran A, Fuchs H, Bracht H, Murugavel S and Roling B 2007 *Phys. Rev. Lett.* **98** 225901
[333] Nakamura M, Takekawa S, Liu Y W and Kitamura K 2009 *Jpn. J. Appl. Phys.* **48** 020214
[334] Korkishko Y N, Fedorov V A, Baranov E A, Alkaev A N, Morozova T V, Kostritskii S M and Laurell F 2001 *Ferroelectrics* **264** 1983-8
[335] Kim Y, Cho Y, Hong S, Buhlmann S, Park H, Min D K, Kim S H and No K 2006 *Appl. Phys. Lett.* **89** 172909
[336] Kim Y, Cho Y, Hong S, Buhlmann S, Park H, Min D K, Kim S H and No K 2006 *Appl. Phys. Lett.* **89** 162907
[337] Tybell T, Paruch P, Giamarchi T and Triscone J M 2002 *Phys. Rev. Lett.* **89** 097601
[338] Cho Y S, Hashimoto S, Odagawa N, Tanaka K and Hiranaga Y 2005 *Appl. Phys. Lett.* **87** 232907
[339] Tayebi N, Narui Y, Franklin N, Collier C P, Giapis K P, Nishi Y and Zhang Y G 2010 *Appl. Phys. Lett.* **96** 023103
[340] Molotskii M and Winebrand E 2005 *Phys. Rev. B* **71** 132103
[341] Morozovska A N and Eliseev E A 2006 *Physica B* **373** 54-63
[342] Durkan C, Welland M E, Chu D P and Migliorato P 1999 *Phys. Rev. B* **60** 16198-204
[343] Emelyanov A Y 2005 *Phys. Rev. B* **71** 132102
[344] Kalinin S V, Gruverman A, Rodriguez B J, Shin J, Baddorf A P, Karapetian E and Kachanov M 2005 *J. Appl. Phys.* **97** 074305
[345] Molotskii M I and Shvebelman M M 2005 *Philos. Mag.* **85** 1637-55
[346] Kalinin S V, Rodriguez B J, Jesse S, Chu Y H, Zhao T, Ramesh R, Choudhury S, Chen L Q, Eliseev E A and Morozovska A N 2007 *Proc. Natl. Acad. Sci. USA.* **104** 20204-9
[347] Chen L Q and Shen J 1998 *Comput. Phys. Commun.* **108** 147-58
[348] Son J Y, Bang S H and Cho J H 2003 *Appl. Phys. Lett.* **82** 3505-7
[349] Kim Y, Hong S, Kim S H and No K 2006 *J. Electroceram.* **17** 185-8
[350] Son J Y, Kyhm K and Cho J H 2006 *Appl. Phys. Lett.* **89** 092907
[351] Kim Y, Kim J, Buhlmann S, Hong S, Kim Y K, Kim S H and No K 2008 *Phys. Status Solidi-R* **2** 74-6

[352] Kim J, Kim Y, No K, Buhlmann S, Hong S, Nam Y W and Kim S H 2006 *Integ. Ferroelectr.* **85** 25-30
[353] Jang D M, Heo J, Yi I S and Chung I S 2002 *Jpn. J. Appl. Phys. Part 1 - Regul. Pap. Short Notes Rev. Pap.* **41** 6739-42
[354] Abplanalp M, Fousek J and Gunter P 2001 *Phys. Rev. Lett.* **86** 5799-802
[355] Buhlmann S, Colla E and Muralt P 2005 *Phys. Rev. B* **72** 214120
[356] Kim Y, Buhlmann S, Hong S, Kim S H and No K 2007 *Appl. Phys. Lett.* **90** 072910
[357] Brugere A, Gidon S and Gautier B 2011 *J. Appl. Phys.* **110** 024102
[358] Li Q A, Liu Y, Schiemer J, Smith P, Li Z R, Withers R L and Xu Z 2011 *Appl. Phys. Lett.* **98** 092908
[359] Kan Y, Bo H F, Lu X M, Cai W, Liu Y F and Zhu J S 2008 *Appl. Phys. Lett.* **92** 172910
[360] Lilienblum M and Soergel E 2011 *J. Appl. Phys.* **110** 052012
[361] Paruch P, Giamarchi T, Tybell T and Triscone J M 2006 *J. Appl. Phys.* **100** 051608
[362] Rodriguez B J, Nemanich R J, Kingon A, Gruverman A, Kalinin S V, Terabe K, Liu X Y and Kitamura K 2005 *Appl. Phys. Lett.* **86** 012906
[363] Kholkin A L, Bdikin I K, Shvartsman V V and Pertsev N A 2007 *Nanotechnology* **18** 095502
[364] Tong S, Jung I W, Choi Y Y, Hong S and Roelofs A 2016 *ACS Nano* **10** 2568-74
[365] Choi Y Y, Tong S, Ducharme S, Roelofs A and Hong S 2016 *Sci. Rep.* **6** 25087
[366] Hong S, Tong S, Park W I, Hiranaga Y, Cho Y S and Roelofs A 2014 *Natl. Acad. Sci. USA.* **111** 6566-9
[367] Tong S, Park W I, Choi Y Y, Stan L, Hong S and Roelofs A 2015 *Phys. Rev. Appl.* **3** 014003
[368] Lee K Y, Kim S K, Lee J H, Seol D, Gupta M K, Kim Y and Kim S W 2016 *Adv. Funct. Mater.* **26** 3067-73
[369] Ievlev A V, Morozovska A N, Eliseev E A, Shur V Y and Kalinin S V 2014 *Nat. Commun.* **5** 4545
[370] Shen J, Zeng H Z, Wang Z H, Lu S B, Huang H D and Liu J S 2006 *Appl. Surf. Sci.* **252** 8018-21
[371] Gleick J 2008 *Chaos: Making a New Science* (Penguin Books)
[372] Strogatz S H 2001 *Nonlinear Dynamics And Chaos: With Applications To Physics, Biology, Chemistry, And Engineering* (Westview Press)
[373] Agronin A, Molotskii M, Rosenwaks Y, Rosenman G, Rodriguez B J, Kingon A I and Gruverman A 2006 *J. Appl. Phys.* **99** 104102
[374] Morozovska A N, Svechnikov S V, Eliseev E A and Kalinin S V 2007 *Phys. Rev. B* **76** 054123
[375] Morozovska A N, Eliseev E A and Kalinin S V 2007 *J. Appl. Phys.* **102** 014109
[376] Morozovska A N, Ievlev A V, Obukhovskii V V, Fomichov Y, Varenyk O V, Shur V Y, Kalinin S V and Eliseev E A 2016 *Phys. Rev. B* **93** 165439
[377] Alikin D O, Ievlev A V, Turygin A P, Lobov A I, Kalinin S V and Shur V Y 2015 *Appl. Phys. Lett.* **106** 182902
[378] Altorfer H B a F 1994 *Acta Cryst.* **B50** 405-14
[379] S.C. Abrahams J M R, J.L. Bernstein 1966 *J. Phys. Chem. Solids* **27** 997-1012
[380] Yōichi Shiozaki T M 1963 *J. Phys. Chem. Solids* **24** 1057-61
[381] Ishibashi Y 1979 *J. Phys. Soc. Jpn.* **46** 1254-7
[382] Molotskii M, Agronin A, Urenski P, Shvebelman M, Rosenman G and Rosenwaks Y 2003 *Phys. Rev. Lett.* **90** 107601
[383] Kim Y, Vrejoiu I, Hesse D and Alexe M 2010 *Appl. Phys. Lett.* **96** 202902
[384] Garrity K, Kolpak A M, Ismail-Beigi S and Altman E I 2010 *Adv. Mater.* **22** 2969-73
[385] Zhao M H, Bonnell D A and Vohs J M 2008 *Surf. Sci.* **602** 2849-55
[386] Zhao M H, Bonnell D A and Vohs J M 2009 *Surf. Sci.* **603** 284-90
[387] Sturman V I B a B I 1980 *Sov. Phys. Usp.* **23** 199-223
[388] ForsberghJr P W 1956 *Encyclopedia of Physics* **4** 264-392
[389] Bolk T R, Grekov A A, Kosonogo.Na, Rodin A I and Fridkin V M 1971 *Sov. Phys, Cryst.* **16** 198

[390] Fridkin V M, Nitsche R, Korchagina N, Kosonogov N A, Magomadov R, Rodin A I and Verkhovskaya K A 1979 *Phys. Status Solidi A-Appl. Res.* **54** 231-7
[391] Warren W L and Dimos D 1994 *Appl. Phys. Lett.* **64** 866-8
[392] Kholkin A L, Iakovlev S O and Baptista J L 2001 *Appl. Phys. Lett.* **79** 2055-7
[393] Mamin R F 1997 *J. Exp. Theor. Phys.* **84** 808-13
[394] Gruverman A, Rodriguez B J, Nemanich R J and Kingon A I 2002 *J. Appl. Phys.* **92** 2734-9
[395] Shao R, Nikiforov M P and Bonnell D A 2006 *Appl. Phys. Lett.* **89** 112904
[396] Yang S Y, Seidel J, Byrnes S J, Shafer P, Yang C H, Rossell M D, Yu P, Chu Y H, Scott J F, Ager J W, Martin L W and Ramesh R 2010 *Nat. Nanotech.* **5** 143-7
[397] Choi T, Lee S, Choi Y J, Kiryukhin V and Cheong S W 2009 *Science* **324** 63-6
[398] Alexe M and Hesse D 2011 *Nat. Commun.* **2** 256
[399] Alexe M 2012 *Nano Lett.* **12** 2193-8
[400] Giocondi J L and Rohrer G S 2001 *J. Phys. Chem. B* **105** 8275-7
[401] Giocondi J L and Rohrer G S 2001 *Chem. Mater.* **13** 241-2
[402] Hanson J N, Rodriguez B J, Nemanich R J and Gruverman A 2006 *Nanotechnology* **17** 4946-9
[403] Liu X Y, Ohuchi F and Kitamura K 2008 *Funct. Mater. Lett.* **1** 177-82
[404] Dunn S, Cullen D, Abad-Garcia E, Bertoni C, Carter R, Howorth D and Whatmore R W 2004 *Appl. Phys. Lett.* **85** 3537-9
[405] Kalinin S V, Bonnell D A, Alvarez T, Lei X, Hu Z, Ferris J H, Zhang Q and Dunn S 2002 *Nano Lett.* **2** 589-93
[406] Haussmann A, Milde P, Erler C and Eng L M 2009 *Nano Lett.* **9** 763-8
[407] Liu X Y, Ohuchi F and Kitamura K 2008 *Funct. Mater. Lett.* **1** 177-82
[408] Seidel J, Maksymovych P, Batra Y, Katan A, Yang S Y, He Q, Baddorf A P, Kalinin S V, Yang C H, Yang J C, Chu Y H, Salje E K H, Wormeester H, Salmeron M and Ramesh R 2010 *Phys. Rev. Lett.* **105** 197603
[409] Wu W D, Guest J R, Horibe Y, Park S, Choi T, Cheong S W and Bode M 2010 *Phys. Rev. Lett.* **104** 217601
[410] Eliseev E A, Morozovska A N, Gu Y J, Borisevich A Y, Chen L Q, Gopalan V and Kalinin S V 2012 *Phys. Rev. B* **86** 085416
[411] Jo J Y, Chen P, Sichel R J, Baek S H, Smith R T, Balke N, Kalinin S V, Holt M V, Maser J, Evans-Lutterodt K, Eom C B and Evans P G 2011 *Nano Lett.* **11** 3080-4
[412] Kalinin S.V. S B G, Archibald R.K. 2015 *Nat. Mater.* **14** 973